\documentclass[onecolumn,nofootinbib,preprintnumbers, sort&compress, floatfix,a4paper,superscriptaddress,10pt]{revtex4-1}
\usepackage[framemethod=TikZ]{mdframed}
\usepackage[most]{tcolorbox}
\usepackage[utf8x]{inputenc}
\usepackage{longtable}
\usepackage{mathcomp}
\usepackage{pifont}
\usepackage{xcolor}
\usepackage{amsthm}
\usepackage{pgf}
\usepackage{pgfpages}
\usepackage{amsmath}
\usepackage{amsfonts}
\usepackage{amssymb}
\usepackage{graphicx}
\usepackage{bm}
\usepackage{hyperref}
\usepackage{lipsum}
\usepackage{array}
\usepackage{booktabs}
\usepackage{tikz}
\usepackage{titlesec}
\hypersetup{bookmarksnumbered}
\def\be{\begin{equation}}
\def\ee{\end{equation}}
\def\bi{\bibitem}
\usepackage{amsmath}
\usepackage{empheq}
\usepackage[most]{tcolorbox}
\newtcbox{\mymath}[1][]{%
    nobeforeafter, math upper, tcbox raise base,
    enhanced, colframe=blue!30!black,
    colback=blue!30, boxrule=1pt,
    #1}
\usepackage{mdframed}
\usepackage{lipsum}
\usepackage{PGF}
\usepackage{pgfkeys}
\usepackage{pgffor}
\usepackage{pgfcalendar}
\usepackage{pgfpages}
\usepackage{tikz}
\usetikzlibrary{calc}
\usepackage{array}
\usepackage{booktabs}
\usepackage{multirow}
\usepackage{latexsym}
\usepackage[english]{babel}
\allowdisplaybreaks
\vbox{\hbox{1000}}
\hypersetup{colorlinks=true,linkcolor=blue!100,citecolor=red!100}
\newcounter{theo}[section] 
\renewcommand{\thetheo}{\arabic{section}.\arabic{theo}}

\newcounter{prf}[section]
\renewcommand{\theprf}{\arabic{section}.\arabic{prf}}

\newcounter{lem}[section]
\renewcommand{\thelem}{\arabic{section}.\arabic{lem}}

\titleformat{\section}
{\color{black}\normalfont\Large\bfseries}
{\color{black}\thesection}{1em}{}
\renewcommand\thesection{\arabic{section}}

\newcommand*{\Rrmybox}[2]{\colorbox[rgb]{1.00,0.47,1.00}{\parbox{.99\linewidth}{#2}}}
\newcommand*{\hrmybox}[2]{\colorbox[rgb]{0.70,0.70,1.00}{\parbox{.99\linewidth}{#2}}}

\newcommand*{\pqrmybox}[2]{\colorbox[rgb]{0.00,0.98,0.25}{\parbox{.99\linewidth}{#2}}}

\allowdisplaybreaks
\usepackage{booktabs}% http://ctan.org/pkg/booktabs

\begin{document}
%\fontsize{10.9}{12}\selectfont\boldmath
\title{\textcolor[rgb]{0.00,0.40,0.80}{\textsf{\huge\boldmath
Phase space of multi-fluid universe in \(F(T)\)-gravity and some enhancements for the oscillating interaction model}}}
\author{‪Behzad Tajahmad}
\email{behzadtajahmad@yahoo.com}
\author{Hossein Motavalli}
\affiliation{Faculty of Physics, University of Tabriz, Tabriz, Iran}

%\affiliation{Research Institute for Astronomy and Astrophysics of Maragha (RIAAM)-Maragha, Iran, P.O. Box: 55134-441\\}
\begin{center}

\begin{abstract}
\begin{tcolorbox}[breakable,colback=white,
colframe=cyan,width=\dimexpr\textwidth+0mm\relax,enlarge left by=-17mm,enlarge right by=-6mm ]%\normalsize
\text{ }\\
\large{\textbf{\textsf{Abstract:}}
Recently, a Friedmann-Robertson-Walker universe filled with various cosmological fluids has been considered by S.D. Odintsov et al. in~\cite{od-asli} from phase space vantage point where various expressions for the Equation-of-State (EoS) parameter were studied. Since these types of EoS parameters are generative of appreciable results in the Hilbert-Einstein model, hence we intend to investigate all the cases in a homogeneous $F(T)$-gravity ($T$ is the torsion) through phase space analysis in precise detail.
In short, three viable models of interaction between dark matter and dark energy, including usual-type, power-law type, and oscillating type, are investigated comprehensively.
It is indicated that the power-law interaction in the related dynamical systems should be of increasing nature with time to get more critical points. Due to the failure of the oscillating model of ref.~\cite{od-asli} in $F(T)$-gravity, four modified models are suggested and examined in both $F(T)$ and Hilbert-Einstein models. As to be seen, the modified models not only are generative of critical points equivalent to that of ref.~\cite{od-asli}, but also give rise to further critical points covering crucial stages of the evolution of the universe. In the context of these four models, such as the old one, at early times the interactions are negligible and they commence to grow as the cosmic time approaches the late-time in which the unification of early inflation and late acceleration is obtained. Using an indirect method, it is shown that the oscillating models have substantial roles in transitions between eras.}
\end{tcolorbox}
\end{abstract}
\maketitle
\end{center}
%\begin{center}
%\begin{tcolorbox}[breakable,colback=white,
%colframe=cyan,width=\dimexpr\textwidth+0mm\relax,enlarge left by=-17mm,enlarge right by=-6mm ]%\normalsize
%\text{ }\\
%\large{\textbf{\textsf{CRediT authorship contribution statement:}}\\
%\textbf{B. Tajahmad:} Definer of work; Conceptualization and ideas; Calculations; Formal analysis; Software; Methodology; Writing - original draft; Writing - review \& editing.\\
%\textbf{H. Motavalli:} Supervision.\\ %Supervisor of the thesis.\\
%\textit{\textbf{I, B. Tajahmd, declare that if there is a problem in this work, it returns to me. Since it is law in Iran that the name of the supervisor of the thesis must be added to the paper, hence the name of H. Motavalli has been added to this paper while he had no effect on this work.}}
%}
%\end{tcolorbox}
%\end{center}
%\break
\hrule \hrule \hrule
\textbf{\textcolor[rgb]{0.00,0.00,0.00}{\tableofcontents}}
\text{ }
\hrule \hrule \hrule
%\break
%\noindent\hrmybox{}{\section{Introduction\label{sec:intro}}}\vspace{5mm}
%\noindent\hrmybox{}{\section{The model\label{2.2.2}}}\vspace{5mm}
%\noindent\rmybox{}{\subsection{ \label{}}}\vspace{5mm}
%\noindent\pqrmybox{red}{\textbf{$\bullet$} \textbf{CSSS-Trick (Combination of Sub-symmetries through Special Selections):}}\\
%%%%%%%%%%%%%%%%%%%%%%%%%%%%%%%%%%%%%%%%%%%%%%%%%%%%%%%%%%%
%%%%%%%%%%%%%%%%%%%%%%%%%%%%%%%%%%%%%%%%%%%%%%%%%%%%%%%%%%%
%%%%%%%%%%%%%%%%%%%%%%%%%%%%%%%%%%%%%%%%%%%%%%%%%%%%%%%%%%%
%\noindent\rmybox{}{\section{Introduction\label{sec:intro}}}\vspace{5mm}
%\noindent\Frmybox{}{\section{Introduction\label{sec:intro}}}\vspace{5mm}
%\noindent\qrmybox{}{\section{Introduction\label{sec:intro}}}\vspace{5mm}
%\noindent\pqrmybox{}{\section{Introduction\label{sec:intro}}}\vspace{5mm}
%\noindent\goldmybox{}{\section{Introduction\label{sec:intro}}}\vspace{5mm}
%\noindent\rgoldmybox{}{\section{Introduction\label{sec:intro}}}\vspace{5mm}
%\noindent\Rrmybox{}{\section{Introduction\label{sec:intro}}}\vspace{5mm}
%\noindent\hrmybox{}{\section{Introduction\label{sec:intro}}}\vspace{5mm}
%\noindent\rmybox{}{\section{Introduction\label{sec:intro}}}\vspace{5mm}
%%%%%%%%%%%%%%%%%%%%%%%%%%%%%%%%%%%%%%%%%%%%%%%%%%%%%%%%%%%%%%%%%%%%%%%%
%%%%%%%%%%%%%%%%%%%%%%%%%%%%%%%%%%%%%%%%%%%%%%%%%%%%%%%%%%%%%%%%%%%%%%%%%
%%%%%%%%%%%%%%%%%%%%%%%%%%%%%%%%%%%%%%%%%%%%%%%%%%%%%%%%%%%%%%%%%%%%%%%%%
\text{ }\\
\text{ }\\
%\pagebreak
\noindent\hrmybox{}{\section{Introduction\label{sec:intro}}}\vspace{5mm}
According to several observational data including supernova type Ia~\cite{km2,km1}, weak lensing~\cite{km4}, CMB studies~\cite{km3}, large-scale structure~\cite{km6}, and baryon acoustic oscillations\cite{km5}, the current status of the expansion of the universe is on the accelerating phase. The explanation of this acceleration is one of the major challenges for cosmologists because it is inconsistent with the standard Einstein's general relativity. In order to explain the late-time-accelerated expansion two main classes of approaches have been suggested up to now. In the first approach, the universe is assumed to be made up of an exotic liquid, the so-called `dark energy' which comprises $70\%$ of the universe.
Einstein's cosmological constant was historically the first model of the accelerated expansion of the Universe. To avoid its `fine-tuning' and `cosmic coincidence' problems~\cite{beh-l-7} and unclear physical nature the alternative models were proposed~\cite{rref1}, called as ’dark energy’ models:
the tachyon field~\cite{beh-l-21}, quintessence~\cite{beh-l-18,beh-l-19}, quintom~\cite{beh-l-14,beh-l-17}, phantom field~\cite{beh-l-10,beh-l-13}, etc. The second approach is to modify Einstein's general relativity by making the action of the theory dependent upon a function of the curvature scalar $R$. As one expects, the theory reduces to general relativity in certain limits of the parameters. Besides $f(R)$-gravity, various theories such as scalar-tensor theories, $F(T)$-gravity ($T$ is the torsion), $F(T)$-gravity with an unusual term~\cite{beh-l-25}, and etcetera have recently been proposed with fruitful results.\\

As is known, there are several tools for studying cosmological models amongst which the Noether symmetry approach, reconstruction methods, and phase space analysis are mentionable.\\
In the Noether symmetry approach, one finds the exact solutions of the model, which carry some conserved currents. Besides conserved currents, presenting the form of unknown functions is the most important feature of this tool. Indeed, the Noether symmetry approach is a powerful standard way for acquiring the forms of unknown functions of alternative theories of gravity. It is important to mention that some hidden conserved currents may not be obtained by this approach~\cite{beh-l-70,beh-l-71}. Carrying more conserved currents as well as new forms of unknown functions are the main motivations for enhancing the method to the `Beyond Noether Symmetry approach' (B.N.S.-approach)~\cite{beh-bns} and `Combination of Sub-symmetries through Special Selections' (CSSS-approach)~\cite{beh-csss} that have recently been proposed with noteworthy outcomes.\\
There are several ways to reconstruct a model of alternative theory such as reconstruction by parameterizing the Hubble parameter in terms of the redshift~\cite{beh-j-62}, reconstruction using the scalar field~\cite{beh-j-77,beh-j}, reconstruction using the Raychaudhuri equation~\cite{beh-e-1,beh-e}, and etcetera. Some new collections of the forms of unknown functions may also be achieved through the reconstruction method especially when one does the reconstruction through the scalar field. A general prescription to reconstruction based on the scalar field has been suggested in ref.~\cite{beh-j} using the approach proposed in ref.~\cite{beh-j-77}. A general powerful approach, the `Boundary function method' ($\mathfrak{B}$-function method), to analyze the results of this approach has recently been suggested~\cite{beh-annals}. The $\mathfrak{B}$-function method can be applied to phase space analysis and exact solutions of alternative theories of gravity. It also can be regarded as a new reconstruction method.\\
The phase space analysis is also a powerful tool for investigating a model. In this approach, the equations of our system are converted into first-order differential equations through defining some dimensionless variables. Although this process could be challenging in some cases, several approaches have been proposed for those challenges. In phase space analysis, we find Critical Points (CPs) of a model along with its eigenvalues and possibly some further helpful parameters. Indeed, one can use these CPs to analyze the model. In general, the nature of a CP can be attractor, saddle, repeller, and null according to its eigenvalues. Needless to say, this approach has attracted cosmologists to take it as a prospective in analyzing gravitational theories.\\

The phase space of a Friedmann-Robertson-Walker universe which is filled with various
cosmological fluids (with and without interactions) has recently been considered in detail in ref.~\cite{od-asli}. Several forms of the equation of state parameter, especially arisen from viscous fluid and a new class of interaction between dark energy and dark matter namely cosine type, have been studied in ref.~\cite{od-asli} successfully. In this paper, we examine all the EoS suggested in~\cite{od-asli} and some further improved interaction types in an \(F(T)\)-model in detail via phase space analysis.\\

The paper is organized as follows: The model is presented in sect.~\ref{sect-0}. In sect.~\ref{sect-II}, some well-known features of the dynamical system approach in cosmological context with generalized fluids suggested in~\cite{od-asli}, are considered. In sect.~\ref{sect-III}, we add dark matter, which interacts with dark energy candidate, to our assumed components. The generalized formalism of sect.~\ref{sect-III} is carefully considered in sect.~\ref{sect-IV} in two main modes. Section~\ref{sect-V} is devoted to studying some interacting models between dark matter and dark energy. Furthermore, some enhancements to cosine-type of interaction are suggested and examined successfully. Finally, the conclusions are provided at the end of the paper.\\

\noindent\hrmybox{}{\section{The model\label{sect-0}}}\vspace{5mm}
Let us start with a \(F(T)\) gravitational action of the form~\cite{faraoni}
\begin{equation}\label{action}
S=\frac{1}{2 \kappa^{2}} \int d^{4}x \; e \; \left(T+F(T)+L_{M} \right),
\end{equation}
where $e=\mathrm{det}(e^{i}_{\nu})=\sqrt{-g}$ with $e^{i}_{\nu}$ being a vierbein (tetrad) basis, $\kappa^{2}=8\pi G$, $T$ is the scalar torsion, and $L_{M}$ is the matter Lagrangian.\\
Adopting flat FRW metric with the signature $(+,-,-,-)$, the Friedmann equations for (\ref{action}) would be
\begin{align}
\frac{3}{\kappa^{2}}H^{2} &=\frac{1}{1+2F_{,T}} \left(\rho -\frac{F}{2\kappa^{2}} \right), \label{Fri1}\\
\frac{-2}{\kappa^{2}} \dot{H} &=\frac{\rho+p}{1+F_{,T}+2TF_{,TT}}, \label{Fri2}
\end{align}
where $\rho$ and $p$ are the total density and pressure, respectively, $H$ is the Hubble parameter, the dot denotes differentiation with respect to the cosmic time, $F_{,T}=\mathrm{d}F/\mathrm{d}T$, and $F_{,TT}=\mathrm{d}^{2}F/\mathrm{d}T^{2}$.
These equations describe the dynamical evolution of the model. The total energy conservation equation for the flat FRW universe, namely
\begin{equation}\label{conservation}
\dot{\rho}+3H(\rho+p)=0,
\end{equation}
is taken to be true for different groups of components that are studied in this paper. However, $F(T)$ is an unknown function of the scalar torsion, but in order to avoid computation problems, let us proceed with a well-known special form as~\cite{mainref} $F(T)=c_{1}+2c_{2}\sqrt{-T}+\alpha T$, where $c_{1}$, $c_{2}$, and $\alpha$ are arbitrary constants. Throughout this paper, we assume that $\alpha \neq -1$, for the left-hand-side of Einstein's equations cannot be produced when one takes $\alpha =-1$.
In $F(T)$-gravity, the constant and the non-linear terms have correspondence with the cosmological constant equation of state~\cite{beh-u-1}. Furthermore, as it has been discussed in ref.~\cite{beh-u-2}, the power-law form of $F(T)$ (i.e. $F(T) \sim T^{n} (n>1)$) such as $T^{2}$ may remove the finite-time future singularity. When $n=0$, $F(T)$ may be regarded as a cosmological constant. The case $n=1/2$ is helpful in understanding power-law inflation and also in elucidating the little-rip and pseudo-rip cosmology~\cite{beh-u-2}, hence making our form more appealing than other options.\\

\noindent\hrmybox{}{\section{Standard approach on dynamical systems and cosmological dynamics\label{sect-II}}}\vspace{5mm}
In this section, a simple case in which the universe is filled with radiation, $\rho_{\mathrm{r}}$, and a perfect fluid with a non-trivial EoS of the form\footnote{The subscript ``$\mathrm{v}$'' refers to viscous type of the fluid.} $p_{\mathrm{v}}=-\rho_{\mathrm{v}}+f(\rho_{\mathrm{v}})+G(H)$~\cite{od-50}, is considered. These types of non-trivial EoS parameters may be deemed as some sort of viscous fluid or generalized EoS fluid~\cite{od-51} as well as an effective fluid presentation of some modified gravity theory~\cite{od-3,od-4}. The motivation of the appearance of the term $G(H)$ which might seem unconventional, has clearly been described in ref.~\cite{od-asli}. \\
The corresponding continuity equations are as follows:
\begin{align}
\rho^{\prime}_{\mathrm{r}} &=-4\rho_{\mathrm{r}}, \label{conservation01}\\
\rho^{\prime}_{\mathrm{v}} &=-3 \left[f+G \right],\label{conservation02}
\end{align}
where the prime denotes differentiation with respect to the e-folding number $N=\ln(a)$ in which $a$ is the scale factor of the FRW universe.
In order to rewrite the cosmological equations in terms of dimensionless variables, we define dimensionless density parameters as
\begin{equation}\label{dim1}
x \equiv \frac{\kappa^{2} \rho_{\mathrm{v}}}{3H^{2}}, \quad y \equiv \frac{\kappa^{2} \rho_{\mathrm{r}}}{3H^{2}}.
\end{equation}
Therefore, dynamical equations turn out to be
\begin{align}
x^{\prime}=&\frac{3x}{1+\alpha}\left(\frac{f+G}{\rho_{\mathrm{v}}}x+\frac{4}{3}y\right)
-\frac{3(f+G)}{\rho_{\mathrm{v}}}x,\label{dse01}\\
y^{\prime}=&\frac{3y}{1+\alpha}\left(\frac{f+G}{\rho_{\mathrm{v}}}x+\frac{4}{3}y\right)
-4y.\label{dse02}
\end{align}
Like ref~\cite{od-asli}, a simple case of EoS is considered in this section. To this end, we assume $f=\rho_{\mathrm{v}}(1+w_{0})$ and $G=w_{1}H^{2}$. Therefore, equations (\ref{dse01})--(\ref{dse02}) take the following forms:
\begin{align}
x^{\prime} =& \frac{3x}{1+\alpha}\left(x+w_{0}x+\frac{4}{3}y+\frac{\kappa^{2}}{3}w_{1} \right)  -3(1+w_{0})x-\kappa^{2}w_{1}, \label{dse03}\\
y^{\prime} =& \frac{3y}{1+\alpha}\left(x+w_{0}x+\frac{4}{3}y+\frac{\kappa^{2}}{3}w_{1} \right)-4y. \label{dse04}
\end{align}
Equating the left-hand sides of (\ref{dse03})--(\ref{dse04}) to zero, three critical points are found:
\begin{align}
&\text{CP1:} \; x_{1}=1+\alpha, \quad y_{1}=0; \label{cp1}\\
&\text{CP2:} \; x_{2}= \frac{-\kappa^{2} w_{1}}{3(w_{0}+1)}, \quad y_{2}=0;\label{cp2}\\
&\text{CP3:} \; x_{3}=\frac{-\kappa^{2}w_{1}}{3w_{0}-1},\quad
y_{3}=\frac{\kappa^{2}w_{1}+3\alpha w_{0}+3w_{0}-\alpha -1}{3w_{0}-1}.\label{cp3}
\end{align}
$\blacktriangledown$ \textbf{CP1:}\\
CP1 is in the physical region when $\alpha > -1$. It corresponds to a universe filled with a viscous fluid only (i.e. without radiation).\\
$\blacktriangledown$ \textbf{CP2:}\\
If one sets $(w_{1}/(1+w_{0}))\leq 0$, then CP2 will be in the physical era. There is a singularity at $w_{0}=-1$ besides $w_{1}=\infty=w_{0}$. Apart from the special case $w_{1}=0$, that refers to the vacuum case, for $w_{1} \neq 0$ this point such as CP1 shows a universe filled with a viscous fluid without radiation.\\
$\blacktriangledown$ \textbf{CP3:}\\
If the condition $(w_{1}/(3w_{0}-1)) \leq 0$ is adopted, then according to the relation $y_{3}=-x_{3}+1+\alpha$, it is sufficient to take $\alpha \geq -(1-x_{3})$ for putting CP3 in the physical region. The sum of the densities of radiation and viscous fluid is controlled by the value of $\alpha$.
There is a singularity at $w_{0}=1/3$. The special cases $w_{1}=0$ and $\{x_{3}=1+\alpha \; \& \; \alpha \neq -1\}$ correspond to pure radiation and pure viscous cases, respectively. And when $\{w_{1}=0 \; \& \; w_{0}=1/3 \}$, the background of the model will be the vacuum. $\blacktriangle$\\

The eigenvalues of the Jacobian matrix for these CPs are as follows:
\begin{align}%
& \text{CP1:}\; \left\{
         \begin{array}{ll}
           \lambda_{1-cp1} &=\gamma_{0}+3w_{0}-1; \\
           \lambda_{2-cp1} &=\gamma_{0}+3w_{0}+3,
         \end{array}
       \right.\\
& \text{CP2:}\; \left\{
         \begin{array}{ll}
           \lambda_{1-cp2} &=-4; \\
           \lambda_{2-cp2} &=-\gamma_{0}-3(1+w_{0}),
         \end{array}
       \right.\\
& \text{CP3:}\; \left\{
         \begin{array}{ll}
           \lambda_{1-cp3} &=-\gamma_{0}-3w_{0}+1; \\
           \lambda_{2-cp3} &=+4,
         \end{array}
       \right.
\end{align}
where $\gamma_{0}=\kappa^{2} w_{1}/(1+\alpha)$.\\
$\blacktriangledown$ \textbf{CP1:}\\
Because $\lambda_{2-cp1}=\lambda_{1-cp1}+4$, hence CP1 can be stable or unstable depending upon the values of constant parameters:
\begin{equation*}
\left\{
\begin{array}{ll}
\text{if}\;\; \lambda_{1-cp1}< -4 \;\;&\Longrightarrow \;\;  \hbox{CP1 is an attractor point;} \\
\text{if}\;\; -4< \lambda_{1-cp1}< 0 \;\;&\Longrightarrow \;\;  \hbox{CP1 is a saddle point;} \\
\text{if}\;\; \lambda_{1-cp1} >0 \;\;&\Longrightarrow \;\; \hbox{CP1 is a repeller point.}
  \end{array}
\right.
\end{equation*}
\text{ }\\
$\blacktriangledown$ \textbf{CP2:}\\
CP2 is an attractor along $y$-direction, while at the next direction it can be stable or unstable based on the chosen values of parameters.\\
Pursuant to the condition of staying CP2 in the physical region, $w_{0}$ and $w_{1}$ must belong to one of the following sets:\\
\textbf{Set-1:} $\{w_{1}<0, \; \& \; w_{0}>-1\}$;\\
\textbf{Set-2:} $\{w_{1}>0, \; \& \; w_{0}<-1\}$.\\
In the case that $w_{0}$ and $w_{1}$ belong to `Set-1', if $\frac{\kappa^{2} |w_{1}|}{1+\alpha}<3(1+w_{0})$ is satisfied, then CP2 will be an attractor otherwise it will be a saddle point. In the case that $w_{0}$ and $w_{1}$ belong to `Set-2', if $\frac{\kappa^{2} |w_{1}|}{1+\alpha}>3|(1+w_{0})|$ is satisfied, then CP2 will be an attractor, otherwise, it will be a saddle point.\\
$\blacktriangledown$ \textbf{CP3:}\\
In general, CP3 is called unstable. More precisely, along $x$-direction, it has stability while along $y$-direction its nature depends upon the amounts of constant parameters.\\
According to the physical conditions of CP3, besides $0\leq x_{3} \leq (1+\alpha)$, $w_{0}$ and $w_{1}$ must belong to one of the following sets:\\
\textbf{Set-3:} $\{w_{1}< 0, \; \& \; w_{0}>1/3 \}$;\\
\textbf{Set-4:} $\{w_{1} > 0, \; \& \; w_{0}<1/3 \}$.\\
If $w_{0}$ and $w_{1}$ belong to `Set-3', then CP3 will be a saddle point, while if they belong to `Set-4', it will be a completely unstable point (i.e. all trajectories will be repelled).\\

\begin{figure*}
	\includegraphics[width=6.5 in]{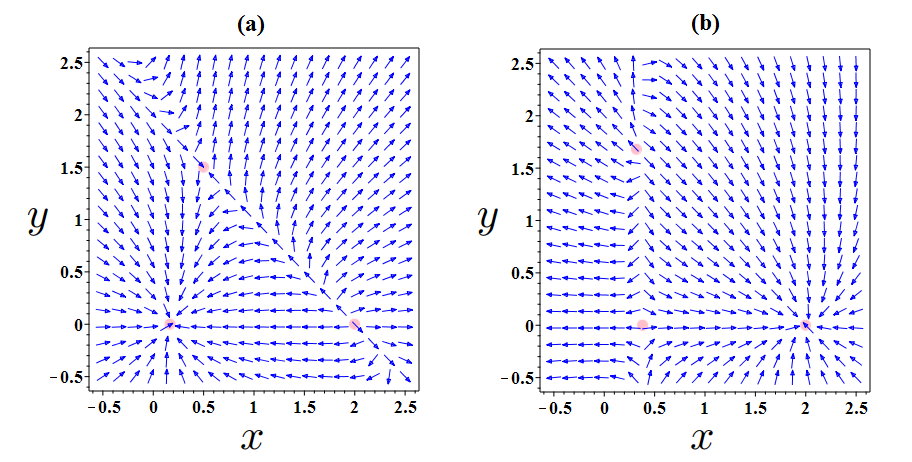}
    \caption{This figure illustrates two different phase-portraits of the system~(\ref{dse03})--(\ref{dse04}). Plot `a' indicates the phase diagram when the values of the constant parameters are selected as: $\kappa^{2}=1$, $w_{0}=+1$, $w_{1}=-1$, and $\alpha=+1$ (i.e. the chosen values belong to `Set-1' and `Set-3'), while plot `b' shows the phase plane when we pick up the values of constant parameters from `Set-2' and `Set-4' as $w_{0}=-8$, $w_{1}=+8$, $\alpha=+1$, and $\kappa^{2}=1$. The pink points refer to the critical points.}
    \label{fig1}
\end{figure*}

As an example, two phase portraits have been illustrated in fig.~\ref{fig1}. The above-mentioned explanations are clear in this figure.\\

\noindent\hrmybox{}{\section{Models with dark matter interacting to dark energy \label{sect-III}}}\vspace{5mm}
In this section, beside the radiation, $\rho_{\mathrm{r}}$, and a viscous fluid, $\rho_{\mathrm{v}}$, the dark matter, $\rho_{\mathrm{dm}}$, is also added so that it interacts with dark energy via a usual\footnote{This type of interaction is the simplest candidate that one can assume.} interaction model. The dynamical system for this case may be taken as
\begin{align}
\rho_{\mathrm{tot.}} &=\rho_{\mathrm{r}}+\rho_{\mathrm{v}}+\rho_{\mathrm{dm}}, \label{3eq1}\\
\rho^{\prime}_{\mathrm{r}} &=-4\rho_{\mathrm{r}}, \label{3eq2} \\
\rho^{\prime}_{\mathrm{v}} &=-3 \left[f+G \right]-\frac{Q}{H}, \label{3eq3} \\
\rho^{\prime}_{\mathrm{dm}} &=-3 \rho_{\mathrm{dm}}+\frac{Q}{H}-3(-3H\zeta), \label{3eq4}
\end{align}
where $Q$ is the coupling parameter that describes the interaction and $-3H\zeta$ corresponds to the bulk viscous pressure of the dark-matter fluid. Let us assume that the bulk viscous coefficient is of the form~\cite{od-17,od-54,od-55}
\begin{align}
\zeta=\frac{\zeta_{0}}{\sqrt{3\kappa^{2}}}\sqrt{\rho_{\mathrm{tot.}}}
=\frac{1}{\kappa^{2}}H\zeta_{0}.
\end{align}
The relation between $\zeta$ and total density is estimated astronomically; see ref.~\cite{od-56}. However, the relation which we use here is $\zeta \sim \sqrt{\rho_{\mathrm{tot.}}}$ instead of the general relation $\zeta \sim \rho^{\lambda}_{\mathrm{tot.}}$, which is due to the fact that this ansatz yields a good fit between astronomical constraints and the fluid approach~\cite{od-asli}.\\
The dimensionless variables for this case are introduced as follows:
\begin{align}
x \equiv \frac{\kappa^{2} \rho_{\mathrm{v}}}{3H^{2}}, \;\; y \equiv \frac{\kappa^{2} \rho_{\mathrm{dm}}}{3H^{2}}, \;\; z \equiv \frac{\kappa^{2} \rho_{\mathrm{r}}}{3H^{2}}, \;\;
q \equiv \frac{\kappa^{2} Q}{3H^{3}}. \label{3eq5}
\end{align}
Therefore, by taking the previous EoS, $f(\rho_\mathrm{v})=\rho_\mathrm{v}(1+w_{0})$ and $G(H)=w_{1}H^{2}$, (\ref{3eq1})--(\ref{3eq4}) in terms of the independent dynamical variables turn out to be
\begin{align}
x^{\prime}=& \left(\frac{3x}{1+\alpha} \right) \left[(1+w_{0})x+y+\frac{4}{3}z+\frac{\kappa^{2}}{3}w_{1} \right] \nonumber \\
&-3(1+w_{0})x-w_{1}\kappa^{2}-q, \label{3eq6} \\
y^{\prime}=& \left(\frac{3y}{1+\alpha} \right) \left[(1+w_{0})x+y+\frac{4}{3}z+\frac{\kappa^{2}}{3}w_{1} \right] \nonumber \\
&-3y+q+3\zeta_{0}, \label{3eq7} \\
z^{\prime}=& \left(\frac{3z}{1+\alpha} \right) \left[(1+w_{0})x+y+\frac{4}{3}z+\frac{\kappa^{2}}{3}w_{1} \right]-4z. \label{3eq8}
\end{align}
Although this system is analytically solvable, the results are unpresentable because they are very long excluding one CP. Thanks to the abundance of such solutions, let us use the acronym `\textbf{VLT}' for such \textbf{V}ery \textbf{L}ong \textbf{T}erms, for the sake of brevity. It means that the related term is very long so that it is somewhat unpresentable, or even if it is, it does not lead to any clear outcome.\\
For the above system, we may present only one CP as:
\begin{align}\label{01cpIII}
&x=\frac{q+w_{1}\kappa^{2}}{1-3w_{0}}, \quad y=-q-3\zeta_{0}, \nonumber \\
&z=\frac{(12q+12\alpha+27\zeta_{0}+12)w_{0}+4w_{1}\kappa^{2}
-4\alpha-9\zeta_{0}-4}{4(3w_{0}-1)}.
\end{align}
Discussing this CP is not so interesting as the parameters cause confusion. Indeed, all the related eigenvalues are `VLT'. About other CPs, we can say that all of them have a common property: $z|_{\text{all other CPs}}=0$ which means that for the remained CPs, the density of radiation is zero and they are on the interaction plane, namely $x-y$ plane. In short, they are not on the radiation-dominated era. Therefore, the coordinates of these remained CPs can be used to reach some fixed relations between the densities of dark matter and dark energy.\\

To avoid `VLT'-type solutions, it seems that there are two approaches:\\
1. Assuming some relations among constant parameters;\\
2. Specifying the values of all or some constant parameters.\\
Examples clarifying these approaches will be presented in the next sections.\\

\noindent\hrmybox{}{\section{Generalized form of cosmological fluid\label{sect-IV}}}\vspace{5mm}
In order to include a generalized form of the equation-of-state, the case studied in Sect.~\ref{sect-II} is extended in this section. A universe filled with radiation and a perfect fluid with a non-trivial EoS, $p_{\mathrm{v}}=-\rho_{\mathrm{v}} +f(\rho_{\mathrm{v}})+G(H)$ \cite{od-50}, is considered here.
To convert to an autonomous system of first-order differential equations, the set of dimensionless variables are defined as
\begin{align*}
x \equiv \frac{\kappa^{2} \rho_{\mathrm{v}}}{3H^{2}}, \quad
y \equiv \frac{\kappa^{2} \rho_{\mathrm{r}}}{3H^{2}}, \quad
E \equiv \frac{1}{\kappa^{2}H^{2}},
\end{align*}
from which we obtain
\begin{align}
x^{\prime} &=\left(\frac{x}{1+\alpha} \right) \left[(f+G)\kappa^{4}E+4y \right]
-\left[f+G \right]\kappa^{4}E, \label{seciv-eq1}
\end{align}
\begin{align}
y^{\prime} &=\left(\frac{y}{1+\alpha} \right) \left[(f+G)\kappa^{4}E+4y \right]
-4y, \label{seciv-eq2}
\end{align}
\begin{align}
E^{\prime} &=\left(\frac{E}{1+\alpha} \right) \left[(f+G)\kappa^{4}E+4y \right]. \label{seciv-eq3}
\end{align}
According to  the definitions, one has $\rho_{\mathrm{v}}=3x /(\kappa^{4} E)$ and $H^{2}=1/ (\kappa^{2}E)$. Obviously, there are only two independent dimensionless variables, but introducing a further variable, $E$, leads to obtaining more information about the system. This trick was firstly proposed in ref.~\cite{od-57}.\\
It would be also helpful for interpreting the evolution of the system that we introduce the effective EoS. This parameter which is defined as\footnote{Here, $\mathfrak{B}[a,0;H(a)]$ and $\mathfrak{B}^{\star}[N,0;H(N)]$ are the $\mathfrak{B}$ and $\mathfrak{B}^{\star}$ functions (Boundary and Boundary-star functions), respectively; see ref.\cite{beh-annals}. All the phase space analysis of this paper can be performed utilizing the $\mathfrak{B}$-function method, but since this approach is new and the reader may be unfamiliar with, hence the analysis of this paper is presented using the usual analysis. The enthusiastic reader can easily convert all the dynamical equations of this paper in terms of the family of the $\mathfrak{B}$-functions (or $\mathfrak{B}^{\star}$-functions) and carry a precise analysis out.}
\begin{align}
w_{\mathrm{eff.}} \equiv -1-\frac{2 \dot{H}}{3 H^{2}}
=& -1-\frac{2}{3}\mathfrak{B}[a,0;H(a)] \nonumber \\
=& -1-\frac{2}{3}\mathfrak{B}^{\star}[N,0;H(N)],
\end{align}
takes the following form for this case of study:
\begin{align}\label{EoS-m}
w_{\mathrm{eff.}}=-1+\frac{1}{3}\left(\frac{1}{1+\alpha} \right) \left[(f+G)\kappa^{4}E+4y \right].
\end{align}

\noindent\Rrmybox{}{\subsection{A simplified form of EoS}}\vspace{5mm}
Let us first consider the simplest case of EoS namely $p_{\mathrm{v}}=w_{0}\rho_{\mathrm{v}}+w_{1}H^{2}$ implying $f(\rho_{\mathrm{v}})=\rho_{\mathrm{v}}(1+w_{0})$, and $G(H)=w_{1}H^{2}$. Therefore, the equations (\ref{seciv-eq1})--(\ref{seciv-eq3}) would be
\begin{align}
x^{\prime}=& \left(\frac{x}{1+\alpha} \right) \left[3(1+w_{0})x+w_{1}\kappa^{2}+4y \right] \nonumber\\ &-3(1+w_{0})x-w_{1}\kappa^{2}, \label{sim-eq-1}
\end{align}
\begin{align}
y^{\prime}=& \left(\frac{y}{1+\alpha} \right) \left[3(1+w_{0})x+w_{1}\kappa^{2}+4y \right]-4y, \label{sim-eq-2}
\end{align}
\begin{align}
E^{\prime}=& \left(\frac{E}{1+\alpha} \right) \left[3(1+w_{0})x+w_{1}\kappa^{2}+4y \right]. \label{sim-eq-3}
\end{align}
The EoS (\ref{EoS-m}) also turns out to be
\begin{align}\label{EoS-m-1}
w_{\mathrm{eff.}}=-1+\frac{1}{3}\left(\frac{1}{1+\alpha} \right)
\left[3(1+w_{0})x+w_{1}\kappa^{2}+4y \right].
\end{align}
For the above system, we obtain three critical points:
\begin{align}
\text{CP1: } &x=-\gamma_{1}, \quad
y=\gamma_{1}+\alpha +1, \quad E=0; \label{sim-cp-1}\\
\text{CP2: } &x=\frac{-\kappa^2 w_{1}\gamma_{1}}{\kappa^2 w_{1}+4 \gamma_{1}}, \quad
y=0, \quad E= m_{3}; \label{sim-cp-2}\\
\text{CP3: } &x=1+\alpha, \quad y=0, \quad E=0,\label{sim-cp-3}
\end{align}
where $\gamma_{1}=\kappa^{2} w_{1}/(3w_{0}-1)$ and $m_{3}$ is an arbitrary value. The eigenvalues of the Jacobian matrix for these CPs are found as follows:
\begin{align}
\text{CP1: } & \lambda_{1}= \lambda_{2}=4, \; \lambda_{3}=-(\gamma_{2}+3w_{0}-1); \label{eig-cp-1}\\
\text{CP2: } &\lambda_{1}=-(\gamma_{2}+3w_{0}+3), \; \lambda_{2}=0, \; \lambda_{3}=-4; \label{eig-cp-2}\\
\text{CP3: } &\lambda_{1}=\gamma_{2}+3w_{0}-1, \; \lambda_{2}=\lambda_{3}=\gamma_{2}+3w_{0}+3, \label{eig-cp-3}
\end{align}
where $\gamma_{2}=\kappa^{2}w_{1}/(1+\alpha)$. The corresponding values of the EoS parameter for these CPs take the following forms:
\begin{align}
\text{CP1: }\quad &w_{\mathrm{eff.}}=\frac{1}{3}; \label{sim-eos-1}\\
\text{CP2: }\quad &w_{\mathrm{eff.}}=-1; \label{sim-eos-2} \\
\text{CP3: }\quad &w_{\mathrm{eff.}}=\frac{\gamma_{2}+3w_{0}}{3}. \label{sim-eos-3}
\end{align}
$\blacktriangledown$ \textbf{CP1}:\\
CP1 is in the physical region if $0< -\gamma_{1} <1+\alpha$. Generally, it is an unstable point because $\lambda_{1}=\lambda_{2}=+4$ --- if $(\gamma_{2}+3w_{0}-1)<0$, then it is completely unstable and it means that all trajectories of phase-portrait are repelled away from it, while if $(\gamma_{2}+3w_{0}-1)>0$, then it will be a saddle point (i.e. the trajectories of $y$-direction are attracted while the trajectories of $x$ and $E$ directions are repelled; indeed, the eigenvalues of $x$ and $E$ are $+4$). The positive eigenvalue of $E$-direction indicates the expansion of the universe and decreasing nature of the Hubble parameter. Under every condition, this point belongs to the radiation-dominated era (see (\ref{sim-eos-1})) and for this reason, obviously, the saddle option should be ascribed to this point in the numerical/qualitative study of phase portrait. In the special case namely $(\gamma_{2}+3w_{0}-1)=0$, however, the point is still unstable but let us study it deeply as it should be a saddle point. Under the condition $(\gamma_{2}+3w_{0}-1)=0$, CP1 has a one-dimensional center manifold. First of all, it is useful to consider the system at CP1--origin. It means that we perform a transformation as $(x,y,E)|_{\mathrm{CP1}} \longrightarrow (X,Y,R)=(0,0,0)$. In order to convert the system to a standard form, three new variables $(u,v,s)$ that are connected with $(X,Y,R)$ via
\begin{align}
X=v-s, \quad Y=s, \quad R=u,
\end{align}
are introduced. The center manifold is represented in the form
\begin{align}
W^{c}=\left\{(u,v,s) \left| \; \left| s \right|< \epsilon,\; u=h_{1}(s), \; v=h_{2}(s) \right. \right\},
\end{align}
where we assume that the mappings $h_{1}$ and $h_{2}$ are of the forms
\begin{align}
h_{1}(s) &=k_{1}s^{2}+k_{2}s^{3}+k_{3}s^{4}+\mathcal{O}(s^{5}),\\
h_{2}(s) &=k_{4}s^{2}+k_{5}s^{3}+k_{6}s^{4}+\mathcal{O}(s^{5}),
\end{align}
where $k_{i}$s are constant parameters, and they satisfy the well-known relations of the center manifold theorem. Computations culminate in the following dynamical system which is topologically equivalent to the previous system
\begin{align}
u^{\prime} &=4u,\\
v^{\prime} &=4v,\\
s^{\prime} &=\left(\frac{1-3w_{0}}{1+\alpha}\right)s^{2}+\mathcal{O}(s^{5}).
\end{align}
It is clearly observed that the point is completely unstable.
Therefore, under the condition $(\gamma_{2}+3w_{0}-1)=0$, CP1 is not a good option for the period dominated by radiation.

It is concluded that CP1 is a suitable candidate for the radiation-dominated era only under the following conditions:
\begin{align}
0< - \gamma_{1}<1+\alpha, \quad \& \quad (\gamma_{2}+3w_{0}-1)>0.
\end{align}

$\blacktriangledown$ \textbf{CP2:}\\
According to~(\ref{sim-cp-2}), CP2 is called a physical point if it belongs to one of the following sets:
\begin{align}
&\textbf{Set-1}: \{w_{1}<0, \; \gamma_{1}<0 \};\\
&\textbf{Set-2}: \{w_{1}\gamma_{1}<0, \; (w_{0}+1)\gamma_{1}>0 \}.
\end{align}
Indeed CP2 is not a critical point, but it is a Critical Line (CL) because $E=\text{an arbitrary value} \equiv m_{3}$.
This CL is so interesting as according to its effective EoS, $w_{\mathrm{eff.}}=-1$, it demonstrates the current status of the universe with acceptable accuracy. Specifically speaking, $E=$an arbitrary value$\equiv m_{3}$ can be fixed exactly to the current time by tuning $m_{3}=(\kappa^{2}H_{0}^{2})^{-1}$ in which $H_{0}$ is the present value of the Hubble parameter. Indeed, the current status of the universe is a point of this CL. Therefore, if one desires to suppose this CP as the current time, then all of its eigenvalues should be negative\footnote{However, some authors state that a CP representing current stage must be attractor, but, according to the recent discoveries of the modern physics, our universe is unstable, hence from this perspective, the saddle case may also be acceptable.\label{fon2}}. According to (\ref{eig-cp-2}), even by putting $(\gamma_{2}+3w_{0}+3)>0$, we still cannot distinguish its stability status and it needs to be studied by the center manifold theorem. If one assumes $(\gamma_{2}+3w_{0}+3)>0$, then the related center manifold is one-dimensional. We first consider this case:\\
Like the previous case, first, the system is translated to the origin, $(x,y,E)|_{\mathrm{CP2}} \longrightarrow (X,Y,R)=(0,0,0)$, and then utilizing the following relations, we get the new coordinate $(u,v,s)$:
\begin{align}
X =&\; \frac{-1}{3E_{0}}\left(\frac{\kappa^{2}w_{1}}{1+w_{0}}+3(1+\alpha)\right)u
+\left(\frac{4\gamma_{1}}{3E_{0}(1+w_{0})} \right)v,\\
Y =&\; \frac{\gamma_{1}+\alpha+1}{-E_{0}} v,\\
R =&\; u+v+s.
\end{align}
Using two mappings,
\begin{align}
h_{1}(s) &=k_{1}s^{2}+k_{2}s^{3}+k_{3}s^{4}+\mathcal{O}(s^{5}),\\
h_{2}(s) &=k_{4}s^{2}+k_{5}s^{3}+k_{6}s^{4}+\mathcal{O}(s^{5}),
\end{align}
where $k_{i}$s are constant parameters, we finally get the following topologically equivalent system:
\begin{align}
u^{\prime} =& -(\gamma_{2}+3w_{0}+3)u, \label{CP2-1D-e1}\\
v^{\prime} =& -4v,\\
s^{\prime} =& \; 0+\mathcal{O}(s^5).
\end{align}
Since the condition $(\gamma_{2}+3w_{0}+3)>0$ has been assumed at first, this system indicates an attractor nature for CP2, as desired. We had this anticipation that the evolution on the center manifold will reduce to $s^{\prime} = 0+\mathcal{O}(s^5)$ (i.e. $s$ is approximately constant on the center manifold) because the equation $E^{\prime}=0$ at CP2 is satisfied by every arbitrary value of $E$.\\
In passing, we found that the current status of the universe can be achieved by CP2.\\
The special case $(\gamma_{2}+3w_{0}+3)=0$, for which the related center manifold is two dimensional, implies that the coordinates of CP2 be $(x,y,E)=(1+\alpha,0,m_{3})$ which obviously has a strong relation with CP3.
Indeed, under the aforementioned condition, one has `CP2 = CP3', provided $m_{3}=0$. Let us now consider this special status of CP2. For simplicity and for being in conformity with CP3, let us take $m_{3}=0$. Transforming the system to the origin, i.e. $\left.(x,y,E)\right|_{\mathrm{CP2}} \longrightarrow (X,Y,R)=(0,0,0)$, and going to a new coordinate $(u,v,s)$ using
\begin{align}
X=& s-u,\\
Y=&u,\\
R=&v,
\end{align}
the system is cast in the standard form. Utilizing a two-dimensional mapping as
\begin{align}
u=h(v,s)=k_{1}v^{2}+k_{2}vs+k_{3}s^{2}+\cdots,
\end{align}
where $k_{i}$s are constant parameters, one may reach a topologically equivalent system having the following form in the polar coordinate, $(s,v)=(r \cos (\theta)$, $r \sin (\theta))$,
\begin{align}
u^{\prime}=&-4u,\\
r^{\prime}=& \left(\frac{2(1+w_{0})\cos (\theta)}{1+\alpha} \right) r^{2}+\mathcal{O}\left|r \right|^{3},\\
\theta^{\prime}=& 0.
\end{align}
In short, only in one direction we have attraction and our system is totally unstable, stemming from the utter instability of its center manifold. Therefore, under the condition $(\gamma_{2}+3w_{0}+3)=0$, CP2 is a saddle point.
Pursuant to the justification in footnote~[\ref{fon2}], this behavior may also be acceptable for the present status of the universe.

$\blacktriangledown$ \textbf{CP3:}\\
According to the effective EoS parameter of CP3, (\ref{sim-eos-3}), four options are of interest for investigating:
\begin{enumerate}
	\item $\gamma_{2}+3w_{0}=1$:\\
This condition leads to the radiation-dominated era ($w_{\mathrm{eff.}}=1/3$) and it is a special form of CP1 --- when CP1 has one-dimensional center manifold, that is
\begin{align*}
&(x,y,E)\left|_{\mathrm{CP3}}\right.
=(1+\alpha,0,0)=(x,y,E)\left|_{\mathrm{CP1}}\right., \\ &\lambda_{1}\left|_{\mathrm{CP3}}\right.=0=\lambda_{3}\left|_{\mathrm{CP1}}\right.,
\\
&\lambda_{2}\left|_{\mathrm{CP3}}\right.=\lambda_{3}\left|_{\mathrm{CP3}}\right.=+4
=\lambda_{1}\left|_{\mathrm{CP1}}\right.=\lambda_{2}\left|_{\mathrm{CP1}}\right..
\end{align*}
For having this CP in the physical region, it is sufficient to set $\alpha > -1$, but, as corroborated already, in this situation the related CP is completely unstable and repeller. Therefore, a saddle point expectation makes this option not an apt candidate for the radiation dominated era.
	\item $\gamma_{2}+3w_{0}=-3$:\\
Under this condition, the special position of CP2 is recovered, viz.,
\begin{align*}
&(x,y,E)\left|_{\mathrm{CP3}}\right.=(1+\alpha,0,0)
=(x,y,E)\left|_{\mathrm{CP2}}\right.,\\
&\lambda_{1}\left|_{\mathrm{CP3}}\right.=-4
=\lambda_{3}\left|_{\mathrm{CP2}}\right.,\\
&\lambda_{2}\left|_{\mathrm{CP3}}\right.
=\lambda_{3}\left|_{\mathrm{CP3}}\right.=0
=\lambda_{1}\left|_{\mathrm{CP2}}\right.=\lambda_{2}\left|_{\mathrm{CP2}}\right..
\end{align*}
Thus, in this case, CP3 would be a saddle point --- just in one direction we have attraction and the two-dimensional center manifold is completely unstable at the CP3-origin. To wrap the argument, according to $w_{\mathrm{eff.}}=-1$ and the aforementioned discussion, one may accept this point as a candidate for the present time.
	\item $\gamma_{2}+3w_{0}=-3.09$:\\
This condition is an interesting option as it yields\footnote{Planck data~\cite{p2018p} show that the value of the effective EoS is $-1.03 \pm 0.03$.} $w_{\mathrm{eff}}=-1.03$ with three negative eigenvalues indicating a complete attraction:
\begin{align}
\lambda_{1}=-4.09, \quad \lambda_{2}=\lambda_{3}=-0.09.
\end{align}
However, the condition $\gamma_{2}+3w_{0}=-3.09$ makes CP3 a good candidate for the present time but this condition causes that CP1 (with $w_{\mathrm{eff}}|_{\mathrm{CP1}}=1/3$) be a repeller point that is not acceptable since it must be a saddle CP.
	\item $\gamma_{2}+3w_{0}=0$:\\
This condition corresponds to the matter-dominated era ($w_{\mathrm{eff.}}=0$). The related eigenvalues turn out to be
\begin{align}
\lambda_{1}=-1, \quad \lambda_{2}=\lambda_{3}=+3.
\end{align}
Therefore, CP3 is a saddle critical point that is completely acceptable behavior for a CP of the matter-dominated era.\\
Now, let us take other CPs into account using the aforementioned condition:\\ Utilizing the condition $\gamma_{2}+3w_{0}=0$, a special form of CP2 (when it has a one-dimensional center manifold) is recovered. Inserting $\gamma_{2}+3w_{0}=0$ into~(\ref{CP2-1D-e1}), the related system can be cast in the following form which is topologically equivalent to the main system:
\begin{align}
u^{\prime} =& -3u,\\
v^{\prime} =& -4v,\\
s^{\prime} =&\; 0+\mathcal{O}(s^{5}),
\end{align}
where $s^{\prime}=0+\mathcal{O}(s^{5})$ is the center manifold which is constant (Its meaning and reason have been explained before). Therefore, CP2, under the aforementioned condition, is an attractor and is suitable for the current position of the universe. However, under the condition $\gamma_{2}+3w_{0}=0$, CP3 and CP2 are very good candidates for the matter-dominated era and the current time, respectively, but this condition leads to a peculiarity: Due to the positivity of eigenvalues of CP1, the CP will be of repelling nature, however, we were anticipating a saddle point because of the radiation dominance.
\end{enumerate}

\begin{figure*}
	\includegraphics[width=6.5 in]{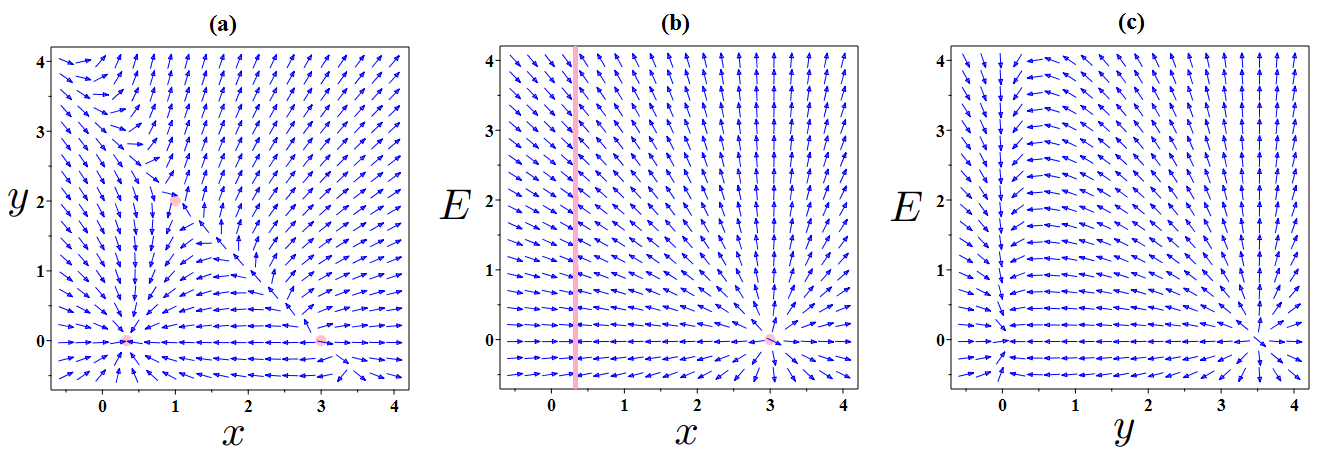}
    \caption{This figure illustrates three different Poincar\'{e} sections of the phase portrait of system (\ref{sim-eq-1})--(\ref{sim-eq-3}) when the values of the constant parameters are as: $w_{0}=+1$, $\kappa^{2}w_{1}=-2$, $\alpha=+2$. The critical points and line have been indicated by pink color.}
    \label{f(3-1)}
\end{figure*}

\begin{figure*}
	\includegraphics[width=6.5 in]{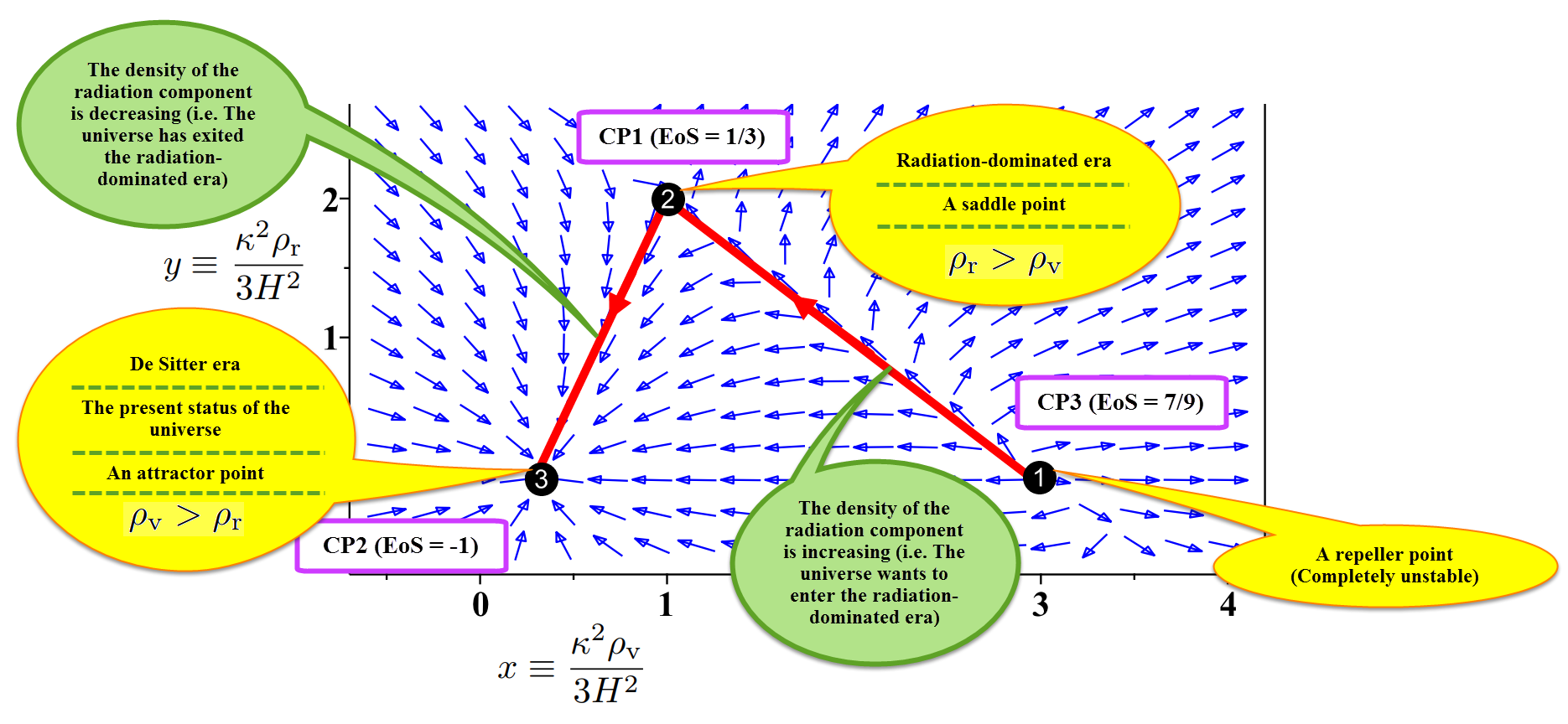}
    \caption{This figure is an explanation of the phase-portrait of system (\ref{sim-eq-1})--(\ref{sim-eq-3}) in Poincar\'{e} section $x\mathbf{-}y$.}
    \label{f(3-2)}
\end{figure*}

According to all discussions and detailed analysis performed above, now it is clear that there are many options for matching with observational data and plotting. But, among them, there are only some cases with no anomalies.
It is not hard to show that we cannot tune all CPs in a way that all the last three main stages of the universe be recovered,\footnote{We cannot tune the achieved CPs so that they form the following picture:
\begin{itemize}
	\item A saddle point with $w_{\mathrm{eff.}}=1/3$;
	\item A saddle point with $w_{\mathrm{eff.}}=0$;
	\item An attractor or a saddle point with $w_{\mathrm{eff.}}=-1$.
\end{itemize}} for it needs to hold the following conditions simultaneously which is impossible:
\begin{align}
&\gamma_{2}+3w_{0}>1, \label{ndo1}\\
&\gamma_{2}+3w_{0}=0. \label{ndo2}
\end{align}
In the best situation, one may be able to keep two of the three aforementioned stages.\\
Among some presentable cases, we single out the constant parameters as
\begin{align}
w_{0}=+1, \quad \kappa^{2}w_{1}=-2, \quad \alpha=+2,
\end{align}
which lead to sufficiently good cases:
\begin{align}
&\text{CP1}:\; \;(x,y,E)=(1,2,0), \quad w_{\mathrm{eff.}}=1/3, \nonumber\\
&(\lambda_{1},\lambda_{2},\lambda_{3})=(4,4,-4/3)\; \longrightarrow \; \text{``A saddle CP''};\\
&\text{CP2}:\; \;(x,y,E)=(1/3,0,m_{3}), \quad w_{\mathrm{eff.}}=-1, \nonumber\\
&(\lambda_{1},\lambda_{2},\lambda_{3})=(-16/3,0,-4)\; \longrightarrow \; \text{``An attractor CP''};\\
&\text{CP3}:\; \;(x,y,E)=(3,0,0), \quad w_{\mathrm{eff.}}=7/9, \nonumber\\
&(\lambda_{1},\lambda_{2},\lambda_{3})=(4/3,16/3,16/3)\; \longrightarrow \; \text{``A repeller CP''},
\end{align}
where $m_{3}$ is an arbitrary constant. Three Poincar\'{e} sections $x-y$, $x-E$, and $y-E$ have been presented in fig.~\ref{f(3-1)}. The physical region in these phase portraits are the positive parts:
\begin{align}
\left\{(x_{i},x_{j}) \;| \; x_{i},x_{j} \in \mathbb{R}^{\geq 0}, \; \& \; i \neq j \right\}; \quad i,j=1,2,3,
\end{align}
where $(x_{1},x_{2},x_{3})=(x,y,E)$. Thanks to the critical line $E=m_{3}$, three critical points can be observed in the phase portrait of $x-y$ plane. This critical line has clearly been illustrated in $x-E$ Poincar\'{e} section together with CP3. As is clear from this phase portrait, this critical line is an attractor --- a hole line. The present position of the universe is on this line (i.e. the point $(\kappa^{2}H_{0}^{2})^{-1}$). The trajectory of the evolution of the universe tends to this point. In $y-E$ phase portrait none of CPs are observable.
Note that the vector fields in fig.~\ref{f(3-1)}(c) may give us the incorrect idea that there are two CPs, $(0,0)$ and $(7/2,0)$, of the system under consideration. In fact, this fallacy follows from just a visual error.\\
In fig.~(\ref{f(3-2)}), it has been tried to illustrate a historical picture based on the studied model. We have presented it in a two-dimensional plane to present a better view. In what follows, we explain its form in a three-dimensional space. The studied model (i.e. ``A simplified form of EoS") with our selections to the values of constant parameters, indicates \textit{some parts} of the universe evolution:\\
Let us assume that the universe arrived at the unstable CP3 (or at least, a neighborhood of this CP) and then got repelled from\footnote{Note that, without loss of generality, we can set the initial conditions so that the system could start from CP3. Hence, its nature (i.e. completely unstable feature) does not cause an anomaly.}.
This repulsion is somewhat tenable, for no stage of the evolution of the universe includes a fluid with such EoS dominance.
During this repulsion and before getting to the next CP (i.e. CP1), as is clear in fig.~\ref{f(3-2)}, the amount of $y$ has increased while $x$ has decreased.
Because $y$ is proportional to the density of the radiation component, hence it makes the universe ready to enter to the radiation-dominated era. After this process, the universe reaches the radiation-dominated era at CP1 with $w_{\mathrm{eff.}}=1/3$. At this CP, the density of the radiation component is greater than the density of the viscous fluid that confirms the domination of the radiation component. This CP is of a saddle nature, hence, beside attraction property, it has a repulsion feature that is needed to get out of this area. Overtime, the effect of this repulsion becomes obvious and the universe not only exits this CP but also the density of the radiation components decreases, then our universe is attracted to CP2 with $w_{\mathrm{eff.}}=-1$ (i.e. the present time). Note that CP2 is on the critical line whose one of its points is presented in this figure. Indeed, the final CP that has been shown in this figure, is a projection of the point being the height $E=+(\kappa^{2}H_{0}^{2})^{-1}$. Since all the eigenvalues related to this point are negative, hence it is an attractor and the universe will be stable at this CP. The density of the viscous fluid is greater than the density of the radiation component, indicating the domination of dark energy at this CP. Surely, if this diagram for the universe evolution be acceptable, then between CP1 and CP2 our universe must pass a CP with $w_{\mathrm{eff.}}=0$. Sadly, our system yields no such CP.\\

\noindent\pqrmybox{}{\subsubsection{An important point: ``Lack of matter-dominated era in the model''}}\vspace{5mm}
We pose ``Why don't we have a CP representing the matter-dominated era?''\\
Because (\ref{ndo1})-(\ref{ndo2}) do not hold simultaneously, so reaching a perfect picture of the last three
stages of the evolution of the universe is impossible.
Assuming the $F(T)$-gravity and the interaction models are correct theories, then two propositions may be examined to see if they could resolve this problem or not:\\
1- Adding baryonic matter component that has $\rho_{\mathrm{m}} \neq 0$ and $P_{\mathrm{m}}=0$.\\
2- Enhancing the action through a scalar field part.\\
In this paper, we are content with the first suggestion merely.
The second one is deferred to another paper.\\
In what follows, it is shown that neither with $E$ nor without $E$, the first suggestion does not work. Both only lead to $w_{\mathrm{eff.}}=-1$ and $w_{\mathrm{eff.}}=0$ and one further EoS which cannot be fixed to a radiation-dominated era:\\
$\bullet$ Without $E$:\\
Introducing
\begin{align}
x \equiv \frac{\kappa^{2} \rho_{\mathrm{v}}}{3H^{2}}, \quad y \equiv \frac{\kappa^{2} \rho_{\mathrm{r}}}{3H^{2}}, \quad z \equiv \frac{\kappa^{2} \rho_{\mathrm{m}}}{3H^{2}},
\end{align}
the following dynamical system is obtained:
\begin{align}
x^{\prime}=& \left(\frac{x}{1+\alpha} \right) \left[ 3(1+w_{0})x+3y+4z+w_{1}\kappa^{2} \right] \nonumber \\ &-3(1+w_{0})x-w_{1}\kappa^{2},
\end{align}
\begin{align}
y^{\prime}=& \left(\frac{y}{1+\alpha} \right) \left[ 3(1+w_{0})x+3y+4z+w_{1}\kappa^{2} \right]-3y,
\end{align}
\begin{align}
z^{\prime}=& \left(\frac{z}{1+\alpha} \right) \left[ 3(1+w_{0})x+3y+4z+w_{1}\kappa^{2} \right]-4z,
\end{align}
with the following EoS
\begin{align}
w_{\mathrm{eff.}}=-1+\frac{1}{3}\left(\frac{1}{1+\alpha} \right)
\left[ 3(1+w_{0})x+3y+4z+w_{1}\kappa^{2} \right].
\end{align}
The fixed points with their associated eigenvalues and EoS parameters are as follows:
\begin{align}
\blacktriangledown & \text{CP1:} \nonumber \\
& x=\frac{- w_{1} \kappa^{2}}{3w_{0}}, \quad y=\frac{3 \chi_{1} }{25 w_{0}}, \quad z=\frac{4 \chi_{1} }{25w_{0}}, \quad w_{\mathrm{eff.}}|_{\mathrm{CP1}}=0, \nonumber\\
&\lambda_{1}=\frac{- \chi_{1} }{1+\alpha }, \quad \lambda_{2}=\lambda_{3}=+3;
\end{align}
\begin{align}
\blacktriangledown & \text{CP2:} \nonumber \\
& x=1+\alpha , \quad y=0, \quad z=0, \quad w_{\mathrm{eff.}}|_{\mathrm{CP2}}=\frac{\chi_{1} }{3(1+\alpha )}, \nonumber\\
& \lambda_{1}=\frac{\chi_{1} }{1+\alpha }, \quad \lambda_{2}=\lambda_{3}=3+\frac{\chi_{1} }{1+\alpha };
\end{align}
\begin{align}
\blacktriangledown & \text{CP3:} \nonumber \\
& x=\frac{-w_{1} \kappa^{2}}{3(1+w_{0})} , \quad y=0, \quad z=0, \quad w_{\mathrm{eff.}}|_{\mathrm{CP3}}=-1, \nonumber\\
& \lambda_{1}=0, \quad \lambda_{2}=-3, \quad \lambda_{3}=-3-\frac{\chi_{1} }{1+\alpha },
\end{align}
where $\chi_{1} =3 w_{0} +3 \alpha w_{0} +w_{1} \kappa^{2} $. As observed, CP1 and CP3 may be fixed to the matter-dominated era and current stage of the universe, respectively. If one tries to set CP2 as the radiation-dominated era, the condition
\begin{align}
w_{0}=\frac{1}{3}-\frac{w_{1}\kappa^{2}}{3(1+\alpha )}
\end{align}
must be imposed.
This condition makes unacceptable nature (repeller) for CP2 while other CPs are appealing candidates for the corresponding eras:
\begin{align}
&\text{CP1:}\; w_{\mathrm{eff.}}=0\; \text{(Matter-dominated era)}, \nonumber\\
&(\lambda_{1},\lambda_{2},\lambda_{3})=(-1,+3,+3) \Rightarrow \text{Acceptable} \; \checkmark ;
\end{align}
\begin{align}
&\text{CP2:}\; w_{\mathrm{eff.}}=\frac{1}{3}\; \text{(Radiation-dominated era)}, \nonumber\\
&(\lambda_{1},\lambda_{2},\lambda_{3})=(+1,+4,+4) \Rightarrow \text{Unacceptable} \; \times ;
\end{align}
\begin{align}
&\text{CP3:}\; w_{\mathrm{eff.}}=-1\; \text{(De-Sitter era)}, \nonumber\\
&(\lambda_{1},\lambda_{2},\lambda_{3})=(0,-3,-4) \Rightarrow \text{Acceptable} \; \checkmark .
\end{align}
It is worthwhile to mention that the radiation-dominated era cannot be reached even approximately. In detail, let us exert the following condition which contains some fluctuations through $\epsilon$:
\begin{align}
w_{0}=\frac{1}{3}-\frac{w_{1}\kappa^{2}}{3(1+\alpha)}+\epsilon,
\end{align}
where $\epsilon$ is an arbitrary constant. It leads to
\begin{align}
&\blacktriangleright \;\; \text{CP1:}\;\;
w_{\mathrm{eff.}}=0\; \text{(Matter-dominated era)}, \nonumber\\
&(\lambda_{1},\lambda_{2},\lambda_{3})=(-(3\epsilon +1),+3,+3);
\end{align}
\begin{align}
&\blacktriangleright \;\; \text{CP2:}\;\; w_{\mathrm{eff.}}=\frac{1}{3}+\epsilon \; \nonumber\\
&(\lambda_{1},\lambda_{2},\lambda_{3})=((3\epsilon +1),(3\epsilon +4),(3\epsilon +4));
\end{align}
\begin{align}
&\blacktriangleright \;\; \text{CP3:}\;\;
w_{\mathrm{eff.}}=-1\; \text{(De-Sitter era)}, \nonumber\\
&(\lambda_{1},\lambda_{2},\lambda_{3})=(0,-3,-(3\epsilon +4)).
\end{align}
It is clear that there is no real value for $\epsilon$ leading to true behavior (i.e. saddle) for the radiation-dominated era. Indeed, if we want to keep the matter-dominated era, then we lose the radiation-dominated era and the vice-versa. Both should be saddle simultaneously which is unattainable through this system.\\
$\bullet$ With $E$:\\
For the generalized case, we define
\begin{align}
x \equiv \frac{\kappa^{2} \rho_{\mathrm{v}}}{3H^{2}}, \quad y \equiv \frac{\kappa^{2} \rho_{\mathrm{r}}}{3H^{2}}, \quad z \equiv \frac{\kappa^{2} \rho_{\mathrm{m}}}{3H^{2}}, \quad E \equiv \frac{1}{\kappa^{2} H^{2}}.
\end{align}
These yield the following dynamical system:
\begin{align}
x^{\prime}=& \left(\frac{x}{1+\alpha} \right) \left[ 3(1+w_{0})x+3y+4z+w_{1}\kappa^{2} \right] \nonumber \\ &-3(1+w_{0})x-w_{1}\kappa^{2},
\end{align}
\begin{align}
y^{\prime}=& \left(\frac{y}{1+\alpha} \right) \left[ 3(1+w_{0})x+3y+4z+w_{1}\kappa^{2} \right]-3y,
\end{align}
\begin{align}
z^{\prime}=& \left(\frac{z}{1+\alpha} \right) \left[ 3(1+w_{0})x+3y+4z+w_{1}\kappa^{2} \right]-4z,
\end{align}
\begin{align}
E^{\prime}=& \left(\frac{E}{1+\alpha}\right) \left[ 3(1+w_{0})x+3y+4z+w_{1}\kappa^{2} \right].
\end{align}
This system leads to the following CPs with their associated EoS and eigenvalues:
\begin{align}
\blacktriangledown & \text{CP1:} \nonumber \\
& x=\frac{- w_{1} \kappa^{2}}{3w_{0}}, \quad y=\frac{3 \chi_{1} }{25 w_{0}}, \quad z=\frac{4 \chi_{1} }{25w_{0}}, \quad E=0, \nonumber\\ & w_{\mathrm{eff.}}|_{\mathrm{CP1}}=0, \nonumber\\
&\lambda_{1}=\frac{- \chi_{1} }{1+\alpha }, \quad \lambda_{2}=\lambda_{3}=+3, \quad \lambda_{4}=+1;
\end{align}
\begin{align}
\blacktriangledown & \text{CP2:} \nonumber \\
& x=1+\alpha , \quad y=0, \quad z=0, \quad E=0, \nonumber\\ & w_{\mathrm{eff.}}|_{\mathrm{CP2}}=\frac{\chi_{1} }{3(1+\alpha )}, \nonumber\\
& \lambda_{1}=\frac{\chi_{1} }{1+\alpha }, \quad
\lambda_{2}=\lambda_{3}=3+\frac{\chi_{1} }{1+\alpha }, \quad \lambda_{4}=+1;
\end{align}
\begin{align}
\blacktriangledown & \text{CP3:} \nonumber \\
& x=\frac{-w_{1} \kappa^{2}}{3(1+w_{0})} , \quad y=0, \quad z=0, \quad E=m_{3}, \nonumber\\ & w_{\mathrm{eff.}}|_{\mathrm{CP3}}=-1, \nonumber\\
& \lambda_{1}=0, \quad \lambda_{2}=-3, \quad
\lambda_{3}=-3-\frac{\chi_{1} }{1+\alpha } \quad \lambda_{4}=+1,
\end{align}
where $\chi_{1} =3 w_{0} +3 \alpha w_{0} +w_{1} \kappa^{2} $ and $m_{3}$ is an arbitrary constant.\\
It is clear that both cases (i.e. `with $E$' and `without $E$') are equivalent. Hence, its explanations and justifications for the inefficiency of adding a pressureless matter part are the same as the previous case. Therefore, we conclude that adding further dimension, $E$, does not present anything special, but two properties:\\
By comparison with the previous case, it is easily understood that the added dimension, $E$, behaves as a repeller direction in general --- the eigenvalue of this direction is $+1$ in all CPs of this subsection. This special repulsion refers to the expansion of the universe. Furthermore, the positive eigenvalues indicate that the Hubble parameter decreases with time.\\

\noindent\pqrmybox{}{\subsubsection{Discussion about $E=\kappa^{-2}H^{-2}=0$}}\vspace{5mm}
In most of this paper, we encounter $E=(\kappa^{2} H^{2})^{-1}=0$ for CPs. Most of these CPs belong to the radiation, matter, and current stages of the universe. Hence, one may argue that the corresponding CPs are not good candidates for the associated eras. But, note that in most cases of interest, this interpretation is wrong. This boils down to the fact that this dependent variable is just an imposition.
When our dynamical system and EoS parameter are of the form \footnote{Here, a four-dimensional system has been explained. The $D$-dimensional systems can be discussed in the same manner.}
\begin{align}
x^{\prime}=&\; k_{1}x\left[F_{1}(x,y,z) \right]+F_{2}(x,y,z),\label{df1}\\
y^{\prime}=&\; k_{2}y\left[F_{1}(x,y,z) \right]+F_{3}(x,y,z),\label{df2}\\
z^{\prime}=&\; k_{3}z\left[F_{1}(x,y,z) \right]+F_{4}(x,y,z),\label{df3}\\
E^{\prime}=&\; k_{4}E\left[F_{1}(x,y,z) \right],\label{df4}\\
w_{\mathrm{eff.}}=&\;-1+\frac{k_{4}}{3}\left[F_{1}(x,y,z) \right],
\end{align}
where $k_{i}$s are constant parameters and $F_{i}$s are functions of $x$, $y$, and $z$, all CPs with $E=0$ may be discussed without serious attention to $E$. Note that $F_{1}(x,y,z)$ is common among them. Furthermore, if we are to keep $E \neq 0$ (which yields $F_{1}=0$ and $w_{\mathrm{eff.}}=-1$), then for reaching CPs, we must have $F_{2}(x,y,z)=F_{3}(x,y,z)=F_{4}(x,y,z)=0$ which does not tend to any CP in most cases of interest. Hence, in most cases, we have $F_{1}(x,y,z) \neq 0$ and it causes $E=0$.
If this additional coordinate $E$ were not defined, then CPs would not be in the position they are now.
Therefore, we can ignore the value of coordinate $E$ in interpreting provided existing above conditions. The benefit of adding such extra dimension has been explained above: The positivity of the eigenvalues of this extra dimension not only refers to the expanding nature of the evolution of the universe but also indicates the decreasing nature of the Hubble parameter with time ($E \propto H^{-2}$).
Due to the fact that $E$ does not explicitly appears in (\ref{df1})-(\ref{df3}), eq.~(\ref{df4}) can be solved analytically at a given CP. For example, for a $\mathrm{CP_{i}}$ one has:
\begin{align*}
&E^{\prime}=\underbrace{\left(k_{4}F_{1}|_{\mathrm{CP_{i}}}\right)}_{\text{A constant}} E \equiv C^{*} E,\\
&\Rightarrow E=E_{0} \exp \left[C^{*}N \right] \Rightarrow H(N)=\sqrt{\frac{1}{\kappa^{2}E_{0}}}
\exp \left[\frac{-C^{*}N}{2} \right],
\end{align*}
or put differently
\begin{align*}
H(t) =H_{1} \frac{2C^{*}}{t},
\end{align*}
where $E_{0}$ and $H_{1}$ are the constants of integration.
The positive eigenvalue of $E$ is equivalent to the positive value of $C^{*}$.
The system cannot reach the zero value of $E$ because at a critical point the corresponding to $E$ eigenvalue is always positive. So $H$ goes over time from infinite value to zero or some constant value.\\
Note the form of EoS parameter and compare it with the equation $E^{\prime}$, then you will find that $E = \textit{const.}$ can occur only for the de-Sitter expansion. However, in order to comply with other CPs, one may also take $E=0$ for such CPs as well, but since this constant can be fixed to the current stage of the universe, hence we would like to keep it non-zero. We know that the effective value of EoS parameter is not exactly $-1$, but it is close to $-1$, so, for example, if, according to Plank data~\cite{p2018p}, we have $w_{\mathrm{eff.}}=-1.03$ at the CP of the present time, then again we obtain $E=0$.\\

Besides the above discussions, it can be predicted for other types of systems like
\begin{align}
x^{\prime}=& k_{1}x\left[F_{1}(x,y,z,E) \right]+F_{2}(x,y,z),\\
y^{\prime}=& k_{2}y\left[F_{1}(x,y,z,E) \right]+F_{3}(x,y,z),\\
z^{\prime}=& k_{3}z\left[F_{1}(x,y,z,E) \right]+F_{4}(x,y,z),\\
E^{\prime}=& k_{4}E\left[F_{1}(x,y,z,E) \right],\\
w_{\mathrm{eff.}}=&-1+\frac{k_{4}}{3}\left[F_{1}(x,y,z,E) \right],
\end{align}
that if the power of $E$ in $F_{1}$ is positive, then all CPs will have $E=0$ excluding those points which belong to the de sitter era (i.e. $w_{\mathrm{eff.}}=-1$). In the cases in which the power of $E$ is negative (e.g. $1/E^{|n|}$), then there will be only one CP that belongs to the de sitter era (i.e. $w_{\mathrm{eff.}}=-1$) with $E = \textit{const.}$ \\
In the following, sections we encounter such systems.\\

\noindent\Rrmybox{}{\subsection{More complicated forms of the EoS parameter}}\vspace{5mm}
In this sub-section, some complicated forms of the EoS parameter are studied. What we are going to do is that in these forms of EoS the powers of $\rho_{\mathrm{v}}$ and $H$ in $f$ and $G$, are increased and decreased. Generally, they may be cast in the form\footnote{Note that $\alpha \neq \alpha_{1}$; $\alpha$ is the coefficient of the torsion in the model while $\alpha_{1}$ is the power of the density.} $f(\rho_{\mathrm{v}})+G(H)=A \rho_{\mathrm{v}}^{\alpha_{1}}+B H^{2\beta}$. It has been demonstrated that this EoS yields finite-time singularities of a cosmological system~\cite{od-50}.
Using $f(\rho_{\mathrm{v}})+G(H)=A \rho_{\mathrm{v}}^{\alpha_{1}}+B H^{2\beta}$ we arrive at:
\begin{align}
x^{\prime}=& \left(\frac{x-1-\alpha}{1+\alpha}\right) \left\{\kappa^{4}E \left[A \left(\frac{3x}{\kappa^{4}E} \right)^{\alpha_{1}}+B \left(\frac{1}{\kappa^{2}E} \right)^{\beta} \right] \right\}
\nonumber \\
&+\frac{4xy}{1+\alpha}, \label{mo-eq-1}
\end{align}
\begin{align}
y^{\prime}=&\frac{y}{1+\alpha} \left\{\kappa^{4}E \left[A \left(\frac{3x}{\kappa^{4}E} \right)^{\alpha_{1}}+B \left(\frac{1}{\kappa^{2}E} \right)^{\beta} \right]+4y \right\}
\nonumber \\&-4y, \label{mo-eq-2}
\end{align}
\begin{align}
E^{\prime}=&\frac{E}{1+\alpha} \left\{\kappa^{4}E \left[A \left(\frac{3x}{\kappa^{4}E} \right)^{\alpha_{1}}+B \left(\frac{1}{\kappa^{2}E} \right)^{\beta} \right]+4y \right\}. \label{mo-eq-3}
\end{align}
The corresponding EoS parameter would also be
\begin{align}
w_{\mathrm{eff.}}=& -1+\frac{1}{3}\left(\frac{1}{1+\alpha} \right)  \nonumber \\ &
\times \left\{\kappa^{4}E \left[A \left(\frac{3x}{\kappa^{4}E} \right)^{\alpha_{1}}+B \left(\frac{1}{\kappa^{2}E} \right)^{\beta} \right]+4y \right\}.
\end{align}
Due to the presence of many unknown constant parameters especially the powers $\alpha_{1}$ and $\beta$, this case may not be considered analytically. If both $\alpha_{1}$ and $\beta$ are given, then analyzing this system is feasible enough. Hence, let us consider this system by choosing some values for $\alpha_{1}$ and $\beta$.\\
Furthermore, two cases seem to be notable in scrutinizing all the cases of this sub-section: $B \neq 0$ and $B=0$.\\

\noindent\pqrmybox{}{\subsubsection{Case-1:}}\vspace{5mm}
As the first case, we take $\alpha_{1}=1$ and $\beta=2$. With these choices the cosmological system reads:
\begin{align}
x^{\prime} &=\frac{x-(1+\alpha)}{1+\alpha} \left(3Ax+\frac{B}{E} \right)+\left( \frac{4}{1+\alpha}\right)xy, \label{case1eq1}\\
y^{\prime} &=\frac{y}{1+\alpha} \left(3Ax+\frac{B}{E}+4y \right)-4y, \label{case1eq2}\\
E^{\prime} &=\frac{E}{1+\alpha} \left(3Ax+\frac{B}{E}+4y \right). \label{case1eq3}
\end{align}
Under the above selections, the effective equation of state would then be:
\begin{align}
w_{\mathrm{eff.}}=-1+\frac{1}{3(1+\alpha)}\left[3Ax+\frac{B}{E}+4y \right].
\end{align}
As observed, unlike the previous case (``A simplified case''), the amount of EoS here is affected by all three coordinates $x$, $y$, and $E$.\\
$\clubsuit$ Case $B \neq 0$:\\
In general, the system presents a two-dimensional critical curve (i.e. $xE=C$ where $C=-B/(3A)$) for a given value of $B/A$:
\begin{align}
x=\frac{-B}{3Am_{3}}, \quad y=0, \quad E=\text{an arbitrary value}\equiv m_{3}.
\end{align}
Thus, for a fixed time of the evolution of the universe, i.e. for a given $H$, the system has only one CP because $E=(\kappa^{2} H^{2})^{-1}$. The conditions $(AB)<0$ and $A \neq 0$ are required to put this critical curve in the physical region. On this curve, the density of the radiation component is zero while the density of the viscus fluid depends upon the values of $A$, $B$, and $H$.\\
Correspondingly, the eigenvalues for the system become:
\begin{align}
\lambda_{1}=\frac{3Am_{3}+B}{-m_{3}}, \quad \lambda_{2}=-4, \quad \lambda_{3}=0.
\end{align}
It is clear that only one of the eigenvalues is affected by the stage of the universe. In the limit $m_{3} \to \infty$ (or equivalently $H \to 0$), one has $\lambda_{1}=-3A$.
So, if we take $A>0$ and $B<0$, then it means that our universe will be attracted to this point in the future.\\
Due to $\lambda_{3}=0$, the related center manifold is at least one-dimensional.
In the case that $3Am_{3}+B=0$, the corresponding center manifold would be two-dimensional.\\
The value of the EoS for the critical curve is
\begin{align}
w_{\mathrm{eff.}}=-1,
\end{align}
which corresponds to de sitter era and the present time.\\

\begin{figure*}
	\includegraphics[width=6.5 in]{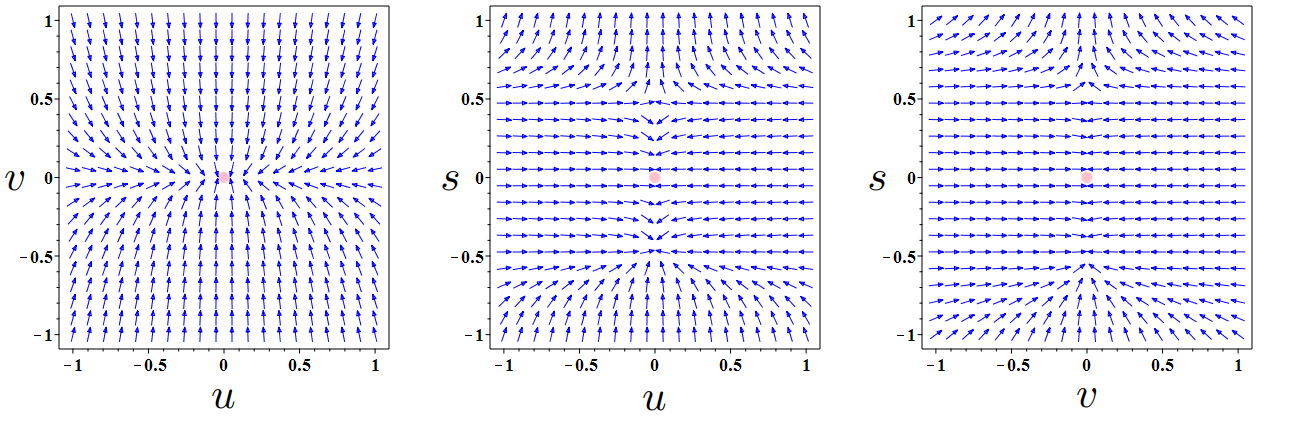}
    \caption{This figure indicates three different Poincar\'{e} sections of the phase portrait of the system~(\ref{epa1-1})--(\ref{epa1-3}). In plotting these diagrams, we have assumed $m_{3}=1$, $A=2/3$, and $B=-1$. The pink points show the critical points.}
    \label{pa1}
\end{figure*}

Assuming $\lambda_{1} \neq 0$, the system has a one-dimensional center manifold. Regardless of whether $\lambda_{1}$ is negative or not, we study it and finally discuss the stability and these problems.\\
Again, let us consider the problem in CP-origin (i.e. the system is transformed into a coordinate $(X,Y,R)$ so that the position of the CPs be $(0,0,0)$ in it). Then using\footnote{Here, we have assumed that $A \neq 0$, $A \neq 4/3$, and $m_{3} \neq 0$.}
\begin{align}
X=& \left(\frac{6Am_{3}+B}{-3Am_{3}^{2}} \right)u +\left(\frac{B(3A+4)}{3Am_{3}^{2}(3A-4)} \right)v+\left(\frac{B}{3Am_{3}^{2}} \right)s,\\
Y=& \left(\frac{6Am_{3}+2B-8m_{3}}{m_{3}^{2}(4-3A)} \right)v,\\
R=& u+v+s,
\end{align}
the system is presented in a new coordinate $(u,v,s)$. After a tedious computational process, we get the following system which is topologically equivalent to the system at the CP-origin:
\begin{align}
u^{\prime}=& \left(\frac{3Am_{3}+B}{-m_{3}} \right)u, \label{epa1-1}\\
v^{\prime}=& -4v,\label{epa1-2}\\
s^{\prime}=& \left[\frac{B(6Am_{3}+B)}{4m_{3}^{2}(3Am_{3}+B)} \right]s^{2} \nonumber \\ &+\left[\frac{B^{4}+12AB^{3}m_{3}-216A^{3}Bm_{3}^{3}}{16m_{3}^{4}
(3Am_{3}+B)^3} \right]s^{4}+\mathcal{O}(s^{5}).\label{epa1-3}
\end{align}
It is observed that the center manifold $s$ is unstable, hence on each point of this critical curve, this system is totally unstable and the critical curve is a saddle curve. Based on the value of $\lambda_{1}$, the number of the direction of attraction and repulsion will change; if $\lambda_{1}>0$, then in two directions we have repulsion, while for $\lambda_{1}<0$ just in one direction (on the center manifold) we have repulsion.
Tuning\footnote{$H_{0}$ is the present value of the Hubble parameter.} $m_{3}=(\kappa^{2} H_{0}^{2})^{-1}$ and $\lambda_{1}<0$, the CP may be a good candidate for the current status of the universe as our universe is unstable. Note that, however, both $\lambda_{1}<0$ and $\lambda_{1}>0$ are acceptable, but setting $\lambda_{1}<0$ speeds up the attraction to this point (indicating acceleration) and it slows down the exit from this era.\\%  causes that the speed of variation decreases.\\
As an example, three different Poincar\'{e} sections of the phase-portrait of the above system have been presented in fig.~\ref{pa1} which illustrates our discussions for a negative value of $\lambda_{1}$.

$\clubsuit$ Case $B = 0$:\\
In this case, indeed, the viscous part is removed.\\
The critical points and their corresponding eigenvalues and EoS parameters are as follows:
\begin{align}
\blacktriangleright \;& \text{CP1:} \;(x,y,E)=(0,0,m_{3}), \nonumber \\ &(\lambda_{1},\lambda_{2},\lambda_{3})=(-4,-3A,0),\quad w_{\mathrm{eff.}}|_{\mathrm{CP1}}=-1;
\end{align}
\begin{align}
\blacktriangleright \;& \text{CP2:} \;(x,y,E)=(1+\alpha ,0,m_{3}), \nonumber \\ &(\lambda_{1},\lambda_{2},\lambda_{3})=(3A-4,3A,3A),\quad w_{\mathrm{eff.}}|_{\mathrm{CP2}}=A-1;
\end{align}
\begin{align}
\blacktriangleright \;& \text{CP3:} \;(x,y,E)=(0,1+\alpha ,m_{3}), \nonumber \\ &(\lambda_{1},\lambda_{2},\lambda_{3})=(4-3A,4,4),\quad w_{\mathrm{eff.}}|_{\mathrm{CP3}}=\frac{1}{3},
\end{align}
where $m_{3}$ is an arbitrary value.\\
$\blacktriangledown$ \textbf{CP1:}\\
CP1 is not a critical \textit{point}, but it is a critical \textit{line} and its EoS parameter is $-1$. Hence, at first sight, it seems better to set $m_{3}=(\kappa^{2}H_{0}^{2})^{-1}$, and try to make this CP as an attractor/a saddle. But, according to the zero densities of viscous fluid and radiation, it is not a good candidate for the present status of the universe.\\
Anyway, the system is topologically equivalent to the following system at CP1-origin:
\begin{align}
u^{\prime}=& -3Au,\\
v^{\prime}=& -4v,\\
s^{\prime}=& \; 0+\mathcal{O}(s^{5}).
\end{align}
It means that if $A>0$, then in two directions we have attraction. On the center manifold the related variable, $s$, is approximately constant and it may be for the point that $E$ imposes a center manifold. Because its equation is satisfied automatically by $x$ and $y$, not $E$, so it makes $E^{\prime}=0$ for all values of $E$.\\
In the case $A<0$, the corresponding CP would be a saddle point for all values of $m_{3}$.\\

$\blacktriangledown$ \textbf{CP2:}\\
In order to put CP2 in the physical region, one must set $\alpha >-1$ and $m_{3} > 0$.
According to the EoS of CP2, it has this potential to be in the matter/radiation/current stage of the universe. Because $E=m_{3}$, thus $m_{3}$ should be fixed to the era in which we desire to insert this point.

$\bullet$ If it belongs to the matter-dominated era, then we must set $A=1$ which leads to $(\lambda_{1},\lambda_{2},\lambda_{3})=(-4,3,3)$. For sure, putting $A=1$ makes this CP a saddle point which is completely acceptable for a matter-dominated era. Note that, here with the assumed conditions, namely $\alpha_{1}=1$, $A=1$, and $B=0$, we have $p_{\mathrm{v}}=-\rho_{\mathrm{v}}+f+G=0$. Indeed, under the aforementioned conditions, the viscous fluid is converted to a pressureless matter. Furthermore, $\rho_{\mathrm{v}}>\rho_{\mathrm{r}}$ at this CP makes it a good candidate for the matter-dominated era.

$\bullet$ If a radiation-dominated era is desired, then $A=4/3$ must be adopted. It is worthy to note that at this CP, the density of the radiation component is zero, but under the conditions $A=4/3$, $B=0$, and $\alpha_{1}=1$, the viscous fluid is converted to the radiation, namely $p_{\mathrm{v}}=-\rho_{\mathrm{v}}+f+G=\rho_{\mathrm{v}}/3$. As a matter of fact, at this CP, the universe is filled only with the radiation.
This selection, $A=4/3$, yields $(\lambda_{1},\lambda_{2},\lambda_{3})=(0,4,4)$. It has a one-dimensional center manifold. If the system be transformed as $(x,y,E)|_{CP2} \longrightarrow (X,Y,R)=(0,0,0)$ and then we use the following transformations,
\begin{align}
X=v-s, \quad Y=s, \quad R=u,
\end{align}
we arrive at the following topologically equivalent system
\begin{align}
u^{\prime}=&\; 4u,\\
v^{\prime}=&\; 4v,\\
s^{\prime}=&\; 0+\mathcal{O}(s^{5}),
\end{align}
where we exploited the following mappings:
\begin{align}
u=& h_{1}(s)=k_{1}s^{2}+k_{2}s^{3}+k_{3}s^{4}+\mathcal{O}(s^{5}),\\
v=& h_{2}(s)=k_{4}s^{2}+k_{5}s^{3}+k_{6}s^{4}+\mathcal{O}(s^{5}).
\end{align}
As is clear, the center manifold, $s$, is approximately constant and this CP is a repeller. Therefore, this CP is not a good point for the radiation era.

$\bullet$ For considering CP2 as a candidate for the current status of the universe, it is better to take $A=-0.03$ instead of $A=0$ because of high accuracy\footnote{Note that, according to Planck data, we have $w_{\mathrm{eff.}}=-1.03 \pm 0.03$ at the present time.}. The case $A=-0.03$ leads to $w_{\mathrm{eff.}}|_{\mathrm{CP2}}=-1.03$ and $(\lambda_{1},\lambda_{2},\lambda_{3})=(-4.09,-0.09,-0.09)$. Therefore, it is an attractor fixed point which is a good option for the present status of our universe, especially with regards to the point that $\rho_{\mathrm{v}}>\rho_{\mathrm{r}}$, representing the domination of dark energy for the current position of the universe.\\
It is easily shown that $A=0$ for CP2 is also a good candidate for the present time.
The case $A=0$ yields $w_{\mathrm{eff.}}|_{\mathrm{CP2}}=-1$ and $(\lambda_{1},\lambda_{2},\lambda_{3})=(-4,0,0)$.
Since one of the eigenvalues is negative, hence, without the use of the center manifold theorem, one can state that this point is a saddle or an attractor point. Both are justifiable for the current status of the universe. Therefore, it is a good candidate for the current stage of the universe as well. However, the case $A=-0.03$ is better than $A=0$.\\
$\blacktriangledown$ \textbf{CP3:}\\
For this point, the conditions of being in the physical region are $\alpha >-1$ and $m_{3}>0$. According to the value of the EoS, $1/3$, this point belongs to a radiation-dominated era and thus it must be a saddle point, mandating $A>(4/3)$ to yield $(\lambda_{1},\lambda_{2},\lambda_{3})=(\text{`A negative number'},4,4)$. In addition, $m_{3}$ can be fixed to the radiation era (i.e. $m_{3}=(\kappa^{2} H^{2}_{\mathrm{r}})^{-1}$ in which $H_{\mathrm{r}}$ is the Hubble parameter corresponding to the radiation-dominated era). Under the aforementioned conditions, this point may be a good candidate for the radiation era, especially, adherent to the fact that at this CP we have $\rho_{\mathrm{r}}>\rho_{\mathrm{v}}$ which indicates the domination of radiation component.

\begin{figure*}
	\includegraphics[width=6.5 in]{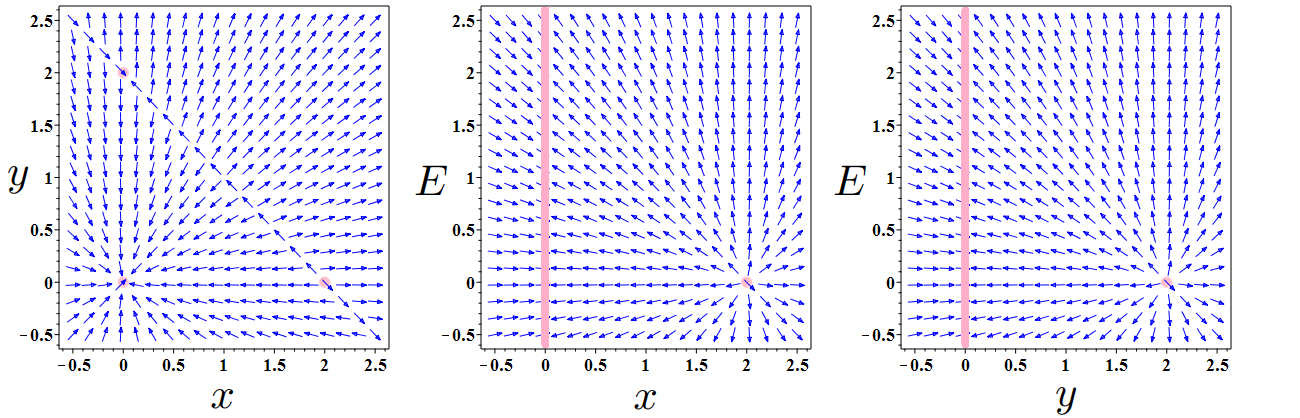}
    \caption{This figure illustrates three different Poincar\'{e} sections of system~(\ref{case1eq1})--(\ref{case1eq3}) when $B=0$. In plotting these phase-portraits, we have set $A=1.35$, and $\alpha =1$. The pink points (lines) refer to the critical points (lines).}
    \label{pa2}
\end{figure*}

In fig.~\ref{pa2}, three Poincar\'{e} sections of phase space of our system have been presented by setting $A=1.35$, and $\alpha =1$.\\

\noindent\pqrmybox{}{\subsubsection{Case-2:}}\vspace{5mm}
As the second example, let us consider the system for the selections $\alpha_{1}=2$ and $\beta=1$. The resulting dynamical system and the associated EoS parameter turn out to be
\begin{align}
x^{\prime} &=\frac{x-(1+\alpha)}{1+\alpha} \left(\frac{9A x^{2}}{\kappa^{4}E}+B \right)+\frac{4}{1+\alpha}xy, \label{case2eq1}
\end{align}
\begin{align}
y^{\prime} &=\frac{y}{1+\alpha} \left(\frac{9A x^{2}}{\kappa^{4}E}+B+4y \right)-4y, \label{case2eq2}
\end{align}
\begin{align}
E^{\prime} &=\frac{E}{1+\alpha} \left(\frac{9A x^{2}}{\kappa^{4}E}+B+4y \right); \label{case2eq3}
\end{align}
\begin{align}
w_{\mathrm{eff.}}=-1+\frac{1}{3(1+\alpha)}\left[\frac{9A x^{2}}{\kappa^{4} E}+B+4y \right].
\end{align}
$\clubsuit$ The case $B \neq 0$:\\
In this case, instead of a critical point, we have a critical curve (i.e. $E=Cx^{2}$ where $C=(-9A/(B\kappa^{2}))>0$) for a given amount of $A/B$:
\begin{align}
x= m_{1}, \quad y=0, \quad E=\frac{-9A m_{1}^{2}}{\kappa^{4} B},
\end{align}
where $m_{1}$ is an arbitrary constant.\\
Interestingly, all CPs of this curve yield
\begin{align}
w_{\mathrm{eff.}}=-1.
\end{align}
In order to put these CPs in the physical region, three conditions are required: $m_{1}>0$, $B \neq 0$, and $AB<0$. The corresponding eigenvalues take the forms:
\begin{align}
\lambda_{1}=\frac{B(2\alpha +2-m_{1})}{m_{1}(1+\alpha)} , \quad \lambda_{2}=0, \quad \lambda_{3}=-4.
\end{align}
The analysis of the CPs requires the use of the center manifold theorem because $\lambda_{2}=0$. Assuming $m_{1} \neq 2(1+\alpha)$, this system has a one-dimensional center manifold. Performing a transformation as $(x,y,E)|_{\mathrm{CPs}} \longrightarrow (X,Y,R)=(0,0,0)$ and then applying the transformation
\begin{align}
X=& \left[\frac{B\kappa^{2}(1+\alpha -m_{1})}{9Am_{1}^{2}} \right]u \nonumber \\ &-\left[\frac{B\kappa^{2}(B+4m_{1})}{18Am_{1}(B+2m_{1})} \right]v-\left[\frac{B \kappa^{2}}{18Am_{1}} \right]s,\\
Y=& \left[\frac{B\kappa^{2}(2\alpha B-Bm_{1}+4\alpha m_{1}+2B+4m_{1})}
{18Am_{1}^{2}(B+2m_{1})} \right]v,\\
R=& \; u+v+s,
\end{align}
our system will be expressed in a new coordinate $(u,v,s)$. By the use of two mappings
\begin{align}
u=&\;h_{1}(s)= k_{1}s^{2}+k_{2}s^{3}+k_{3}s^{4}+\mathcal{O}(s^{5}),\\
v=&\;h_{2}(s)= k_{4}s^{2}+k_{5}s^{3}+k_{6}s^{4}+\mathcal{O}(s^{5}),
\end{align}
we arrive at a system that is topologically equivalent to the original system at its CPs:
\begin{align}
u^{\prime}=& \left[\frac{B(2\alpha +2-m_{1})}{m_{1}(1+\alpha)} \right]u,\label{e-pa-3-1}\\
v^{\prime}=& -4v, \label{e-pa-3-2}\\
s^{\prime}=& \left[\frac{2B^{2}\kappa^{4}(1+\alpha -m_{1})}{36m_{1}^{2}(1+\alpha )(2+2\alpha -m_{1})} \right]s^{2} \nonumber \\ &+\left[\frac{5B^{4}\kappa^{12}(m_{1}-1-\alpha
)}{23328 A^{3}m_{1}^{6}(1+\alpha)(2+2\alpha -m_{1})^{3}} \right]s^{4}+\mathcal{O}(s^{5}).\label{e-pa-3-3}
\end{align}
The center manifold is repeller and hence it makes the CPs unstable and saddle. Setting
\begin{align*}
\frac{B(2\alpha +2-m_{1})}{m_{1}(1+\alpha)} < 0,
\end{align*}
engenders attraction property in two directions and hence under this condition and according to the unstable nature of our universe, one may conclude that the CPs in which we have set $B \kappa^{2} H_{0}^{2}+9Am_{1}^{2}=0$ (for putting them in the current time) are good candidates for the present status of the universe. It is worthwhile to mention that for non-zero values of $m_{1}$, one has $\rho_{\mathrm{v}}>\rho_{\mathrm{r}}$ which refers to the domination of dark energy.\\

\begin{figure*}
	\includegraphics[width=6.5 in]{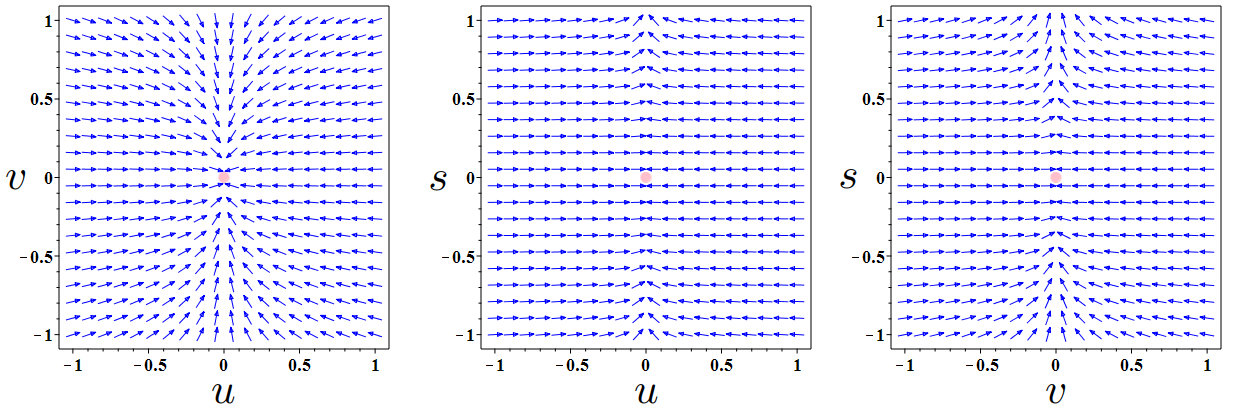}
    \caption{This figure illustrates three different Poincar\'{e} sections of phase space of the system~(\ref{e-pa-3-1})--(\ref{e-pa-3-3}). In plotting these diagrams, we have assumed $\kappa^{2}=1$, $A=1$, $B=-9$, $m_{1}=1$, and $\alpha =1$. The pink points show the critical points.}
    \label{pa3}
\end{figure*}

As an example, three different Poincar\'{e} sections of the phase space of the above system have been presented in fig.~\ref{pa3} by setting $\kappa^{2}=1$, $A=1$, $B=-9$, $m_{1}=1$, and $\alpha =1$.\\

$\clubsuit$ The case $B = 0$:\\
Setting $B=0$, reduces the system to the following straight line of critical points:
\begin{align}
x=0, \quad y=0, \quad E=\text{an arbitrary value}\equiv m_{3},
\end{align}
with the EoS parameter
\begin{align}
w_{\mathrm{eff.}}|_{CPs}=-1,
\end{align}
and these eigenvalues
\begin{align}
\lambda_{1}=-4, \quad \lambda_{2}=\lambda_{3}=0.
\end{align}
According to the value of EoS parameter, this case may be set to the current time when one takes $m_{3}=(\kappa^{2}H_{0}^{2})^{-1}$, but this CP would not be a good case for the present status of the universe because $\rho_{\mathrm{v}}=\rho_{\mathrm{r}}=0$. Indeed, these CPs correspond to a vacuum universe.\\
Anyway, since $\lambda_{1}=-4<0$, so this critical line has an attraction in one direction, implying that all CPs of this line would be attractor or saddle points. Let's pause and consider this problem deeply using the center manifold theorem. Clearly, the corresponding center manifolds are two-dimensional. Using two transformations, first $(x,y,E)|_{CPs} \longrightarrow (X,Y,R)=(0,0,0)$ and then
\begin{align}
X=&\; s,\\
Y=&\; \left(\frac{\alpha +1}{-m_{3}} \right)u,\\
R=&\; u+v,
\end{align}
one may get the topologically equivalent system:
\begin{align}
u^{\prime}=& \; -4u, \label{e-pa-4-1}\\
v^{\prime}=& \; \left[\frac{-9A}{\kappa^{4} (1+\alpha )} \right]
\left(\frac{(1+\alpha)s^{2}+s^{3}}{m_{3}+v} \right)+\cdots, \label{e-pa-4-2}\\
s^{\prime}=& \; \left(\frac{9A}{\kappa^{4} (1+\alpha )} \right)s^{2}+\cdots, \label{e-pa-4-3}
\end{align}
where we have used the mapping
\begin{align*}
u=k_{1}v^{2}+k_{2}vs+k_{3}s^{2}+\cdots.
\end{align*}
As seen, the plane of the center manifold, $v-s$, is completely unstable and all CPs are saddle.\\

\begin{figure*}
	\includegraphics[width=6.5 in]{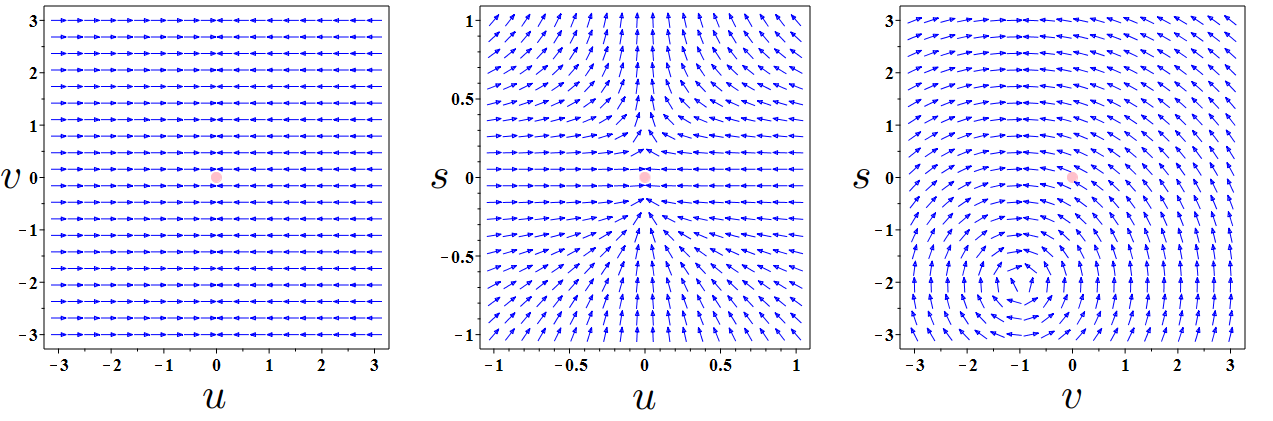}
    \caption{This figure indicates three different Poincar\'{e} sections of phase space of the system~(\ref{e-pa-4-1})--(\ref{e-pa-4-3}). In plotting these portraits, we have taken $\kappa^{2}=1$, $A=1$, $m_{3}=1$, and $\alpha =1$. The pink points represent the critical points.}
    \label{pa4}
\end{figure*}

As an example, three different Poincar\'{e} sections of the phase space of the above system have been presented in fig.~\ref{pa4} in which we have set $\kappa^{2}=1$, $A=1$, $m_{3}=1$, and $\alpha =1$.\\

\noindent\pqrmybox{}{\subsubsection{Case-3:}}\vspace{5mm}
As the third case, let us consider the system with the choices $\alpha_{1}=2$ and $\beta=2$, then the dynamical system takes the following form:
\begin{align}
x^{\prime} &=\frac{x-(1+\alpha )}{1+\alpha } \left(\frac{9A x^{2}}{\kappa^{4}E}+\frac{B}{E} \right)+\frac{4}{1+\alpha }xy, \label{case3eq1}\\
y^{\prime} &=\frac{y}{1+\alpha } \left(\frac{9A x^{2}}{\kappa^{4}E}+\frac{B}{E}+4y \right)-4y, \label{case3eq2}\\
E^{\prime} &=\frac{E}{1+\alpha } \left(\frac{9A x^{2}}{\kappa^{4}E}+\frac{B}{E}+4y \right). \label{case3eq3}
\end{align}
The associated EoS parameter also reads:
\begin{align}
w_{\mathrm{eff.}}=-1+\frac{1}{3(1+\alpha )}\left[\frac{9A x^{2}}{\kappa^{4}E}+\frac{B}{E}+4y  \right].
\end{align}

$\clubsuit$ The case $B \neq 0$:\\
In this case, all the points of $x-E$ plane can be the fixed point to our system:
\begin{align}
x= \epsilon \kappa^{2}\; \sqrt{\frac{-B}{9A}} \equiv m_{1}, \quad y=0, \quad
E= m_{3},
\end{align}
where $m_{3}$ is an arbitrary constant, and $\epsilon=\pm 1$. However, both $\epsilon=+1$ and $\epsilon= -1$ are acceptable from a mathematical point of view, but only $\epsilon=+ 1$ is acceptable from a physical perspective. Besides this condition, one must have $AB \leq 0$, $A \neq 0$, and $m_{3} > 0$ to maintain these CPs in the physical region. The corresponding EoS parameter at the CP-plane turns out to be
\begin{align}
w_{\mathrm{eff.}}|_{\mathrm{CPs}}=-1.
\end{align}
The eigenvalues of the Jacobian matrix at the CPs would be
\begin{align}
\lambda_{1}=\frac{18Am_{1}(m_{1}-1-\alpha )}{m_{3} \kappa^{4} (1+\alpha )}, \quad \lambda_{2}=-4, \quad \lambda_{3}=0.
\end{align}
According to the value of the EoS parameter, the conditions $\lambda_{1}<0$, $m_{1}>0$, and $m_{3}=(\kappa^{2} H_{0}^{2})^{-1}$ ($H_{0}$ is the current value of the Hubble parameter) should be adopted. In fact, $\lambda_{1}<0$ enhances the number of attractor directions, $m_{1}>0$ guarantees the domination of dark energy, and the condition $m_{3}=(\kappa^{2} H_{0}^{2})^{-1}$ fixes the CP at the current time.
Under these conditions, the related CP would be a good case for the present time.\\

Let us consider this CP deeply using the center manifold theorem. First, we transform the system into the CP-origin (i.e. $(x,y,E)|_{\mathrm{CP}} \longrightarrow (X,Y,R)=(0,0,0)$). Then utilizing the relations
\begin{align}
X=& \left[\frac{m_{1}-1-\alpha}{m_{3}} \right]u + \left[\frac{2\kappa^{4} m_{1}}{2\kappa^{4} m_{3}-9Am_{1}} \right]v, \\
Y=& \left[\frac{2\kappa^{4}m_{3}\alpha +2\kappa^{4}m_{3}-9Am_{1}\alpha +9Am_{1}^{2}-9Am_{1}}{9Am_{1}-2\kappa^{4} m_{3}} \right]v,\\
R=& \; u+v+s,
\end{align}
our system can be transformed into a new coordinate $(u,v,s)$. Using two mappings $h_{1}$ and $h_{2}$ as
\begin{align}
u=& h_{1}(s)=k_{1}s^{2}+k_{2}s^{3}+k_{3}s^{4}+\mathcal{O}(s^{5}),\\
v=& h_{2}(s)=k_{4}s^{2}+k_{5}s^{3}+k_{6}s^{4}+\mathcal{O}(s^{5}),
\end{align}
one arrives at the following topologically equivalent system
\begin{align}
u^{\prime}=& \left[\frac{18Am_{1}(m_{1}-1-\alpha )}{m_{3} \kappa^{4} (1+\alpha )} \right]u, \label{e-pa-5-1}\\
v^{\prime}=& \; -4v, \label{e-pa-5-2}\\
s^{\prime}=& \; 0+\mathcal{O}(s^{5}). \label{e-pa-5-3}
\end{align}
Clearly, for the CP under the aforementioned conditions (i.e. $\lambda_{1}<0$, $\&$ $m_{1}>0$, $\&$ $m_{3}=(\kappa^{2} H_{0}^{2})^{-1}$), this system indicates attractor behavior for the present stage of the universe which is justifiable. The center manifold, $s$, is approximately constant, hence it is neither attractor nor repeller.
Indeed, a particle on the center manifold is in the rest status so that there is no force on it and the particle continues to move at the speed it has arrived at this point.\\

\begin{figure*}
	\includegraphics[width=6.5 in]{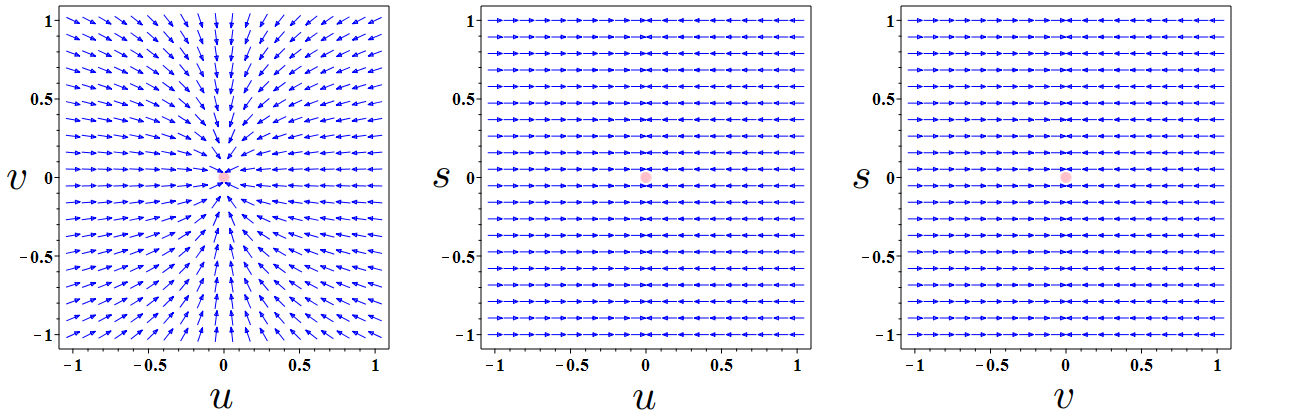}
    \caption{This figure shows three different Poincar\'{e} sections of phase space of the system~(\ref{e-pa-5-1})--(\ref{e-pa-5-3}). In plotting these sections of phase-portrait, we have selected $\kappa^{2}=1$, $A=1$, $B=-9$, $m_{1}=1$, $m_{3}=1$, and $\alpha =1$. The pink points represent the critical points.}
    \label{pa5}
\end{figure*}

As an example, three different Poincar\'{e} sections of the phase space of the above system have been illustrated in fig.~\ref{pa5} in which we have chosen $\kappa^{2}=1$, $A=1$, $B=-9$, $m_{1}=1$, $m_{3}=1$, and $\alpha =1$.\\

$\clubsuit$ The case $B=0$:\\
Under the condition $B=0$, the special case, $B=0$, of ``\textit{case-2}'' is recovered.\\

\noindent\pqrmybox{}{\subsubsection{An important conclusion of ``More complicated forms of the EoS'' and a remedy for its main problem}}\vspace{5mm}
\begin{table*}
\caption{A short review of some features of ``More Complicated forms of the EoS''.} \label{tableokh}
\centering
\begin{tabular}{|p{1.6 in} |p{1.2 in} |p{1.2 in} |p{1.2 in}|}
\toprule[1.7pt]
\textbf{Case} &\textbf{The value of $B$} & \textbf{The number of CPs} & \textbf{The values of the effective EoS of the CPs}
\\
\toprule[1.4pt]
\multirow{2}{*}{Case-1 ($\alpha_{1}=+1$; \; $\beta=+2$)} & $B \neq 0$ & $1$ & $-1$ \\
& $B = 0$ & $3$ & $-1$ \; ; \; $1/3$ \; ; \; $A-1$
  \\ \toprule[1pt]
\multirow{2}{*}{Case-2 ($\alpha_{1}=+2$; \; $\beta=+1$)} & $B \neq 0$ & $1$ & $-1$ \\
& $B = 0$ & $1$ & $-1$
  \\ \toprule[1pt]
  \multirow{2}{*}{Case-3 ($\alpha_{1}=+2$; \; $\beta=+2$)} & $B \neq 0$ & $1$ & $-1$ \\
& $B = 0$ & $1$ & $-1$
\\ \toprule[1.7pt]
\end{tabular}
\end{table*}
First of all, check Table~\ref{tableokh}. In case $B \neq 0$, the critical points of `\textit{case-1}' for different values of the constant parameters could span a plane (i.e. $x-E$), but in that case, all the points of that plane had $w_{\mathrm{eff.}}=-1$. Therefore, if one tunes the constant parameters for the present time, the system gives us only one critical point and it may describe only one stage of the universe. In the case $B=0$ we saw that it could act in a better way with more fixed points and stages. On the other hand, this occurred only for `\textit{case-1}'.  Perhaps, one may argue that such a behaviour stems from the viscous part of the system restricting the behavior of the system and not generating the last stages of the universe in the current model. But, the power of density, $\alpha_{1}$, also may have such behavior if it is greater than $1$. For sure, $f(\rho)=A \rho^{\alpha_{1} =+1}$ is better than $f(\rho)=A \rho^{\alpha_{1} =+2}$ in the current model of study.\\
According to the above achievements, it is inspired that the absence of some CPs and stages of the evolution of the universe may be due to the fact that we have considered those cases that their $\alpha_{1}$s are positive.
Nevertheless, $\alpha_{1}$ with negative values could accentuate the significance of the viscous part. \textit{Note that the negative values of $\beta$ cannot be related to the main problem, because the viscous part, $H^{2\beta}$, is highly likely to be the case at early times, so we deduce that $\beta >0$ would be its representative.}\\
Let us delve deeper into the subject by specifying some values for the constant parameters first.\\
Let $\alpha_{1} =-1$ and $\beta =+1$. Fortunately, these selections lead to an analytically solvable case without further quantifying the constant parameters. The CPs and corresponding eigenvalues and associated EoS parameters are as follows:
\begin{align}
\blacktriangledown \; &\text{CP1:} \nonumber\\ &(x,y,E)=\left(\frac{-A \kappa^{6}m_{3}^{2}}{3B},0,m_{3} \right),\quad
w_{\mathrm{eff.}}|_{\mathrm{CP1}}=-1, \nonumber\\
& (\lambda_{1},\lambda_{2},\lambda_{3})=\left(-4, \frac{3B^{2}+3B^{2}\alpha -AB m_{3}^{2}\kappa^{6}}{Am_{3}^{2}\kappa^{4}(1+\alpha)},0 \right);
\end{align}
\begin{align}
\blacktriangledown \; & \text{CP2:} \nonumber\\ & (x,y,E)=\left(\frac{B \kappa^{2}}{4},1+\alpha - \frac{B \kappa^{2}}{4},0 \right),\nonumber\\ & w_{\mathrm{eff.}}|_{\mathrm{CP2}}=\frac{1}{3},
\quad (\lambda_{1},\lambda_{2},\lambda_{3})=\left(4- \frac{B \kappa^{2}}{1+\alpha },4,4 \right);
\end{align}
\begin{align}
\blacktriangledown \; & \text{CP3:} \nonumber\\ & (x,y,E)=\left(1+\alpha ,0,0 \right),\quad w_{\mathrm{eff.}}|_{\mathrm{CP3}}=\frac{1}{3}\left(\frac{B \kappa^{2}}{1+\alpha } \right)-1, \nonumber\\
& (\lambda_{1},\lambda_{2},\lambda_{3})
=\frac{1}{1+\alpha }\left( B \kappa^{2}, B \kappa^{2}-4(1+\alpha ), B \kappa^{2}  \right).
\end{align}
Clearly, the role of the viscous part is significant. Especially, in CP3, this impression is very obvious as it can be used to put this CP in an arbitrary stage of the evolution of the universe. Furthermore, as is observed, $\alpha$ (the coefficient of torsion in the model) acts as assistance for tuning.\\
$\blacktriangledown$ \textbf{CP1:}\\
CP1 can be fixed to the current stage of the universe by adopting the following conditions which guarantee almost all aspects of it (domination of dark energy, fitting to the current stage, attractor property, and being in the physical region):
\begin{align}
m_{3}=\frac{1}{\kappa^{2}H^{2}_{0}}, \quad B \neq 0, \quad AB<0, \quad \lambda_{2}<0.
\end{align}
$\blacktriangledown$ \textbf{CP2:}\\
CP2 should be regarded as a candidate for the radiation-dominated era because of the value of the EoS parameter. The conditions reassuring the presence of this CP in the physical region are:
\begin{align}
B>0, \quad \& \quad 4(1+\alpha)> B \kappa^{2}.
\end{align}
Under the condition
\begin{align}
2(1+\alpha)> B \kappa^{2},
\end{align}
the domination of the radiation component is satisfied, and the condition
\begin{align}
4 < \frac{B \kappa^{2}}{1+\alpha},
\end{align}
makes this CP as a saddle point. The common domain of all these conditions tends to a good CP for the radiation era.\\
$\blacktriangledown$ \textbf{CP3:}\\
By setting $\alpha > -1$, this point falls into a physical region in which dark energy dominates.
The status of CP3 is affected by the conditions imposed on CP1 and CP2. After tuning the constant parameters, the situation of this CP is clarified.\\
According to the EoS of this CP, it seems that it is flexible, but one must note that this point should be in a dark-energy-dominated era because of $\rho_{\mathrm{v}}>\rho_{\mathrm{r}}$.
Furthermore, the viscous fluid could not be converted to other types of components (i.e. matter and radiation) due to the imposed conditions.

According to the above explanations and examinations, the main problem has been solved by switching to negative values of $\alpha_{1}$. \textit{This change means that the Chaplygin is a better candidate to couple with the viscous part than Van der Waals gas.} Last but not least, since the negative values of $\alpha_{1}$ have strong correspondence with Chaplygin gas, the best value for $\alpha_{1}$ may be in the range $-1.8 \leq \alpha_{1} \leq -0.2$. In ref.~\cite{beh-j}, it was indicated that the constant power in the generalized Chaplygin gas model, namely $\nu$ in $P=-\sigma_{2}/ \rho^{\nu}$, might not be absolutely constant, but could
be slowly varying with cosmic time. The reason for this theory was to reach the unification theory. More precisely, the power of the density is a time-dependent variable falling so slowly from an initial constant value within $0.2 \leq \nu \leq 1.8$ to zero. However, it is better to take the aforementioned initial domain of $\nu$ as $| \nu | \lesssim 0.05$ when we focus only on the dark sector~\cite{beh-j}.\\
These discussions guide us to develop the EoS of viscous fluid by making $\alpha_{1}$ and $\beta$ dependent upon the cosmic time:
\begin{equation*}
P_{\mathrm{v}}=\sigma_{1} \rho_{\mathrm{v}}-\frac{\sigma_{2}}{\rho^{\nu(t)}_{\mathrm{v}}}
+BH^{\beta(t)},
\end{equation*}
where $\sigma_{1}$, $\sigma_{2}$, and $B$ are constant parameters.
Considering such intricate generalized cases are beyond the scope of this paper, hence we terminate this section here.\\

\noindent\hrmybox{}{\section{Some interacting models\label{sect-V}}}\vspace{5mm}
In this section, some non-trivial interacting models are studied.
To begin with, using a power-law term a modified type of interaction is investigated in detail. Then, the oscillating type of interaction (`Cosine' type) which has been recently introduced by S.D. Odintsov et al.~\cite{od-asli} is considered widely. Furthermore, some new types of oscillating interactions are suggested instead of the aforementioned type. It is illustrated that through some enhancements and modifications to the oscillating type, more proportionate results are derived in comparison to that of the basic model. Finally, the benefits of the oscillating types of the interacting models are indicated that are: ``Unification between the early inflation and late-time-accelerated expansion'' and ``The crucial role of the oscillating types of interactions in getting out of some special eras that were under some dominations''.\\

\noindent\Rrmybox{}{\subsection{Interaction through modified term (using some power-law terms)}}\vspace{5mm}
As the first non-trivial case, the model which has been studied in sect.~\ref{sect-III} is developed here by utilizing the multiplier $H^{2j}$ where $j$ is a constant~\cite{od-asli}. Let's perform the replacement~\cite{od-asli}
$$
\frac{Q}{H} \longrightarrow \frac{Q}{H} H^{2j}
$$
in the equations of $\rho^{\prime}_{\mathrm{v}}$ and $\rho^{\prime}_{\mathrm{dm}}$.
Again, let us assume the following form for the bulk viscous coefficient $\zeta$:
\begin{align}
\zeta =\frac{\zeta_{0}}{\sqrt{3 \kappa^{2}}} \rho^{1/2}_{\mathrm{tot.}}
=\frac{1}{\kappa^{2}}H \zeta_{0}.
\end{align}
The dynamical system, in this case, is expressed as follows:
\begin{equation}\begin{split}
\frac{3}{\kappa^{2}}H^{2}=&\frac{\sqrt{6}H}{2\sqrt{6}(1+\alpha)H-2c_{1}}\\
&\times \left(\rho-\frac{1}{2\kappa^{2}} \left[-6\alpha H^{2}
+2\sqrt{6} \; c_{1}H+c_{2} \right] \right), \label{sect5eq1}
\end{split}\end{equation}
\begin{align}
\frac{-2}{\kappa^{2}}H H^{\prime} =&\frac{\rho+P}{1+\alpha},\label{sect5eq2}\\
\rho^{\prime}_{\mathrm{r}}=&-4 \rho_{\mathrm{r}} ,\label{sect5eq3}\\
\rho^{\prime}_{\mathrm{v}}=&-3[f+G]-\frac{Q}{H}H^{2j} ,\label{sect5eq4}\\
\rho^{\prime}_{\mathrm{dm}}=&-3\rho_{\mathrm{dm}}+\frac{Q}{H}H^{2j}-3(-3H\zeta)  .\label{sect5eq5}
\end{align}
Defining the dimensionless variables as
\begin{align}
&x \equiv \frac{\kappa^{2} \rho_{\mathrm{v}}}{3H^{2}}, \quad
y \equiv \frac{\kappa^{2} \rho_{\mathrm{dm}}}{3H^{2}}, \quad
z \equiv \frac{\kappa^{2} \rho_{\mathrm{r}}}{3H^{2}}, \nonumber \\
&q \equiv \frac{\kappa^{2} Q}{3H^{3}}, \quad
E \equiv \frac{1}{\kappa^{2} H^{2}},
\end{align}
we get
\begin{align}
x^{\prime} =&\frac{3x}{1+\alpha} U-\kappa^{4}E(f+G)-q\left(\frac{1}{\kappa^{2}E} \right)^{j}, \label{sect5eq6}
\end{align}
\begin{align}
y^{\prime} =&\frac{3y}{1+\alpha} U-3y-q\left(\frac{1}{\kappa^{2}E} \right)^{j}+3\zeta_{0}, \label{sect5eq7}
\end{align}
\begin{align}
z^{\prime} =&\frac{3z}{1+\alpha} U-4z, \label{sect5eq8}
\end{align}
\begin{align}
E^{\prime} =&\frac{3E}{1+\alpha} U, \label{sect5eq9}
\end{align}
in which
\begin{align}\label{sect5eq10}
U=\frac{1}{3}\kappa^{4} E(f+G)+\frac{4}{3}z.
\end{align}
The effective EoS in terms of unknown function $U$ for this case takes the form
\begin{align}\label{sect5eq11}
w_{\mathrm{eff.}}=-1+\frac{1}{1+\alpha}U.
\end{align}
Now, using $f=\rho_{\mathrm{v}}(1+w_{0})$ and $G=w_{1}H^{2}$, the equations of our dynamical system become:
\begin{align}
x^{\prime} =&\frac{x}{1+\alpha}\left[3(1+w_{0})x+3y+4z+w_{1}\kappa^{2} \right]
\nonumber \\ &-3(1+w_{0})x-q \left(\frac{1}{\kappa^{2} E} \right)^{j}-w_{1}\kappa^{2},  \label{sect5eq12}\\
y^{\prime} =&\frac{y}{1+\alpha}\left[3(1+w_{0})x+3y+4z+w_{1}\kappa^{2} \right]
\nonumber \\ &-3y+q \left(\frac{1}{\kappa^{2} E} \right)^{j} +3\zeta_{0}, \label{sect5eq13}\\
z^{\prime} =&\frac{z}{1+\alpha}\left[3(1+w_{0})x+3y+4z+w_{1}\kappa^{2} \right]-4z,  \label{sect5eq14}\\
E^{\prime} =&\frac{E}{1+\alpha}\left[3(1+w_{0})x+3y+4z+w_{1}\kappa^{2} \right]. \label{sect5eq15}
\end{align}
Furthermore, the effective EoS parameter turns out to be:
\begin{align}
w_{\mathrm{eff.}}=-1+\frac{1}{3+3\alpha}
\left[3(1+w_{0})x+3y+4z+w_{1}\kappa^{2} \right].\label{sect5eq16}
\end{align}
Investigating this case without any assumption is challenging. This behavior was predictable according to the simpler case which has been considered in sect.~\ref{sect-III}. Hence, we prefer to go about some special cases. Like ref.~\cite{od-asli}, let us consider three cases: $j=0$, $j=-1$, and $j=+1$. After studying these cases, some discussions about $j \leq 0$ and $j>0$ are performed and the behaviors of these general cases are clarified.\\

\noindent\pqrmybox{}{\subsubsection{The case $j=0$}}\vspace{5mm}
The case $j=0$ corresponds to the usual interaction. In fact, when one ignores $E$, the equations of sect.~\ref{sect-III} are recovered, so those points with $E=0$, are common between this case and the one considered in sect.~\ref{sect-III}. For this case, all CPs and eigenvalues can be reached analytically but they are `VLT' such that we cannot present them excluding one CP.
The presentable CP is as follows:
\begin{align}
&E=0, \quad
x=\frac{q+w_{1}\kappa^{2}}{1-3w_{0}}, \quad y=-q-3\zeta_{0}, \nonumber \\
&z=\frac{(12q+12\alpha+27\zeta_{0}+12)w_{0}+4w_{1}\kappa^{2}
-4\alpha-9\zeta_{0}-4}{4(3w_{0}-1)}.
\end{align}
Pay attention to its correspondence with~(\ref{01cpIII}). This CP belongs to the radiation-dominated era because of $w_{\mathrm{eff.}}|_{\mathrm{CP.}}=1/3$. The eigenvalues of this CP are `VLT' except one eigenvalue $\lambda_{1}=+4$. Other eigenvalues have many constant parameters that may be fixed such that they tend to a saddle nature for this CP. Anyway, according to $\lambda_{1}=+4$, one may conclude instability for this CP --- it can be a saddle or a repeller point (note that saddle is acceptable). It is interesting to note that $\lambda_{1}=+4$ is the eigenvalue of $E$-direction and it refers to the expansion nature of the universe and decreasing nature of Hubble parameter with time. It may be concluded that the expansion feature of the universe causes its unstable nature. \\
Other CPs and their eigenvalues and EoS parameters are `VLT' so that they propose nothing interesting excluding one common property: All other CPs are in $x-y$ plane for which we have $E=z=0$. It means that all the remained CPs belong to the accelerated era and, moreover, they can be used for fixing some relations between $\rho_{\mathrm{v}}$ and $\rho_{\mathrm{dm}}$. Indeed, all other CPs are on the surface which describes interactions between dark matter and dark energy. It is notable that on this surface, these CPs show zero value for the density of the radiation.

The discussiblity of the CPs of the above system comes through fixing some relations among constant parameters or tuning constant parameters numerically merely.\\
As an example, let us present a case with the fixed condition $q=-w_{1} \kappa^{2}$. Under this condition, four CPs are obtained as follows:
\begin{align*}
&\blacktriangledown \text{CP1:}\\
&\left[
   \begin{array}{c}
     x \\
     y \\
     z \\
     E \\
   \end{array}
 \right]=\left[
   \begin{array}{c}
     0 \\
     \kappa^{2}w_{1}-3\zeta_{0} \\
     1+\alpha -\kappa^{2} w_{1}+\frac{9\zeta_{0}}{4} \\
     0 \\
   \end{array}
 \right], \quad w_{\mathrm{eff.}}|_{\mathrm{CP1}}=\frac{1}{3},
\end{align*}
\begin{align*}
&\left[
   \begin{array}{c}
     \lambda_{1} \\
     \lambda_{2} \\
     \lambda_{3} \\
     \lambda_{4} \\
   \end{array}
 \right]=\left[
   \begin{array}{c}
     +4 \\
     1-3w_{0} \\
     \frac{\chi_{2}-\kappa^{2}w_{1}+5(1+\alpha )}{2(1+\alpha )} \\
     \frac{-\chi_{2}-\kappa^{2}w_{1}+5(1+\alpha )}{2(1+\alpha )} \\
   \end{array}
 \right];
\end{align*}
\begin{align*}
&\blacktriangledown \text{CP2:}\\
&\left[
   \begin{array}{c}
     x \\
     y \\
     z \\
     E \\
   \end{array}
 \right]=\left[
   \begin{array}{c}
     0 \\
     \frac{\chi_{2}-\kappa^{2}w_{1}+3(1+\alpha )}{6} \\
     0 \\
     0 \\
   \end{array}
 \right],
\end{align*}
\begin{align*}
&w_{\mathrm{eff.}}|_{\mathrm{CP2}}=
 \frac{\chi_{2}+\kappa^{2}w_{1}-3(1+\alpha )}{6(1+\alpha )},
\end{align*}
\begin{align*}
&\left[
   \begin{array}{c}
     \lambda_{1} \\
     \lambda_{2} \\
     \lambda_{3} \\
     \lambda_{4} \\
   \end{array}
 \right]=\left[
   \begin{array}{c}
     \frac{\chi_{2}}{1+\alpha } \\
     \frac{\chi_{2}+\kappa^{2}w_{1}+3(1+\alpha )}{2(1+\alpha )} \\
     \frac{\chi_{2}+\kappa^{2}w_{1}-5(1+\alpha )}{2(1+\alpha )} \\
     \frac{\chi_{2}+\kappa^{2}w_{1}-3(1+2w_{0})(1+\alpha )}{2(1+\alpha )} \\
   \end{array}
 \right];
\end{align*}
\begin{align*}
&\blacktriangledown \text{CP3:}\\
&\left[
   \begin{array}{c}
     x \\
     y \\
     z \\
     E \\
   \end{array}
 \right]=\left[
   \begin{array}{c}
     \frac{-\kappa^{2}w_{0}w_{1}+3\alpha w^{2}_{0}-\kappa^{2}w_{1}+3\alpha w_{0}+3w^{2}_{0}+3\zeta_{0}+3w_{0}}{3w_{0}(1+w_{0})} \\
     \frac{\kappa^{2}w_{1}-3\zeta_{0}}{3w_{0}} \\
     0 \\
     0 \\
   \end{array}
 \right],
\end{align*}
\begin{align*}
&w_{\mathrm{eff.}}|_{\mathrm{CP3}}= w_{0},
\end{align*}
\begin{align*}
&\left[
   \begin{array}{c}
     \lambda_{1} \\
     \lambda_{2} \\
     \lambda_{3} \\
     \lambda_{4} \\
   \end{array}
 \right]=\left[
   \begin{array}{c}
     3w_{0}-1 \\
     3w_{0}+3 \\
     \frac{\chi_{2}-\kappa^{2}w_{1}+3(1+2w_{0})(1+\alpha )}{2(1+\alpha )} \\
     \frac{-\chi_{2}-\kappa^{2}w_{1}+3(1+2w_{0})(1+\alpha )}{2(1+\alpha )} \\
   \end{array}
 \right];
\end{align*}
\begin{align*}
&\blacktriangledown \text{CP4:} \\
&\left[
   \begin{array}{c}
     x \\
     y \\
     z \\
     E \\
   \end{array}
 \right]=\left[
   \begin{array}{c}
     0 \\
     \frac{-\kappa^{2}w_{1}+3(1+\alpha )-\chi_{2}}{6} \\
     0 \\
     0 \\
   \end{array}
 \right],
\end{align*}
\begin{align*}
&w_{\mathrm{eff.}}|_{\mathrm{CP4}}=
 \frac{-\chi_{2}+\kappa^{2}w_{1}-3(1+\alpha )}{6(1+\alpha )},
\end{align*}
\begin{align*}
&\left[
   \begin{array}{c}
     \lambda_{1} \\
     \lambda_{2} \\
     \lambda_{3} \\
     \lambda_{4} \\
   \end{array}
 \right]=\left[
   \begin{array}{c}
     \frac{-\chi_{2}}{1+\alpha } \\
     \frac{-\chi_{2}+\kappa^{2}w_{1}+3(1+\alpha )}{2(1+\alpha )} \\
     \frac{-\chi_{2}+\kappa^{2}w_{1}-5(1+\alpha )}{2(1+\alpha )} \\
     \frac{-\chi_{2}+\kappa^{2}w_{1}-3(1+2w_{0})(1+\alpha )}{2(1+\alpha )} \\
   \end{array}
 \right],
\end{align*}
where $\chi_{2}=(\kappa^{4}w^{2}_{1}+6\kappa^{2}w_{1}\alpha +6\kappa^{2}w_{1}+9\alpha^{2}-36 \alpha \zeta_{0}+18\alpha -36\zeta_{0}+9 )^{1/2}$.\\
$\blacktriangledown$ \textbf{CP1:}\\
It is clear that this CP belongs to the radiation-dominated era because of its EoS. At this point, dark energy has no any effect. For maintaining this point in the physical region, we must set the following condition:
\begin{align}
\left(1+\alpha -\kappa^{2}w_{1}+\frac{9}{4}\zeta_{0}\right) >\left(\kappa^{2}w_{1}-3\zeta_{0}\right) \geq 0.
\end{align}
Note that after the first part of the above condition, there is ``$ > 0$'' while after the second, we see ``$ \geq 0$''. The reason is that this CP is in the radiation-dominated era hence the associated $\rho_{\mathrm{r}}=0$ is meaningless. Furthermore, this condition provides the domination of the radiation component.\\
We have a fixed relation between dark matter and radiation at this CP:
\begin{align}
\rho_{\mathrm{dm}}=\left(\frac{4\kappa^{2}w_{1}-12\zeta_{0}}{4(1+\alpha )-4\kappa^{2}w_{1}+9\zeta_{0}} \right) \rho_{\mathrm{r}}.
\end{align}
According to $\lambda_{1}=+4$, this CP is unstable. To obtain a physical behavior for CP1, we must take one of the remained eigenvalues less than zero. This leads to saddle nature for CP1 which is desirable.\\
$\blacktriangledown$ \textbf{CP2, CP3, and CP4:}\\
The conditions for having these CPs in physical regions are clear; it is sufficient to keep the non-zero coordinates positive.
According to the coordinates of these CPs, CP3 can be fixed for the dark-energy-dominated era while CP2 and CP4 are neither suitable for the present stage nor for the radiation-dominated era. In fact, CP2 and CP4 belong to an era in which dark matter dominates. If it is desired to convert the dark matter to pressureless matter, then putting $\zeta_{0}=0$ would suffice by which the pressure of dark matter vanishes, and the interaction between dark matter and dark energy becomes the interaction between a pressureless matter and dark energy. For example, by the use of the selections
\begin{align}
&\kappa^{2}=1, \quad \zeta_{0}=0, \quad w_{0}=-1.03, \quad
w_{1}=0, \quad q=0,
\end{align}
one has
\begin{align*}
&\blacktriangleright  \textbf{CP1:}\;\; (x,y,z,E)=(1+\alpha, 0,0, 0),
\quad w_{\mathrm{eff.}}=-1.03, \\
&(\lambda_{1}, \lambda_{2}, \lambda_{3}, \lambda_{4})\approx(-0.09,-0.09,-3.09,-4.09);
\end{align*}
\begin{align*}
&\blacktriangleright  \textbf{CP2:}\;\; (x,y,z,E)=(0, 1+\alpha,0, 0),
\quad w_{\mathrm{eff.}}=0,\\
&(\lambda_{1}, \lambda_{2}, \lambda_{3}, \lambda_{4})\approx(-1,+3.09,+3,+3);
\end{align*}
\begin{align*}
&\blacktriangleright  \textbf{CP3:}\;\; (x,y,z,E)=(0, 0,1+\alpha, 0), \quad w_{\mathrm{eff.}}=1/3,\\
&(\lambda_{1}, \lambda_{2}, \lambda_{3}, \lambda_{4})\approx (+1,+4.09,+4,+4);
\end{align*}
\begin{align*}
&\blacktriangleright  \textbf{CP4:}\;\; (x,y,z,E)=(0, 0,0, m_{4}),
\quad w_{\mathrm{eff.}}=-1,\\
&(\lambda_{1}, \lambda_{2}, \lambda_{3}, \lambda_{4})\approx (+0.09,-3,-4,0),
\end{align*}
where $m_{4}$ is an arbitrary constant.\\
In this example, dark matter is converted to a pressureless matter by choosing $\zeta_{0}=0$. For reaching suitable and physical points, it was preferred to remove the interaction by setting $q=0$.\\
It is clear that, for $\alpha >-1$, CP1 and CP2 are good candidates for the present time and matter-dominated era, respectively. CP3, however, has EoS$=1/3$ and the density of the radiation component is greater than others when $\alpha >-1$, but because it is a repeller point, hence it is not a good option for the radiation era. CP4 corresponds to a vacuum universe which is unacceptable.\\
Thus, two salient stages of the universe were achieved.\\

\begin{figure*}
	\includegraphics[width=6.5 in]{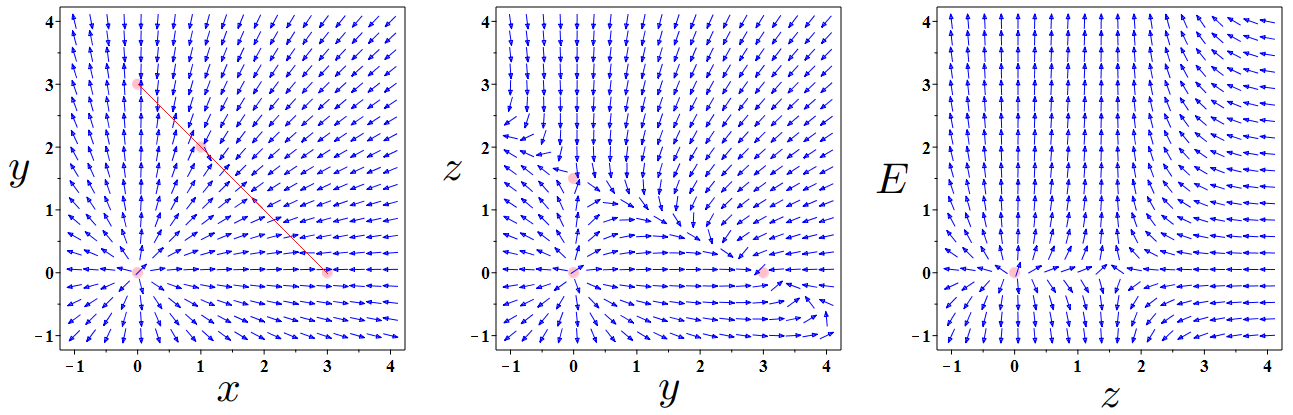}
    \caption{This figure indicates three different phase-portraits of the system~(\ref{sect5eq12})--(\ref{sect5eq15}) when one sets $j=0$, $w_{0}=0$, $\zeta_{0}=-6$, $\alpha =-4$, $q=18$, $\kappa^{2}=1$, and $w_{1}=-18$. The pink points represent the critical points and the thin red line (from $(x,y)=(0,3)$ to $(x,y)=(3,0)$) indicates a physical saddle line. All the points on this line, have $w_{\mathrm{eff.}}=0$. Since this line is thin, hence, schematically, three points have been put on it to make it visible. If we draw it thick, then we lose some vector fields which illustrate the saddle nature of this line.}
    \label{pa6}
\end{figure*}

In the above example, a fixed relation, $q=-w_{1}\kappa^{2}$, caused some presentable solutions. Now let us consider the case $j=0$ by specifying some values for constant parameters.\\
As an example, we present a data set yielding good results:
\begin{align}
&j=0, \quad w_{0}=0, \quad \zeta_{0}=-6, \quad \alpha=-4, \nonumber\\
&\kappa^{2} =1, \quad w_{1}=-18, \quad q=18.
\end{align}
These lead to three CPs as follows:
\begin{align}
&\blacktriangleright \text{CP1:}\;\; (x,y,z,E)=(0,0,0,0), \quad w_{\mathrm{eff.}}|_{\mathrm{CP1}}=+1, \nonumber\\
& \left(\lambda_{1},\lambda_{2},\lambda_{3},\lambda_{4} \right)=(6,2,3,3),\\
&\blacktriangleright \text{CP2:}\;\; (x,y,z,E)=(0,0,3/2,0), \quad w_{\mathrm{eff.}}|_{\mathrm{CP2}}=\frac{1}{3}, \nonumber\\
& \left(\lambda_{1},\lambda_{2},\lambda_{3},\lambda_{4} \right)=(4,-2,1,1),\\
&\blacktriangleright \text{CP3:}\;\; (x,y,z,E)=(m_{1},3-m_{1},0,0), \quad w_{\mathrm{eff.}}|_{\mathrm{CP3}}=0, \nonumber\\
& \left(\lambda_{1},\lambda_{2},\lambda_{3},\lambda_{4} \right)=(0,-3,3,-1).
\end{align}
As observed, CP1 with EoS$=+1$ is a repeller fixed point. At this point, all densities are zero.\\
CP2 is in the radiation-dominated era with saddle nature that is acceptable. Note that at CP2, all densities are zero excluding radiation density ($\rho_{\mathrm{r}} \neq 0$). This indicates the domination of radiation.\\
Adopting the condition $0 < m_{1} < 3$, CP3 can lie in the physical region. According to its EoS, this CP belongs to a matter-dominated era.
The related eigenvalues are indicative of a saddle property which is acceptable.\\
Some Poincar\'{e} sections of the phase space of this system have been presented in fig.~\ref{pa6}.

As another example with the above-mentioned data-set, let us generalize our equations by adding the baryonic matter, $\rho_{\mathrm{m}}$, to the collection of densities. Defining
\begin{align}
&x\equiv \frac{\kappa^{2} \rho_{\mathrm{v}}}{3H^{2}}, \quad y\equiv \frac{\kappa^{2} \rho_{\mathrm{dm}}}{3H^{2}}, \quad z\equiv \frac{\kappa^{2} \rho_{\mathrm{r}}}{3H^{2}},
\nonumber\\
&g\equiv \frac{\kappa^{2} \rho_{\mathrm{m}}}{3H^{2}}, \quad q\equiv \frac{\kappa^{2}Q}{3H^{3}},
\end{align}
we get the following equations
\begin{align}
x^{\prime}=& \left(\frac{x}{1+\alpha} \right) \left[3(1+w_{0})x+3y+3g+4z+\kappa^{2}w_{1}) \right] \nonumber\\
&-3(1+w_{0})x-\kappa^{2}w_{1}-q,\\
y^{\prime}=& \left(\frac{y}{1+\alpha} \right) \left[3(1+w_{0})x+3y+3g+4z+\kappa^{2}w_{1}) \right] \nonumber\\
&-3y+q+3\zeta_{0},\\
g^{\prime}=& \left(\frac{g}{1+\alpha} \right) \left[3(1+w_{0})x+3y+3g+4z+\kappa^{2}w_{1}) \right]-3g\\
z^{\prime}=& \left(\frac{z}{1+\alpha} \right) \left[3(1+w_{0})x+3y+3g+4z+\kappa^{2}w_{1}) \right]-4z.
\end{align}
The effective EoS parameter for this system then reads
\begin{align}
&w_{\mathrm{eff.}}=-1 \nonumber \\ &+\frac{1}{3}\left(\frac{1}{1+\alpha} \right)\left[3(1+w_{0})x+y+4z+3g+\kappa^{2} w_{1} \right].
\end{align}
With the previous selections for parameters, the following CPs with their corresponding eigenvalues and EoS parameters are found:
\begin{align}
&\blacktriangleright \text{CP1:} \; (x,y,z,g)=(0,0,0,0), \quad w_{\mathrm{eff.}}|_{\mathrm{CP1}}=+1, \nonumber\\
&(\lambda_{1},\lambda_{2},\lambda_{3},\lambda_{4})=(2,3,3,3)\\
&\blacktriangleright \text{CP2:} \; (x,y,z,g)=(0,0,3/2,0), \quad w_{\mathrm{eff.}}|_{\mathrm{CP2}}=+1/3, \nonumber\\
&(\lambda_{1},\lambda_{2},\lambda_{3},\lambda_{4})=(-2,1,1,1)\\
&\blacktriangleright \text{CP3:} \; (x,y,z,g)=(m_{1},3-(m_{1}+m_{2}),0,m_{2}), \nonumber\\
&w_{\mathrm{eff.}}|_{\mathrm{CP3}}=\frac{2}{9}[3-(m_{1}+m_{2})], \nonumber\\
&(\lambda_{1},\lambda_{2},\lambda_{3},\lambda_{4})=(0,0,-1,-3).
\end{align}
CP1 and CP2 are similar to the previous example and therefore their interpretations are the same.\\
The effective EoS parameter of CP3 depends upon the values of $x \propto \rho_{\mathrm{v}}$ and $g \propto \rho_{\mathrm{m}}$. Note that the position of $y$ is affected by $x$ and $g$. In fact, the density of dark matter is affected by the amounts of densities of both dark energy and baryonic matter. It is clear that the value of EoS is proportional to the density of dark matter. For being in a physical region, the condition $0 \leq (m_{1}+m_{2}) \leq 3$ must be adopted besides $m_{1} \geq 0$ and $m_{2} \geq 0$. Therefore, the dark matter causes the existence of an upper bound on the summation of densities of dark energy and baryonic matter.\\
According to the elements of coordinates of CP3 and its EoS parameter and eigenvalues, this point cannot be fixed to the radiation-dominated era, but it is a good option for the present time.\\

In the generalized case, namely by defining $E=(\kappa^{2}H^{2})^{-1}$ and adding
\begin{align*}
E^{\prime}=\left(\frac{E}{1+\alpha}\right)[3(1+w_{0})x+3y+3g+4z+\kappa^{2}w_{1})]
\end{align*}
to our equations, the same results are repeated with $E|_{\mathrm{all\; CPs}}=0$ and the following positive eigenvalues, which indicate the expansion nature of the universe evolution and decreasing property of the Hubble parameter:
\begin{align*}
\lambda_{4}|_{\mathrm{CP1}}=+6, \quad \lambda_{4}|_{\mathrm{CP2}}=+4, \quad \lambda_{4}|_{\mathrm{CP3}}=+3.
\end{align*}

Overall, our data analysis indicates that we cannot set \textit{all} the constant parameters so that \textit{all} these CPs belong to \textit{different} crucial stages of the universe --- in matching data, some of CPs do not belong to well-known eras, or some of CPs are repeated, for example, two CPs for the radiation-dominated era, or in most of the cases, some of points ruled out because of strange physical meanings, etc. In a nutshell, after examining many cases under the aforementioned condition (i.e. $q=-w_{1}\kappa^{2}$), it is concluded that we cannot expect to find a data set so that these CPs illustrate a complete schema of the crucial stages of the evolution of the universe.\\

\noindent\pqrmybox{}{\subsubsection{The case $j=-1$}}\vspace{5mm}
For the case $j=-1$ there is a similar situation. However, all of CPs can be found analytically, but they are `VLT' excluding one CP:
\begin{align}
&E=0, \quad x=\frac{-\kappa^{2} w_{1}}{3w_{0}-1}, \quad y=-3\zeta_{0}, \nonumber\\
&z=\frac{4\kappa^{2} w_{1}+12w_{0}\alpha +27w_{0}\zeta_{0}-4\alpha -9\zeta_{0}+12w_{0}-4}{4(3w_{0}-1)}.
\label{commonsolutions}
\end{align}
The EoS parameter of this CP is $1/3$ which indicates that this CP is in the radiation-dominated era, hence it must be a saddle point. One of the associated eigenvalues of this CP is $+4$ showing its unstable nature. Other eigenvalues are `VLT', but they could be tuned so that at least one of them will be negative, making it a saddle point then.\\
Other CPs have a common property: They are on $y-z$ plane because we have $x=E=0$ for all of them. As mentioned earlier, the plane of $y-z$ is the plane of the interaction between dark matter and dark energy. Clearly, at these CPs one can fix some relations between the densities of dark matter and dark energy. Moreover, these CPs are in the accelerated era.

For reaching a presentable set of CPs, again there are two ways like the previous case. \textit{Note that the number of different CPs is affected by our fixed relations among parameters and our selections for the values of constant parameters.}

Let us here we take $w_{0}=-1$. This condition leads to three presentable CPs with their properties as follows:
\begin{align*}
&\blacktriangledown \text{CP1:} \\
&\left[
   \begin{array}{c}
     x \\
     y \\
     z \\
     E \\
   \end{array}
 \right]=\left[
   \begin{array}{c}
     \frac{\kappa^{2} w_{1}}{4} \\
     -3\zeta_{0} \\
     1+\alpha +\frac{9\zeta_{0}-\kappa^{2} w_{1}}{4} \\
     0 \\
   \end{array}
 \right], \quad w_{\mathrm{eff.}}|_{\mathrm{CP1}}=\frac{1}{3}
\end{align*}
\begin{align*}
 &\left[
   \begin{array}{c}
     \lambda_{1} \\
     \lambda_{2} \\
     \lambda_{3} \\
     \lambda_{4} \\
   \end{array}
 \right]=\left[
   \begin{array}{c}
     \frac{\chi_{3}-\kappa^{2} w_{1}+5(1+\alpha )}{2(1+\alpha )} \\
     \frac{-\chi_{3}-\kappa^{2} w_{1}+5(1+\alpha )}{2(1+\alpha )} \\
     4 \\
     4 \\
   \end{array}
 \right];
\end{align*}
\begin{align*}
&\blacktriangledown \text{CP2:} \\
&\left[
   \begin{array}{c}
     x \\
     y \\
     z \\
     E \\
   \end{array}
 \right]=\left[
   \begin{array}{c}
     \frac{2\kappa^{2} w_{1}(1+\alpha )}{\chi_{3}+\kappa^{2}w_{1}+3(1+\alpha )} \\
     \frac{\chi_{3}-\kappa^{2}w_{1}+3(1+\alpha )}{6} \\
     0 \\
     0 \\
   \end{array}
 \right],
\end{align*}
\begin{align*}
w_{\mathrm{eff.}}|_{\mathrm{CP2}}=\frac{\chi_{3}+\kappa^{2}w_{1}-3(1+\alpha )}{6(1+\alpha )},
\end{align*}
\begin{align*}
 &\left[
   \begin{array}{c}
     \lambda_{1} \\
     \lambda_{2} \\
     \lambda_{3} \\
     \lambda_{4} \\
   \end{array}
 \right]=\left[
   \begin{array}{c}
     \frac{\chi_{3}}{1+\alpha} \\
     \frac{\chi_{3}+\kappa^{2}w_{1}-5(1+\alpha )}{2(1+\alpha)} \\
     \frac{\chi_{3}+\kappa^{2}w_{1}+3(1+\alpha )}{2(1+\alpha)} \\
     \frac{\chi_{3}+\kappa^{2}w_{1}+3(1+\alpha )}{2(1+\alpha)} \\
   \end{array}
 \right];
\end{align*}
\begin{align*}
&\blacktriangledown \text{CP3:}\\
&\left[
   \begin{array}{c}
     x \\
     y \\
     z \\
     E \\
   \end{array}
 \right]=\left[
   \begin{array}{c}
     \frac{2\kappa^{2} w_{1}(1+\alpha )}{-\chi_{3}+\kappa^{2}w_{1}+3(1+\alpha )} \\
     \frac{-\chi_{3}-\kappa^{2}w_{1}+3(1+\alpha )}{6} \\
     0 \\
     0 \\
   \end{array}
 \right],
\end{align*}
\begin{align*}
w_{\mathrm{eff.}}|_{\mathrm{CP3}}=\frac{-\chi_{3}+\kappa^{2}w_{1}-3(1+\alpha )}{6(1+\alpha )},
\end{align*}
\begin{align*}
 &\left[
   \begin{array}{c}
     \lambda_{1} \\
     \lambda_{2} \\
     \lambda_{3} \\
     \lambda_{4} \\
   \end{array}
 \right]=\left[
   \begin{array}{c}
     \frac{-\chi_{3}}{1+\alpha} \\
     \frac{-\chi_{3}+\kappa^{2}w_{1}-5(1+\alpha )}{2(1+\alpha)} \\
     \frac{-\chi_{3}+\kappa^{2}w_{1}+3(1+\alpha )}{2(1+\alpha)} \\
     \frac{-\chi_{3}+\kappa^{2}w_{1}+3(1+\alpha )}{2(1+\alpha)} \\
   \end{array}
 \right],
\end{align*}
where $\chi_{3}=(\kappa^{2} w^{2}_{1}-6\kappa^{2}w_{1}\alpha -6\kappa^{2}w_{1}+9\alpha^{2}+36\alpha \zeta_{0}+18\alpha -36\zeta_{0}+9)^{1/2}$.\\
$\blacktriangledown$ \textbf{CP1:}\\
CP1 is an excellent candidate for the radiation-dominated era. Besides maintaining the coordinates of this point greater than zero, one must note that the value of $z$ must be greater than others for providing the domination of the radiation component. Also, we must care about the amounts of eigenvalues; at least one of $\lambda_{1}$ or $\lambda_{2}$ must be negative because this CP should be a saddle point.\\
$\blacktriangledown$ \textbf{CP2 and CP3:}\\
According to the coordinates of CP2 and CP3, these points can be fixed for the accelerated era --- from the onset of acceleration of the universe up to now. For this end, besides tuning the values of the EoS parameters, one must take the points (excluding the present point) as a saddle point, for saddle points can lie on the evolutionary trajectory of the universe. The associated CP of a point scribed to the present time will be of attractor or saddle nature.\\
Furthermore, at a CP of this era, the value of $x$ must be greater than the value of $y$ due to the acceleration feature of expansion and the domination of dark energy.\\

\noindent\pqrmybox{}{\subsubsection{The case $j=+1$}}\vspace{5mm}
For the case $j=+1$, no solution was found. Even, under special conditions like $w_{0}=w_{1}=0$, there is no analytical solution for the model. In $j=+1$ case, $E$ appears in the denominator of the equations of $x$, $y$, and $z$. On the other hand, the evolution equations cannot be satisfied when $E \neq 0$. We know that a solution comes through $E=0$ and since it leads to infinity and singularity for our system, hence this system is not able to provide any CP.\\

\noindent\pqrmybox{}{\subsubsection{Discussion about different values of $j$}}\vspace{5mm}
In brief, it is argued that only for $j \leq 0$ analytical solutions may be found. It is interesting to mention that among the analytical solutions for the case $j < 0$, all solutions are `VLT' excluding one common solution, (\ref{commonsolutions}), which belongs to the radiation-dominated era. The `VLT' solutions are in the interaction plane of dark matter and dark energy namely $x-y$ plane, for we have $E=z=0$. Hence, the critical points can be used to fix some relations between the densities of dark matter and dark energy at these CPs. Furthermore, these CPs are in the accelerated era.\\
In the case $j=0$, three coordinates $x \propto \rho_{\mathrm{v}}$, $y \propto \rho_{\mathrm{dm}}$, and $z \propto \rho_{\mathrm{r}}$ are affected by the interaction part ($q  \propto Q$), while in the case $j<0$, we do not observe such affects. It is due to the fact that in the case $j=0$, the contribution of the interaction term in the equations appears without other variables as coefficients and it causes no effect by the amounts of other variables. It serves in the equations as a shift value, while in the case $j<0$, the interaction term appears as $qE^{|j|}$, and for the satisfaction of the equations we must have $E=0$ (imposed by the equation of $E^{\prime}$ ), hence this term has been removed by the coefficient $E$. Therefore, the interaction does not play any role in the fixed points of the systems with $j<0$.

It is interesting to indicate that the upper bound on $j$ (for the existence of analytical solutions) is increasable from $j \leq 0$ to $j \leq 3/2$ through changes in the definitions of the dimensionless variables:\\
Combine $q$ with the multiplier $H^{2j}$ to get
\begin{equation*}
\frac{Q \kappa^{2n-4}}{3}\; E^{n} \;\; ; \quad n=\frac{3-2j}{2}
\end{equation*}
instead of
\begin{equation*}
q \left(\frac{1}{\kappa^{2}E} \right)^{j},
\end{equation*}
hence the bound $j \leq 0$ is increased to $j \leq 3/2$.\\
\textit{Both bounds demonstrate that the power-law interactions in the related dynamical systems should be of increasing nature with time to get more critical points.}\\
The benefit of defining $q$ instead of the aforementioned combination in our case is that the behaviors of other types of interaction (e.g. oscillating types that are studied in the next sub-section) are predictable. For more merits check the next sub-section.\\

\noindent\Rrmybox{}{\subsection{Oscillating dark energy-dark matter interaction}}\vspace{5mm}
In this sub-section, the oscillating model of interaction is studied. This type of interaction (`Cosine' function in which its argument is proportional to $H^{2}$ or equivalently $E^{-1}$) has recently been introduced in ref.~\cite{od-asli}.\\ Furthermore, some enhancements about this type of interaction are suggested and examined.\\
The dynamical equations of the cosine type of interaction may be easily obtained by the replacement~\cite{od-asli}
\begin{equation*}
-QH^{2j} \; \to \; -Q \cos \left(h_{0}H^{2} \right)
\end{equation*}
in the equations of the previous power-law interaction.
The equations would then be:
\begin{align}
x^{\prime} =&\frac{x}{1+\alpha}\left[3(1+w_{0})x+3y+4z+w_{1}\kappa^{2} \right]
\nonumber \\ &-3(1+w_{0})x-q \cos \left(h_{0}H^{2} \right) -w_{1}\kappa^{2},  \label{sect5eq12}\\
y^{\prime} =&\frac{y}{1+\alpha}\left[3(1+w_{0})x+3y+4z+w_{1}\kappa^{2} \right]
\nonumber \\ &-3y+q \cos \left(h_{0}H^{2} \right) +3\zeta_{0}, \label{sect5eq13}\\
z^{\prime} =&\frac{z}{1+\alpha}\left[3(1+w_{0})x+3y+4z+w_{1}\kappa^{2} \right]-4z,  \label{sect5eq14}\\
E^{\prime} =&\frac{E}{1+\alpha}\left[3(1+w_{0})x+3y+4z+w_{1}\kappa^{2} \right], \label{sect5eq15}
\end{align}
where $\cos \left(h_{0}H^{2} \right)=\cos \left(h_{0} / \kappa^{2} E \right)$. It has been assumed in ref.~\cite{od-asli} that the argument of cosine function changes monotonically from $\pi /2$ to $0$. The starting point, $\pi /2$, is equivalent to the early inflationary era while the endpoint corresponds to the present time. This kind of behavior can be understood by taking sufficiently small values of the parameter $h_{0}$~\cite{od-asli}. Indeed, the cosine model under the above assumptions indicates that at early times (i.e. inflationary era; $H^{2} \sim M_{\mathrm{Pl}}$; $t \to 0$) there is no interaction between dark matter and dark energy, and as the universe ages this interaction starts to grow until the maximum value at $H=0$ (equivalently $t \to \infty$)~\cite{od-asli}. That is why besides the general case, two asymptotic behaviors are of interest:
$\cos \left(h_{0}H^{2} \right)=0$ (The inflationary epoch), and $\cos \left(h_{0}H^{2} \right)=1$ (The present time and late-time acceleration).\\

\noindent\pqrmybox{}{\subsubsection{General Case.}}\vspace{5mm}
Without assuming any condition, the system of evolution may not be solved analytically. This is because the interaction term,
\begin{align*}
\cos \left(h_{0}H^2 \right)=\cos \left(\frac{h_{0}}{\kappa^{2}E}\right)
=\sum_{k=0}^{\infty}\frac{(-1)^{k}(h_{0}H^{2})^{2k}}{(2k)!},
\end{align*}
is like interaction $H^{2j}$. As earlier mentioned, for positive values of $j$, there is no analytical solution, while for $j \leq 0$ the system is analytically solvable.
Hence, by switching
\begin{align}\label{mod-cosine01}
\cos \left(h_{0}H^{2}\right)=
\cos \left(\frac{h_{0}}{\kappa^{2}E}\right) \to \cos \left(\frac{h_{0}}{H^{2}}\right)=\cos (h_{0}\kappa^{2}E),
\end{align}
the problem could be solved. In this case, $\cos (h_{0}\kappa^{2}E)$, all CPs can be achieved analytically. Excluding one CP, all other solutions are `VLT' and they are in $x-y$ plane (i.e. for all CPs we have $E=z=0$ and hence they cannot belong to the radiation-dominated era). The presentable CP with EoS$=1/3$ is as follows:
\begin{align}
&E=0, \quad x=\frac{\kappa^{2}w_{1}+q}{1-3w_{0}}, \quad y=-q-3\zeta_{0}, \nonumber \\
&z=\{4\kappa^{2}w_{1}+12\alpha w_{0}+12qw_{0}+27w_{0}\zeta_{0}-4\alpha-9\zeta_{0} \nonumber \\ &+12w_{0}-4 \} \times (4(3w_{0}-1))^{-1/2}.
\label{cossol}\end{align}
The corresponding eigenvalues are `VLT' except $+4$ case, implying instability.  Due to the existing some constant parameters, this CP can be fixed as a saddle point which is completely acceptable.\\

It is interesting to note another oscillating case namely
\begin{align}\label{mod-cosine02}
\sin \left( h_{0}\kappa^{2}E \right)=\sin \left(\frac{h_{0}}{H^{2}}\right).
\end{align}
In this case, the situation is similar to the previous case. But in this case, the coordinates of CPs are shifted with respect to the above solutions because of the phase difference between `Cosine' and `Sine' functions. For instance, the CP of the radiation era would be:
\begin{align}
&E=0, \quad x=\frac{\kappa^{2}w_{1}}{1-3w_{0}}, \quad y=-3\zeta_{0}, \nonumber \\
&z=\frac{4\kappa^{2}w_{1}+12\alpha w_{0}+27w_{0}\zeta_{0}-4\alpha -9\zeta_{0}+12w_{0}-4 }{4(3w_{0}-1)}.
\label{sinsol}\end{align}
As observed, the interaction term does not affect this fixed point while in the previous case it has an active role. It is due to the phase difference between the two types of interactions. Note that since all CPs in both cases have $E=0$, hence the interaction term only in `Cosine'-type has an effect, not in `Sine'-type, thanks to $\sin(0)=0$ (which removes the interaction term from the coordinate of the CP).\\

Since most of the CPs of the general case were of `VLT' type, hence we have to consider the problem by one of two aforementioned approaches. As an example, let's set $w_{0}=-1$ for the modified `Cosine' type of interaction (\ref{mod-cosine01}). Under this condition, all CPs are presentable:
\begin{align*}
&\blacktriangledown \; \text{CP1:} \nonumber \\
&\left[
   \begin{array}{c}
     x \\
     y \\
     z \\
     E \\
   \end{array}
 \right]=\frac{1}{4}\left[
   \begin{array}{c}
     \kappa^{2}w_{1}+q \\
     -4q-12\zeta_{0} \\
     4+4\alpha +3q+9\zeta_{0}-\kappa^{2}w_{1} \\
     0 \\
   \end{array}
 \right],
\end{align*}
\begin{align*}
w_{\mathrm{eff.}}|_{\mathrm{CP1}}=\frac{1}{3},
\end{align*}
\begin{align*}
&\left[
   \begin{array}{c}
     \lambda_{1} \\
     \lambda_{2} \\
     \lambda_{3} \\
     \lambda_{4} \\
   \end{array}
 \right]=\left[
   \begin{array}{c}
     \frac{5}{2}+\frac{\chi_{4}-\kappa^{2}w_{1}}{2(1+\alpha)} \\
     \frac{5}{2}-\frac{\chi_{4}+\kappa^{2}w_{1}}{2(1+\alpha)} \\
     4 \\
     4 \\
   \end{array}
 \right];
\end{align*}
\begin{align*}
&\blacktriangledown \; \text{CP2 and CP3:} \nonumber \\
&\left[
  \begin{array}{c}
    x \\
    y \\
    z \\
    E \\
  \end{array}
\right]=\left[
  \begin{array}{c}
    \frac{2(\kappa^{2} w_{1}+q)(1+\alpha )}{\kappa^{2}w_{1}+3(1+\alpha ) \pm \chi_{4}} \\
    \frac{3(1+\alpha )-\kappa^{2}w_{1} \pm \chi_{4}}{6} \\
    0 \\
    0 \\
  \end{array}
\right],
\end{align*}
\begin{align*}
w_{\mathrm{eff.}}|_{\mathrm{CP2,3}}=\frac{-1}{2}
+\frac{\kappa^{2}w_{1}}{6(1+\alpha)} \pm \frac{\chi_{4}}{6(1+\alpha)},
\end{align*}
\begin{align*}
& \lambda_{1}=\lambda_{2}=\frac{3}{2}+\frac{\kappa^{2}w_{1} \pm \chi_{4}}{2(1+\alpha )}, \nonumber \\
&\lambda_{3}=\lambda_{4}=\frac{1}{1+\alpha } \left\{\frac{\kappa^{2}w_{1}}{4} \pm \frac{\chi_{4}}{4}-\frac{5}{4}(1+\alpha ) \right. \nonumber \\
& +\frac{1}{4} (2\kappa^{4}w^{2}_{1} \mp 2 \chi_{4}\kappa^{2}w_{1} \pm
10\alpha \chi_{4} \pm 10\chi_{4}-16\alpha \kappa^{2}w_{1}-12q\nonumber \\
&\left.16 \kappa^{2}w_{1} +34\alpha^{2}-12q\alpha -36\alpha \zeta_{0}+68\alpha -36\zeta_{0}+34 )^{1/2} \right\},
\end{align*}
where
\begin{align*}
\chi_{4}=(&\kappa^{4}w^{2}_{1}-6\alpha \kappa^{2}w_{1}-6\kappa^{2}w_{1}+9\alpha^{2} -12q\alpha \\&-36\alpha \zeta_{0} +18 \alpha -12q-36\zeta_{0} +9)^{1/2}.
\end{align*}
Besides $z_{\mathrm{CP1}} > x_{\mathrm{CP1}} \geq 0$ and $z_{\mathrm{CP1}} > y_{\mathrm{CP1}} \geq 0$, at least one of $\lambda_{1}$ or $\lambda_{2}$ must be less than zero to maintain CP1 as a physical CP (a saddle CP) to the radiation-dominated era. According to the coordinates, EoS parameters, and eigenvalues, CP2 and CP3 have this potential to be fixed as candidates for the present status of the universe or even earlier stages up to the inflection point where the acceleration of the expansion of the universe started. Thus, there are many options for these points. But the best seems to be the current situation of the universe (i.e. $w_{\mathrm{eff.}}=-1.03 \pm 0.03$). To guarantee the domination of dark energy for the current stage of the universe, one must set $x|_{\mathrm{CP}}>y|_{\mathrm{CP}}>0$.\\

\noindent\pqrmybox{}{\subsubsection{Asymptotic behavior 1: $\cos(h_{0}H^{2})=0$}}\vspace{5mm}
This case corresponds to the inflationary epoch of the universe.
The results of this case are exactly the results of\\
$\sin(h_{0} \kappa^{2}E)$  (i.e. eq.~(\ref{sinsol}) etc). The reason is that all CPs have $E=0$.
Nonetheless, since the inflationary era is our objective in this sub-section, hence let us single out some parameters and fix some relations to reach this end.\\
If one takes
\begin{align}
&w_{0}=-1, \quad \alpha=\frac{w_{1} \kappa^{2}}{-3}-1, \nonumber\\
&q=-w_{1} \kappa^{2} -3\zeta_{0}+\frac{\eta }{4w_{1}\kappa^{2}},
\end{align}
then the following CPs are found:
\begin{align}
x= \frac{12\zeta_{0}w_{1} \kappa^{2}-\eta}{6\epsilon \sqrt{\eta}}, \quad y=\frac{w_{1}\kappa^{2}}{-3}+\frac{\sqrt{\eta}}{6\epsilon},\quad z=E=0,
\end{align}
where $\epsilon =\pm 1$ and $\eta$ is an arbitrary constant.
The effective EoS parameter of these points would then be
\begin{align}
w_{\mathrm{eff.}}=-1-\frac{\sqrt{\eta}}{2\epsilon w_{1}\kappa^{2}}.
\end{align}
Clearly, at the limit like
\begin{align}
0<\frac{\sqrt{\eta}}{2\epsilon w_{1}\kappa^{2}} \lll 1
\label{ptl}
\end{align}
(i.e. very close to zero) our goal is fulfilled. Depending upon the values of the parameters, both or one of the CPs can be regarded as a candidate for the inflationary epoch.\\
The corresponding eigenvalues are as follows:
\begin{align}
&\lambda_{1}=\lambda_{2}=\frac{-3\sqrt{\eta}}{2\epsilon w_{1} \kappa^{2}}, \nonumber\\
&\lambda_{3}=-2-\frac{3}{2}\lambda_{1}
+\frac{\sqrt{64w^{2}_{1}\kappa^{4}-48\epsilon w_{1}\kappa^{2} \sqrt{\eta}+9\eta}}{4w_{1}\kappa^{2}},\nonumber\\
&\lambda_{4}=-\lambda_{3}-4-3\lambda_{1}.
\end{align}
Respecting the imposed limitation (i.e. (\ref{ptl})), $\lambda_{1}$ and $\lambda_{2}$ will be negative which undertake the attractor nature of the points --- at least, we have attraction in two directions. So, the values of the constant parameters specify the stability/instability of the fixed points according to the values of $\lambda_{3}$ and $\lambda_{4}$.\\

\noindent\pqrmybox{}{\subsubsection{Asymptotic behavior 2: $\cos(h_{0}H^{2})=1$}}\vspace{5mm}
This asymptotic case corresponds to the case $j=0$ in the power-law interaction which has been studied above carefully. Nevertheless, let us consider the possibility of the description of the current status of the universe with non-zero $E$ like ref.~\cite{od-asli}. However, the equation governing the evolution of $E$ is dependent and it is not the original and crucial equation. But let us take it into account to see under which conditions our objective is feasible. To this end, we rewrite the equations in the following form
\begin{align}
x^{\prime}&=\left(\frac{3x}{1+\alpha } \right)\mathcal{U}-3x-w_{0}x-w_{1}\kappa^{2}-q, \\
y^{\prime}&=\left(\frac{3y}{1+\alpha } \right)\mathcal{U}-3y+q+3\zeta_{0}, \\
z^{\prime}&=\left(\frac{3z}{1+\alpha } \right)\mathcal{U}-4z,\\
E^{\prime}&=\left(\frac{3E}{1+\alpha } \right)\mathcal{U}, \\
w_{\mathrm{eff.}}&=-1+\left(\frac{1}{1+\alpha} \right)\mathcal{U},
\end{align}
where $\mathcal{U}=(3(1+w_{0})x+3y+w_{1}\kappa^{2}+4z)/3$.\\
If $\mathcal{U} \to 0$, then one has
\begin{align}
x^{\prime}& \longrightarrow -3x-w_{0}x-w_{1}\kappa^{2}-q,\\
y^{\prime}&\longrightarrow -3y+q+3\zeta_{0}, \\
z^{\prime}&\longrightarrow -4z,\\
E^{\prime}&\longrightarrow 0, \\
w_{\mathrm{eff.}}&\longrightarrow -1.
\end{align}
Therefore, the CP of this system would be
\begin{align}
x\approx \frac{w_{1}\kappa^{2} +q}{-3(1+w_{0})}, \quad y\approx \frac{q}{3}+\zeta_{0}, \quad z\approx 0, \quad E\approx m_{4},
\end{align}
where $m_{4}$ taken to be $m_{4}=(\kappa^{2}H^{2}_{0})^{-1}$, $H_{0}$ being the current value of the Hubble parameter. These coordinate points stipulates $\zeta_{0} \to 0$ because of $\mathcal{U}\to 0$. In fact, the current stage of the universe can be achieved when the pressure of the bulk viscous approaches to zero.\\

\noindent\pqrmybox{}{\subsubsection{Some new types of interaction between dark matter and dark energy}}\vspace{5mm}
As mentioned above, the lack of a general solution makes $\cos(h_{0}H^{2})=\cos(h_{0}/(\kappa^{2}E))$ of no favor.
So, improvements seem necessary.
In what follows, some remedies are suggested and finally, we examine it in the system of ref.~\cite{od-asli} and show some missed CPs which cover other stages of the evolution of the universe.
\begin{enumerate}
  \item Using
      \begin{align*}
       \cos \left(\frac{h_{0}H^{2}}{1+CH^{2}} \right)=\cos \left(\frac{h_{0}}{\kappa^{2}(E+C)} \right)
      \end{align*}
      or
      \begin{align*}
       \sin \left(\frac{h_{0}H^{2}}{1+CH^{2}} \right)=\sin \left(\frac{h_{0}}{\kappa^{2}(E+C)} \right)
      \end{align*}
      where $C$ is an arbitrary constant, instead of
      \begin{align*}
       \cos \left(h_{0}H^{2} \right)=\cos \left(\frac{h_{0}}{\kappa^{2}E} \right).
      \end{align*}
      These modified models have at least two benefits:\\
      \textit{i}. This $C$ allows those CPs with $E=0$ to be included in the category of CPs. Thus, the number of our CPs can be increased through this modification.\\
      \textit{ii}. Unlike ref.~\cite{od-asli}, $C$ discards any restriction on $h_{0}$.
  \item Utilizing $\cos(h_{0}\kappa^{2}E)$ or $\sin(h_{0}\kappa^{2}E)$ instead of $\cos(h_{0}/ \kappa^{2}E)$.
\end{enumerate}
If it is preferred to restrict the evolution range of the argument of cosine and sine functions in the above models such as ref.~\cite{od-asli}, the range of this limitation is not the same as the range of ref.~\cite{od-asli}.
The evolutionary equivalent ranges have been presented in Table~\ref{table1}. \textit{As is clear from Table~\ref{table1}, the asymptotic behaviors of all five models are the same. Therefore, the asymptotic solutions that were considered above, hold true for all five types of interactions. As a result, all five types of interaction are successful in describing both early inflation and late-time-accelerated expansion (i.e. the unification of early and late accelerations is achieved through these models).}\\
\begin{table*}
\caption{The evolution ranges of the interaction models (here we have assumed $E=\kappa^{-2}H^{-2}$).} \label{table1}
\centering
\begin{tabular}{|p{2.1 in} |p{1.2 in} |p{1.4 in} |p{1.4 in}|}
\toprule[1.7pt]
\textbf{The interaction models; $\cos (\theta)$ and $\sin (\theta)$} &\textbf{Inflationary epoch; $\theta_{i}$} & \textbf{The present time; $\theta_{f}$}
\\
\toprule[1.4pt]
$\cos \left(h_{0}H^{2} \right)=\cos \left(\frac{h_{0}}{\kappa^{2}E} \right)$ (Proposed in\cite{od-asli}) & $\frac{\pi}{2}$ & $0$  \\ \toprule[1pt]
$\cos \left(\frac{h_{0}H^{2}}{1+CH^{2}} \right)=\cos \left(\frac{h_{0}}{\kappa^{2}(E+C)} \right)$ & $\frac{\pi}{2}$ & $0$  \\ \toprule[1pt]
%%%%%
$\sin \left(\frac{h_{0}H^{2}}{1+CH^{2}} \right)=\sin \left(\frac{h_{0}}{\kappa^{2}(E+C)} \right)$ & $\pi$ & $\frac{\pi}{2}$  \\ \toprule[1pt]
%%%%%%%%%
$\cos\left(\frac{h_{0}}{H^{2}} \right)=\cos(h_{0}\kappa^{2}E)$ & $\frac{3 \pi}{2}$ & $2 \pi$  \\ \toprule[1pt]
$\sin \left(\frac{h_{0}}{H^{2}} \right)=\sin (h_{0}\kappa^{2}E)$ & $0$ & $\frac{\pi}{2}$ \\
\toprule[1.7pt]
\end{tabular}
\end{table*}
$\bullet$ \textbf{Examination of the models:}\\
Most aspects of our enhancements to the interaction model of ref.~\cite{od-asli} were illustrated in the \(F(T)\) model successfully, but let us examine them in the system of ref.~\cite{od-asli} as well. In this part, we use their notations. In ref.~\cite{od-asli}, by defining the dimensionless variables as
\begin{align}
x\equiv \frac{\kappa^{2} \rho_{\mathrm{v}}}{3H^{2}}, \quad y\equiv \frac{\kappa^{2} \rho_{\mathrm{dm}}}{3H^{2}}, \quad z\equiv \frac{1}{\kappa^{2} H^{2}},
\end{align}
they arrived at the following system~\cite{od-asli}:
\begin{align*}
x^{\prime}=(x-1)(w_{1}\kappa^{2} +3xw_{0}-x)-x(y+3\zeta_{0}) -q\cos \left(\frac{h_{0}}{\kappa^{2}z} \right),
\end{align*}
\begin{align*}
y^{\prime}=&-y^{2}+y(1-3\zeta_{0}+3xw_{0}-x+w_{1}\kappa^{2})+3\zeta_{0}
\\&+q \cos \left(\frac{h_{0}}{\kappa^{2}z} \right),
\end{align*}
\begin{align*}
z^{\prime}=z(4-x-y+3xw_{0}-3\zeta_{0}+w_{1}\kappa^{2}).
\end{align*}
If we study this \textit{generally}, then only one CP is achieved in a closed form:
\begin{align}
&x=\frac{3\zeta_{0}-3-\kappa^{2}w_{1}}{3w_{0}}, \quad
y=1-\frac{3\zeta_{0}-3-\kappa^{2}w_{1}}{3w_{0}}, \nonumber \\
&z=\frac{h_{0}}{\kappa^{2}\left(\pi -\arccos \left(\frac{3\zeta_{0}+3\zeta_{0}w_{0}-3w_{0}-3-\kappa^{2}w_{1}}{q w_{0}} \right) \right)}.\label{mog01}
\end{align}
The effective EoS parameter of this CP is $-1$ indicating the present stage of the universe.\\
$\bullet$ Now, if, for example, we change the model of interaction as
\begin{align}
\cos \left(\frac{h_{0}}{\kappa^{2}z} \right) \longrightarrow \cos \left(h_{0}\kappa^{2}z \right)
\end{align}
then we get to the following CPs:
\begin{align}
&\blacktriangleright \;\; \text{CP1:}\; w_{\mathrm{eff.}}|_{\mathrm{CP1}}=-1,
\nonumber \\
&x=\frac{3\zeta_{0}-3-\kappa^{2}w_{1}}{3w_{0}}, \quad
y=1-\frac{3\zeta_{0}-3-\kappa^{2}w_{1}}{3w_{0}}, \nonumber\\
&z=\frac{\pi -\arccos \left(\frac{3\zeta_{0}+3\zeta_{0}w_{0}-3w_{0}-3-\kappa^{2}w_{1}}{q w_{0}} \right)}{\kappa^{2} h_{0}}; \label{mog02}
\end{align}
\begin{align}
&\blacktriangleright \;\; \text{CP2:}\;\; w_{\mathrm{eff.}}|_{\mathrm{CP2}}=\frac{1}{3}, \nonumber\\
&x=\frac{\kappa^{2}w_{1}+q}{1-3w_{0}}, \quad y=-(3\zeta_{0}+q), \quad z=0;
\end{align}
\begin{align}
&\blacktriangleright \;\; \text{CP3:}\;\; w_{\mathrm{eff.}}|_{\mathrm{CP3}}=w_{0}+\frac{\kappa^{2}w_{1}+q}{3x_{03}},
\nonumber \\
&x=\frac{3\zeta_{0}+3w_{0}-\kappa^{2}w_{1}+\chi_{5}}{6w_{0}} \equiv x_{03}
\nonumber \\
&y=\frac{\kappa^{2}w_{1}-3\zeta_{0}}{3w_{0}}
-\frac{\kappa^{2}w_{1}+q}{3w_{0}x_{03}}, \quad z=0;
\end{align}
\begin{align}
&\blacktriangleright \;\; \text{CP4:}\;\; w_{\mathrm{eff.}}|_{\mathrm{CP4}}=w_{0}+\frac{\kappa^{2}w_{1}+q}{3x_{04}},
\nonumber \\
&x=\frac{3\zeta_{0}+3w_{0}-\kappa^{2}w_{1}-\chi_{5}}{6w_{0}} \equiv x_{04}
\nonumber \\
&y=\frac{\kappa^{2}w_{1}-3\zeta_{0}}{3w_{0}}
-\frac{\kappa^{2}w_{1}+q}{3w_{0}x_{04}}, \quad z=0;
\end{align}
where $\chi_{5}=(\kappa^{4}w^{2}_{1}-6\kappa^{2}\zeta_{0}w_{1}+6\kappa^{2}
w_{0}w_{1}+9\zeta^{2}_{0}+18\zeta_{0}w_{0}+12qw_{0}+9w^{2}_{0})^{1/2}$.\\
Obviously, the number of CPs has been increased and some missed-out stages have been covered in our approach.\\
First of all, note the correspondence between (\ref{mog01}) and (\ref{mog02}). Unlike $x$ and $y$, $z$ is different between both, but these two CPs are almost equivalent due to the existence of $h_{0}$. Thus, the CP of the previous interaction can be recovered here.\\
CP2 can be related to the radiation-dominated era because of its EoS parameter.\\
CP3 and CP4 can be fixed to arbitrary stages. It is interesting to note that the effective EoS parameter and the density of dark matter, in both CP3 and CP4, are affected by the value of the dark energy density. Furthermore, the difference between the effective EoS parameter of CP4 and CP3 is proportional to the difference between their dark-matter densities, viz.,
\begin{align}
w_{\mathrm{eff.}}|_{\mathrm{CP4}} - w_{\mathrm{eff.}}|_{\mathrm{CP3}}
&= - w_{0} \left(y|_{\mathrm{CP4}} - y|_{\mathrm{CP3}}\right); \nonumber\\
\Longrightarrow \Delta w_{\mathrm{eff.}}|_{i \to f} &= -w_{0}\; \Delta y|_{i \to f}.
\label{GRB}
\end{align}
It means that the slop of the straight line between CP3 and CP4 in $w_{\mathrm{eff.}}-y$ plane is `$-w_{0}$'. Note that $w_{0}$ is the EoS of non-viscous dark energy (i.e. $w_{1}=0$) when one takes $f=\rho_{\mathrm{v}}(1+w_{0})$ and $G=w_{1}H^{2}$, thus the negative values for $w_{0}$ are desirable (i.e. $-w_{0}>0$). On the other hand, we know that $w_{\mathrm{eff.}}$ decreases with time and hence according to~(\ref{GRB}), $y$, which is proportional to the density of dark matter, must decrease with time as well. Therefore, if CP3$\neq$CP4 and both CPs lie on the evolution trajectory of the universe, then the density of the dark matter decreases while the density of the dark energy increases\footnote{Note that in the case CP3$\neq$CP4, we have $\Delta x=x|_{\mathrm{CP4}}-x|_{\mathrm{CP3}}=\left(\frac{-2}{6w_{0}}\right)\chi_{5}>0$. Indeed, $\chi_{5}$ is a square root term and in general it must be $\chi_{5} \geq 0$ to have a physical meaning. On the other hand, when CP3$\neq$CP4, then we have $\chi_{5}> 0$ (i.e. CP3$=$CP4 $\Leftrightarrow$ $x_{03} = x_{04}$ (or equivalently $\chi_{5}=0$)). Hence, we get $\Delta x >0$ implying an increase in the density of the dark energy.} as the universe ages which engenders the domination of the dark energy.\\

Furthermore, all CPs can be used to fix some relations between the densities of dark matter and dark energy. Note that all further (extra) CPs (i.e. CP2, CP3, and CP4) have $z=0$. This indicates that all these CPs are in the interaction plane where $z$ behaves like an independent variable and therefore it does not tend to an anomaly.

$\bullet$ If the type of interaction is shifted as
\begin{align}
\cos \left(\frac{h_{0}}{\kappa^{2}z} \right) \longrightarrow \cos \left(\frac{h_{0}}{\kappa^{2}(z-C)} \right),
\end{align}
then we get
\begin{align}
& \blacktriangleright \;\; \text{CP1:}\;\; w_{\mathrm{eff.}}|_{CP1}=-1, \nonumber \\
&x=\frac{3\zeta_{0}-3-w_{1}\kappa^{2}}{3w_{0}}, \quad y=1-\frac{3\zeta_{0}-3-w_{1}\kappa^{2}}{3w_{0}}, \nonumber \\
&z=\frac{h_{0}+C\kappa^{2}\pi -C\kappa^{2}\arccos \left(\frac{3\zeta_{0}+3\zeta_{0}w_{0}-\kappa^{2}w_{1}-3w_{0}-3}{qw_{0}} \right)}{\kappa^{2} \left(\pi - \arccos \left(\frac{3\zeta_{0}+3\zeta_{0}w_{0}-\kappa^{2}w_{1}-3w_{0}-3}{qw_{0}} \right) \right)}; \label{shm1}
\end{align}
\begin{align}
& \blacktriangleright \;\; \text{CP2:}\;\; w_{\mathrm{eff.}}|_{\mathrm{CP2}}=\frac{1}{3}, \nonumber \\
&x=\frac{w_{1}\kappa^{2}+q \cos \left(\frac{h_{0}}{\kappa^{2}C} \right)}{1-3w_{0}} \nonumber \\
&y=-q \cos \left(\frac{h_{0}}{\kappa^{2}C} \right), \quad z=0;
\end{align}
\begin{align}
&\blacktriangleright \;\; \text{CP3:}\;\; w_{\mathrm{eff.}}|_{\mathrm{CP3}}=w_{0}+\frac{w_{1}\kappa^{2}+q\cos \left( \frac{h_{0}}{\kappa^{2}C}\right)}{3x_{0}} \nonumber\\
&x=\frac{3\zeta_{0}+3w_{0}-w_{1}\kappa^{2}+\chi_{6}}{6w_{0}} \equiv x_{0}, \nonumber \\
&y=\frac{\kappa^{2}w_{1}-3\zeta_{0}}{3w_{0}}
-\frac{\kappa^{2}w_{1}+q\cos \left( \frac{h_{0}}{\kappa^{2}C}\right)}{3w_{0}x_{0}}, \quad z=0;
\end{align}
\begin{align}
&\blacktriangleright \;\; \text{CP4:}\;\; w_{\mathrm{eff.}}|_{\mathrm{CP4}}=w_{0}+\frac{w_{1}\kappa^{2}+q\cos \left( \frac{h_{0}}{\kappa^{2}C}\right)}{3x_{0}}\nonumber\\
&x=\frac{3\zeta_{0}+3w_{0}-w_{1}\kappa^{2}-\chi_{6}}{6w_{0}} \equiv x_{0}, \nonumber \\
&y=\frac{\kappa^{2}w_{1}-3\zeta_{0}}{3w_{0}}
-\frac{\kappa^{2}w_{1}+q\cos \left( \frac{h_{0}}{\kappa^{2}C}\right)}{3w_{0}x_{0}}, \quad z=0,
\end{align}
where $\chi_{6}=(\kappa^{4}w^{2}_{1}-6\kappa^{2}\zeta_{0}w_{1}+6\kappa^{2}
w_{0}w_{1}+9\zeta^{2}_{0}+18\zeta_{0}w_{0}+12qw_{0}\cos (h_{0}/(\kappa^{2} C))+9w^{2}_{0})^{1/2}$.\\
As seen, the number of CPs has been raised and some missed-out stages have been achieved in this development as well.\\
Interestingly, the CPs of this type of interaction are similar to the previous type. Hence, the previous discussions are valid here, even the relation (\ref{GRB}) between CP3 and CP4. As a final point, note that thanks to the existence of $h_{0}$ and $C$, CP1, i.e.~(\ref{shm1}), is equivalent to both (\ref{mog01}) and (\ref{mog02}). Indeed, one solution is common among different types of interacting models.\\

According to all examinations performed in this paper, we conclude that not only our proposed models are generative of CPs equivalent to that of ref.~\cite{od-asli} but also give rise to further CPs. Furthermore, as showed in `General case', our models yield CPs that are absent in ref.~\cite{od-asli}.\\
Therefore, the proposed extensions to interactions seem to be useful.\\

\noindent\pqrmybox{}{\subsubsection{A nice feature and benefit of the modified oscillating interactions \label{MAXX}}}\vspace{5mm}
As is clear from eq.~(\ref{cossol}), the interaction between dark matter and dark energy has an effect on radiation density in the radiation-dominated era.
Now, let us go to a neighboring point of this CP that may be achieved approximately as follows:\\
In the previous calculations, we found that in the coordinates of the CP of the radiation-dominated era we have $q$ instead of $q \cos (h_{0}H^{-2})$, but be sure that after passing this point, the cosine function will show itself whenever $q$ exists. Hence, this CP can be regarded as a neighboring CP of the previous CP which has been found for the \textit{radiation} era:
\begin{align}
x \approx & \frac{\kappa^{2} w_{1}+q \cos \left(h_{0}H^{-2} \right)}{1-3w_{0}}, \nonumber\\
y\approx &-q \cos \left(h_{0} H^{-2} \right) -3 \zeta_{0}, \nonumber\\
z\approx & \left(4(3w_{0}-1) \right)^{-1} \times \left(12w_{0}q \cos \left(h_{0}H^{-2} \right)+4\kappa^{2}w_{1} \right. \nonumber \\ & \left.+12\alpha w_{0}+27 \zeta_{0} w_{0} -4\alpha -9 \zeta_{0}+12w_{0}-4 \right).
\end{align}
These elements can be rewritten in the following forms:
\begin{align*}
&x\approx \left[\text{non-oscillating part}\right]+\underbrace{\left(\frac{1}{1-3w_{0}} \right)q \cos \left( h_{0}H^{-2}\right)}_{\text{oscillating part}}, \\
&y\approx \left[ \text{non-oscillating part} \right]+\underbrace{(-1)q \cos \left( h_{0}H^{-2}\right)}_{\text{oscillating part}}, \\
&z\approx \left[ \text{non-oscillating part} \right] +\underbrace{\left(\frac{-3}{1-3w_{0}}\right) q \cos \left( h_{0}H^{-2}\right)}_{\text{oscillating part}}.
\end{align*}
It is clear that the oscillating part has appeared in $x \propto \rho_{\mathrm{v}}$, $y  \propto \rho_{\mathrm{dm}}$, and $z \propto \rho_{\mathrm{r}}$. Existing this part in $x$ and $y$ is normal due to the assumed interaction between dark matter and dark energy. But, ``What are the reasons and benefits of the appearance of this term in $z$?'' To answer this question, first of all, let us single out the values of constant parameters so that the non-oscillating parts become equal among them. To this end, we take
\begin{align*}
&\alpha=\frac{29}{4}, \quad \zeta_{0}=-1, \quad \kappa^{2}=1, \quad
w_{0}=\frac{2}{9}, \quad w_{1}=1.
\end{align*}
These selections yield
\begin{align}
x\approx &+3+(+3)q \cos \left( h_{0}H^{-2}\right),\\
y\approx &+3+(-1)q \cos \left( h_{0}H^{-2}\right),\\
z\approx &+3+(-9)q \cos \left( h_{0}H^{-2}\right).
\end{align}
Therefore, for maintaining this CP with the above selections in the physical region, the oscillating term $q \cos \left( h_{0}H^{-2}\right)$ must be between $-1$ and $1/3$:
$$-1<q \cos \left( h_{0}H^{-2}\right)< \frac{1}{3}.$$
Remember that this CP belongs to the radiation-dominated era.
Now, note the amplitude of this oscillating term in $x$, $y$, and $z$. When the density of radiation becomes less than other components, then we exit the radiation-dominated era, hence, the interaction term, $q \cos \left( h_{0}H^{-2}\right)$, must be positive for reaching this goal. It means that the boundaries should be corrected as
\begin{equation*}
0<q \cos \left( h_{0}H^{-2}\right)<\frac{1}{3}.
\end{equation*}
These bounds clarify one of the benefits of restricting the arguments of interacting models to positive eras. Clearly, as the universe ages, the effect of this oscillating term on the reduction of the density of the radiation component is higher than dark matter because of the coefficients ($-9$ and $-1$). This guarantees exiting the radiation era. On the other hand, the density of dark energy raises with time under the assumed conditions, and it causes the late-time-accelerated expansion and the domination of the dark energy.\\
The same process may be performed for other types of modified interaction introduced in this paper.\\

\noindent\hrmybox{}{\section{Conclusion}}\vspace{5mm}
This paper was devoted to investigating a multi-fluid universe using the dynamical system approach in precise detail. The model of this study is a special and well-known form of \(F(T)\)-gravity in FRW background geometry.\\
First, a universe filled with radiation and a perfect fluid with a non-trivial EoS was studied based on the standard approach of dynamical systems.\\
Then, in sect.~\ref{sect-III}, by adding dark matter to the collection of density-set and assuming usual and well-known interaction between dark matter and dark energy, the system was considered. It is notable that in this study, the bulk viscous pressure of the dark matter fluid was taken into account according to the best fit of observational data. Despite that the interaction was assumed between dark matter and dark energy, the results demonstrate its effect on another component, radiation, as well (It should be noted that this \textit{direct effect} of interaction on all components, was observed in all interactions studied in this paper).
It was stated that excluding one CP which belongs to the radiation-dominated era, other CPs are of VLT-type and all these CPs are on the interaction plane (i.e. they belong to the accelerated era and can be used to fix some relations between the densities).\\
Next, in sect.~\ref{sect-IV}, by employing one extra dimension, $E=\kappa^{-2}H^{-2}$, and considering the effective equation of state parameter,  generalized forms of cosmological fluid were studied in two main parts: `A simplified form of EoS' and `More complicated forms of EoS'. In the former one, indeed, at a special limit, the case of sect.~\ref{sect-III} was recovered.
It was found that we cannot set the constant parameters so that all crucial stages of the universe evolution be given. We also generalized this case by adding baryonic matter to the density-collection and found that the problem still prevails. Also, it was noticed that the eigenvalue of the extra dimension $E$ is positive in all CPs. It refers to the universe expansion and reduction in the Hubble parameter overtime. Furthermore, it was found that the value of $E$ is zero in all CPs excluding those with EoS$=-1$. In fact, if one, according to the observational data, takes the current value of the effective EoS parameter equal to $-1.03$ or any value which is close to it (excluding $-1$), then $0$ will be the value we find for $E$.
Indeed, rather than the value of the variable itself, it is the eigenvalues that represent physical information.\\
In the latter one, a more complicated form of EoS of dark energy candidate, $p_{\mathrm{v}}=-\rho_{\mathrm{v}}+A \rho^{\alpha_{1}}_{\mathrm{v}}+B H^{2\beta}$, was considered. It was indicated that for different positive values of $\alpha_{1}$ and $\beta$, all CPs belong to the present time with EoS$=-1$. Furthermore, these models were also studied for $B=0$. The solutions of $B=0$ were better than $B \neq 0$ since they could cover more stages of the universe evolution. The big question in studying the complicated forms was the lack of other stages in case $B \neq 0$. It was illustrated that by switching to negative values of $\alpha_{1}$ in $B \neq 0$, the problem may be solved. In fact, this change of $\alpha_{1}$ value is equivalent to the shift from Van der Waals to Chaplygin gas. Therefore, more stages of the universe evolution coverage by Chaplygin gas make it a better candidate than Van der Waals to couple with the viscous part.
Furthermore, it was argued that the case in which $\alpha_{1}$ and $\beta$ are time-dependent variables can be taken into account.\\
Then, in sect.~\ref{sect-V}, some interesting models of interaction between dark matter and dark energy were studied. This consideration was done for two types. First, using the multiplier $H^{2j}$ the usual interaction of sect.~\ref{sect-III} was enhanced.
This type of interaction has been suggested in ref.~\cite{od-asli}.
Generally, without any assumption on $j$, studying this case is challenging. Hence, it was split into three main parts; $j=-1$, $j=0$, and $j=+1$. After studying these cases analytically and numerically, it was concluded that only for $j \leq 0$ there are analytical solutions, not for $j >0$, meaning that the power-law interactions in the related dynamical systems should be of increasing nature with time to get more critical points.
The reasons for this event, were discussed in the text. Finally, the cosine type of interaction, $\cos (h_{0}H^{+2})$, that has recently been proposed in ref.~\cite{od-asli} was examined. It was argued that, in the general case (i.e. without any approximation or assumption), there is no analytical solution for this type of interaction in our model of study. After indicating the root of this problem, four new types of interaction to rectify this problem were proposed. The suggested new types of interactions were $\cos (h_{0}H^{2}/(1+CH^{2}))$, $\sin (h_{0}H^{2}/(1+CH^{2}))$, $\cos (h_{0}H^{-2})$, and $\sin (h_{0}H^{-2})$. It was illustrated that these suggestions work fairly well in both $F(T)$ and Hilbert-Einstein models.
The differences and similarities between the four proposed models in this paper and the one suggested in ref.~\cite{od-asli} are as follows:\\
\textbf{Similarities:}
\begin{enumerate}
	\item The behaviors of all five models in asymptotic limits namely `early inflation' and `late-time-accelerated expansion' are the same and the results are equal. In fact, all five types of interaction are successful in describing both early inflation and late-time-accelerated expansion (i.e. the unification between early and late accelerations is achieved through these models).
	\item Thanks to the existence of the free constant parameters, it can be argued that if in a special gravity model, the interacting model of ref.~\cite{od-asli} gives some CPs, we can get the same solutions through our four models as well. More precisely, the solutions obtained by the model of ref.~\cite{od-asli} can be recovered by our four models as well besides some further solutions. Note that, the common solutions are not exactly the same, but they can be converted to each other by tuning the constant parameters.
\end{enumerate}
\textbf{Differences:}
\begin{enumerate}
	\item Unlike the interaction model of ref.~\cite{od-asli}, our four models in \(F(T)\)-gravity yield analytical solutions (i.e. CPs). Moreover, contrary to the model of ref.~\cite{od-asli}, which renders some CPs in some gravity models like Hilbert-Einstein model, ours not only presents the equivalent solutions but also some further CPs.
	\item These four models give more points that recover some serious stages of the universe evolution.
	\item Unlike the model of ref.~\cite{od-asli}, in two of our models namely $\cos (h_{0}H^{2}/(1+CH^{2}))$ and $\sin (h_{0}H^{2}/(1+CH^{2}))$ restricting $h_{0}$ to small values could be negated.
\end{enumerate}
Finally, utilizing an auxiliary approach, one of the motivations of restricting the oscillating interactions to positive domains as well as their crucial role in transitions between eras were clarified.\\

%%%%%%%%%%%%%%%%%%%%%%%%%%%%%%%%%%%%%%%%%%%%%%%%%%%%%%%%%%%%%
%%%%%%%%%%%%%%%%%%%%%%%%%%%%%%%%%%%%%%%%%%%%%%%%%%%%%%%%%%%%%%
%%%%%%%%%%%%%%%%%%%%%%%%%%%%%%%%%%%%%%%%%%%%%%%%%%%%%%%%%%%%%%
%\section*{\noindent\goldmybox{red}{\vspace{3mm} Acknowledgments \vspace{3mm}}}
%
%
%This work has been supported financially by Research Institute for Astronomy $\&$ Astrophysics of Maragha (RIAAM) under research project No. 1/5440-32.\\
%
%
\hrule \hrule \hrule \hrule \hrule \hrule


\begin{thebibliography}{134}
\vspace*{0.1cm}
\begin{tcolorbox}[colback=blue!4,colframe=white,breakable,
enhanced,width=\dimexpr\textwidth+0mm\relax,enlarge left by=-6.5mm, enlarge right by=-7mm,]
\bibitem{km2} S. Perlmutter et al., Nature \textbf{391}, 51 (1998).
\bibitem{km1} D.N. Spergel et al., Astrophys. J. Suppl. S. \textbf{170}, 377 (2007).
\bibitem{km4} B. Jain, A. Taylor, Phys. Rev. Lett. \textbf{91}, 141302 (2003).
\bibitem{km3} C.B. Netterfield, Astrophys. J. \textbf{571}, 604 (2002).
\bibitem{km6} S. Cole et al., Mon. Not. Roy. Astron. Soc. \textbf{362}, 505 (2005).
\bibitem{km5} D.J. Eisentein et al., Astrophys. J. \textbf{633}, 560 (2005).
\bi{beh-l-7} V. Sahni, A. Starobinsky, Int. J. Mod. Phy. D \textbf{9}, 373 (2000).
\bibitem{rref1} C. \'{O}Raifeartaigh, et al. EPJ~H \textbf{43}, 73 (2018).
\bi{beh-l-21} T. Padmanabhan, Phys. Rev. D \textbf{66}, 021301 (2002).
\bi{beh-l-18} B. Ratra, P.J.E. Peebles, Phys. Rev. D \textbf{37}, 3406 (1988).
\bi{beh-l-19} C. Wetterich, Nucl. Phys. B \textbf{302}, 668 (1988).
\bi{beh-l-14} B. Feng, X. Wang, X. Zhang, Phys. Lett. B \textbf{607}, 35, (2005).
\bi{beh-l-17} Z.K. Guo et al., Phys. Lett. B \textbf{608}, 177, (2005).
\bi{beh-l-10} R.R. Caldwell Phys. Lett. B \textbf{545}, 23, (2002).
\bi{beh-l-13} R.R. Caldwell, M. Kamionkowski, N.N. Weinberg, \textbf{91}, 071301 (2003).
\bi{beh-l-25} B. Tajahmad, Eur. Phys. J. C \textbf{77}, 510 (2017).
\bi{beh-l-70} A.K. Sanyal, Phys. Lett. B \textbf{524}, 177 (2002).
\bi{beh-l-71} B. Tajahmad, A.K. Sanyal, Eur. Phys. J. C \textbf{77}, 217 (2017).
\bi{beh-bns} B. Tajahmad, Eur. Phys. J. C \textbf{77}, 211 (2017).
\bi{beh-csss} B. Tajahmad, JHEP \textbf{02}, 084 (2020).
\bi{beh-j-62} T.D. Saini, S. Raychaudhury, V. Sahni, A.A. Starobinsky, Phys. Rev. Lett. \textbf{85}, 1162 (2000).
\bi{beh-j-77} A.Y. Kamenshchik, A. Tronconi, G. Venturi, Phys. Lett. B \textbf{702}, 191 (2011).
\bi{beh-j} B. Tajahmad, ``Reconstruction of \(F(T)\) gravity in homogeneous backgrounds'', (2018), arXiv: 1812.10339.
\bi{beh-e-1} S.N. Choudhury, A. Dasgupta, N. Banerjee, MNRAS \textbf{485}, 5693 (2019).
\bi{beh-e} B. Tajahmad, Eur. Phys. J. C \textbf{80}, 378 (2020).
\bi{beh-annals} B. Tajahmad, Annals Phys. \textbf{420}, 168253 (2020), DOI: 10.1016/j.aop.2020.168253, arXiv: 1812.03317.
\bi{mainref} Y.F. Cai et al., Rep. Prog. Phys. \textbf{79}, 106901 (2016).
\bi{beh-u-1} R. Myrzakulov, Eur. Phys. J. C \textbf{71}, 1752 (2011).
\bi{beh-u-2} K. Bamba, R. Myrzakulov, S. Nojiri, S.D. Odintsov, Phys. Rev. D \textbf{85}, 104036 (2012).
\bi{od-asli} S.D. Odintsov, V.K. Oikonomou, P.V. Tretyakov, Phys. Rev. D \textbf{96}, 044022 (2017).
\bi{faraoni} Y.F. Cai, et al. Reports on Progress in Physics, \textbf{79}, 106901 (2016).
\bi{od-50} S. Nojiri, S.D. Odintsov, Phys. Rev. D \textbf{72}, 023003 (2005).
\bi{od-51} K. Bamba, S. Capozziello, S. Nojiri, S.D. Odintsov, Astrophys. Space Sci. \textbf{342}, 155 (2012).
\bi{od-3} S. Nojiri, S.D. Odintsov, Phys. Rept. \textbf{505}, 59 (2011).
\bi{od-4} S. Nojiri, S.D. Odintsov, eConf C \textbf{0602061}, 06 (2006) [Int. J. Geom. Meth. Mod. Phys. \textbf{4}, 115 (2007)]
\bi{od-17} A. Avelino, Y. Leyva and L. A. Urena-Lopez, Phys. Rev. D \textbf{88}, 123004 (2013).
\bi{od-54} I.H. Brevik and O. Gorbunova, Gen. Rel. Grav. \textbf{37}, 2039 (2005).
\bi{od-55} I.H. Brevik, E. Elizalde, O. Gorbunova, A.V. Timoshkin, Eur. Phys. J. C \textbf{52}, 223 (2007).
\bi{od-56} B.D. Normann, I. Brevik, Entropy \textbf{18}, article 215 (2016).
\bi{od-57} J.D. Barrow, S. Hervik, Phys. Rev. D \textbf{74}, 124017 (2006).
\bi{p2018p} N. Aghanim et al. [Planck Collaboration], ``Planck 2018 results. VI. Cosmological parameters.'', (2018), arXiv: 1807.06209.
\end{tcolorbox}
\end{thebibliography}
\end{document}